\documentclass[12pt]{article}
\RequirePackage[authoryear]{natbib}
\RequirePackage{amsthm,amsmath,amsfonts,amssymb}

\usepackage[english]{babel}
\usepackage{dsfont}
\usepackage{mathdots}
\usepackage{graphicx}
\usepackage{enumerate}
\usepackage{natbib}
\usepackage{multirow}
\usepackage{xr}
\makeatletter

\usepackage{color}
\usepackage{float}
\usepackage{placeins}
\usepackage{url}
\usepackage{comment}
\usepackage{todonotes}
\usepackage{multirow}
\usepackage{diagbox}
\usepackage{mathtools}
\usepackage{ifthen}

\newcommand{\blind}{1}

\newtheorem{thm}{{Theorem}}
\newtheorem{lem}[thm]{{Lemma}}
\newtheorem{rmk}{{Remark}}
\newtheorem{asp}{{Assumption}}

\newtheorem{corollary}[thm]{{Corollary}}

\DeclarePairedDelimiter\floor{\lfloor}{\rfloor}

\newcommand{\var}{\mbox{Var}}
\newcommand{\cov}{\mbox{Cov}}
\newcommand{\cor}{\mbox{Corr}}
\newcommand{\cum}{\mbox{cum}}

\def\eps{\varepsilon}

\newcommand{\gps}[1]{g(#1,p,n,M,\tau)}

\newcommand{\gpbetade}[1]{g_{\beta^{(de)}}(#1,p,n,M,\tau,\spar)}
\newcommand{\gpbias}{g_{Bias}}

\newcommand{\bp}{\widetilde O_p}

\newcommand{\flw}{\widehat f_M}
\newcommand{\fsp}{\widehat f_{M}}
\newcommand{\fspi}[1]{\widehat f_{M;#1}}
\newcommand{\fspc}{\widehat f_{M,c}}

\newcommand{\R}{\mathds{R}}
\newcommand{\C}{\mathds{C}}

\newcommand{\N}{\mathds{N}}
\newcommand{\Z}{\mathds{Z}}
\newcommand{\spar}{s(p)}

\newcommand{\ind}{\mathds{1}}

\newcommand{\fracd}[2]{#1/#2}

\newcommand{\methoda}{\emph{Testing}}
\newcommand{\methodb}{\emph{Regularizing}}
\renewcommand{\Re}{\operatorname{Re}}
\renewcommand{\Im}{\operatorname{Im}}

\newcommand{\Nkernel}{2}

\allowdisplaybreaks
\begin{document}

\def\spacingset#1{\renewcommand{\baselinestretch}%
{#1}\small\normalsize} \spacingset{1}


\if1\blind
{
  \title{\bf Frequency Domain Statistical Inference for High-Dimensional Time Series}
  \author{Jonas Krampe\thanks{{\bf Acknowledgments.} The authors are grateful to the editor, an associate editor and three referees for their valuable and insightful comments that led to an improved manuscript. The research of the first author was supported by the Research Center (SFB) 884 ``Political Economy of Reforms''(Project B6), funded by the German Research Foundation (DFG) and National Institute of Health (grants R01GM135926 and R21NS120227). Furthermore, the first author acknowledges support by the state of Baden-W{\"u}rttemberg through bwHPC. The research of the second author has been supported by a University of Cyprus research grant.}
    \hspace{.2cm}\\
    Cornell University\\
    and \\
    Efstathios  Paparoditis\\
    University of Cyprus 
    }
   
  \maketitle
} \fi

\if0\blind
{
  \bigskip
\thanks{{\bf Acknowledgments.} The authors are grateful to the editor, an associate editor and three referees for their valuable and insightful comments that led to an improved manuscript.}
  \bigskip
  \bigskip
  \begin{center}
    {\LARGE\bf Frequency Domain Statistical Inference for High-Dimensional Time Series}
\end{center}
  \medskip
} \fi

\bigskip
\begin{abstract}
Analyzing time series in the  frequency domain  enables the development of   powerful  tools 
for investigating the second-order characteristics of multivariate  processes. Parameters like the spectral density matrix and its inverse, the coherence or the partial coherence,  encode comprehensively the complex linear relations between the component processes of the multivariate system.  In this paper, we develop inference procedures for such parameters in a high-dimensional, time series setup. Towards this goal, we first focus on the derivation of consistent estimators of the coherence and, more importantly,  of the partial coherence which possess manageable limiting distributions that are suitable for testing purposes. Statistical tests of the hypothesis that the maximum over frequencies of the coherence, respectively, of the partial coherence,  do not exceed a prespecified threshold value are developed. Our approach allows for testing hypotheses for individual coherences and/or partial coherences as well as for multiple testing of large sets of such parameters. In the latter case, a consistent procedure to control the false discovery rate is developed.  
The finite sample performance of the inference procedures introduced 
is investigated by means of  simulations and 
applications to the construction of graphical interaction models for
brain connectivity based on EEG data are presented.
\end{abstract}

\noindent%
\vfill

\newpage
\spacingset{1.9} 


\section{Introduction}
Spectral analysis refers to  a number of powerful tools for analyzing the second-order properties of multiple time series. Parameters like the coherence or the partial coherence, describe comprehensively the linear relations between the components of the vector time series by taking into account all lead and lag relations as well as the distinction between direct and indirect effects. Coherence and partial coherence are process parameters that can be expressed as functions of the spectral density, respectively, of the inverse spectral density matrix; see among other the classical textbooks to multivariate time series analysis by  \cite{hannan1970multiple}, \cite{koopmans1995spectral} and \cite{brillinger2001time}.
More specifically, partial coherence, which is a measure of the strength of the linear relations between two time series after eliminating the indirect linear effects caused by all 
other time series of the system, plays a crucial role, for instance, in extending the concept of graphical models to time dependent data; see  \cite{dahlhaus2000graphical} and \cite{eichler2012graphical}. In graphical models, each time series represents a vertex of a network while edges between two vertices describe conditional (on all other time series) linear dependence of the corresponding time series. If the underlying process is  Gaussian,  then a zero partial coherence even reflects conditional independence between the pair of  time series considered. An edge between two vertices exists if the partial coherence is different from zero. The corresponding network describes then a Gaussian graphical model for the multivariate times series at hand. Clearly, if the underlying distribution is not specified or if it is non-Gaussian, then such statements about independence cannot be made. However, 
even in such cases, visualizing the linear dependence structure of the multivariate time series system can be helpful in providing valuable information about existing    relations. 
Analyzing  linear dependencies is a useful tool in many areas of applied research like, for instance, finance, \citep{gray2014central}, signal processing, \citep{bach2004learning}, or  medicine, where  a rich literature exits devoted to  investigations of brain connectivity problems and spectral analysis of EEG and fMRI data; see among others
\cite{medkour2009graphical}, 
\cite{
fiecas2016modeling} and \cite{schneider2016p
}. 

When the dimension of the time series   is small, statistical inference for frequency-domain parameters like coherence or partial coherence is a well-developed area of multiple time series analysis; see among others,   \cite{hannan1970multiple}, \cite{koopmans1995spectral} and \cite{brillinger2001time}. Testing hypothesis about parameters of the spectral density matrix or of its inverse, also have been considered in the literature; 
see
\cite{eichler2008testing} and  \cite{dette2009bootstrapping}.
\cite{schneider2016p} investigated various combinations of p-values for testing partial coherences and considered their applicability to EEG data. However, when the dimension of the time series is moderate or large compared to the sample size, one needs to somehow restrict dependencies between the component processes in order to make statistical inference possible. 
For this, several (not necessarily exclusive) approaches exist in the literature. 

One approach is to impose sparsity assumptions directly in the frequency domain and  to restrict the number of nonzero elements of the spectral density or of its  inverse. 
In such a sparsity context, \cite{sun2018large} considered estimators of large spectral density matrices, while \cite{fiecas2019spectral}, obtained non-asymptotic results for smoothed (kernel-type) estimators. \cite{zhang2020convergence} established convergence rates of nonparametric estimators  also allowing for a time-varying autocovariance structure. \cite{rosuel2021asymptotic} considered high-dimensional Gaussian time series, 
while 
\cite{jung2015graphical}, \cite{schneider2016partial} and \cite{tugnait2021sparse}, used shrinkage methods to estimate the inverse spectral density matrix and to construct Gaussian graphical models.

An alternative to sparsity  is to use high-dimensional time series models for estimating the spectral density matrix, respectively,  its inverse. 
Factor models or other types of low rank models,  \citep{barigozzi2021algebraic}; sparse vector autoregressive (VAR) models \citep{krampe2020Est}; as well as combinations thereof, \citep{krampe2021dynamic}, have been used in this context. 
Notice  that a sparse inverse spectral density matrix does not necessarily imply a sparse spectral density matrix,  a sparse autocovariance structure, or more general, a sparse time-domain representation of the underlying high-dimensional process. The reverse argument also is true. Hence, depending on the particular structure of the high-dimensional time series  at hand,  different approaches may be used.

This paper develops statistical inference procedures for coherences and partial coherences for  high-dimensional time series  under mild assumptions on the underlying process and focuses on  applications of these procedures in the discovery and identification of dependence structures as well as in the  construction of graphical models. 
Towards this goal, 
we first develop consistent, nonparametric estimators of the parameters of interest, which possess manageable limiting distributions suitable for inferring properties of their population counterparts. Especially for partial coherences, the development of such estimators in the high-dimensional context is challenging and much more involved than in the finite-dimensional setup. This is mainly due to difficulties in deriving    distributional results for  regularized estimators of these parameters.
To overcome these difficulties, we develop so-called de-biased estimators of  partial coherences that use appropriate, regularized regression-type estimators which involve the finite Fourier transform of the vector time series at hand.
Notice that in contexts and for inference problems different to those considered in this paper, de-biased or de-sparsified estimators are useful tools for developing  statistical inference in a high-dimensional set-up; see  \cite{javanmard2014confidence}, \cite{deGeer2014} and  \cite{zhang2014}.

After introducing the estimators of coherence and more importantly of partial coherence used,
the focus  is directed towards the development of powerful testing procedures for the null hypothesis that, within a frequency band of interest,  the aforementioned frequency domain parameters do not exceed some user-specified threshold value. 
For this, a max-type test statistic is introduced which evaluates coherences and partial coherences over a (with sample size increasing) number of frequencies within the frequency band of interest. The testing
procedure developed allows for testing hypotheses for a single pair of component time series as well as for multiple testing, i.e., for testing a large set of such hypotheses. For the latter case, a procedure to control the false discovery rate (FDR) is proposed and theoretically justified. The procedure is based on  a  screening of the test statistics  using an appropriate threshold  and it  adapts to the  high-dimensional time series setup,  thresholding procedures for FDR control proposed for the construction of graphical models for  Gaussian,   i.i.d. data;  see  \cite{liu2013gaussian},   \cite{cai2016large}, and \cite{li2023transfer}.  As already mentioned, the construction of graphical models for high-dimensional time series proposed, uses a thresholded, max-type testing procedure   to determine whether the  partial coherences exceed a pre-specified  value over a frequency band of interest.
 Notice that alternatives to such a testing  approach   exist in the literature. In particular, \cite{jung2015graphical} and \cite{tugnait2021sparse} considered regularization based procedures to directly estimate graphical models.  Testing approaches have, however,  two main advantages. First, they offer  a  greater flexibility due to  the fact that they allow for different   estimators of the (inverse) spectral density matrix used.
 Second, the tuning parameters involved in the direct estimation of graphical models when regularizing procedures are used,
 have a great impact on the number of edges and therefore on the dependencies discovered. Although several data-adaptive approaches to select such tuning parameters exist,
 it is not clear 
 how these parameters should be chosen in order to achieve a desired level of FDR control. 
 We refer to  Section 4.2  for more detailed comparisons. 

The remaining of the paper is organized as follows. Section 2 discusses some useful preliminary concepts and  presents  the estimators of the coherence and of the partial coherence 
used.
 Section 3 deals with testing  
 and focuses on  testing single as well as multiple hypotheses for partial coherences.  The procedure used for FDR  control  is also presented  and theoretically justified in this section. Section 4 investigates via simulations  the finite sample performance  of the testing procedure proposed,   while Section 5 presents the construction of a graphical model for brain connectivity. Section 6 is devoted to asymptotic considerations, states the technical assumptions needed  and derives the limiting distributions of the estimators introduced in Section 2. Auxiliary lemmas as well as technical proofs are deferred 
 to the supplementary material of this paper.   The algorithm is implemented in the R-package \emph{HDSpectralAnalysis} which is
available to download at \url{https://github.com/JKrampe/HDSpectralAnalysis.}

{\bf Convention.} \ Throughout the paper the following  notation is used. For a vector $x\in \R^p$, $ \| x \|_1 = \sum_{j=1}^p |x_j|$, $\| x \|_2^2 = \sum_{j=1}^p |x_j|^2$ and $\|x\|_\infty=\max_j |x_j|$.  For a $r\times s$ matrix $B$ with elements $b_{i,j}$, $i=1,\ldots,r$ and $ j=1,\ldots,s$,  $\|B\|_1=\max_{1\leq j\leq s}\sum_{i=1}^r|b_{i,j}|=\max_j \| B e_j\|_1$, 
$\|B\|_\infty=\max_{1\leq i\leq r}\sum_{j=1}^s|b_{i,j}|=\max_{i} \| e_i^\top B\|_1$ and $\|B\|_{\max}=\max_{i,j} |e_i^\top B e_j|$, where $e_j$
is   the unit  vector of appropriate dimension with the $1$ appearing  in the $j$th position. The largest absolute eigenvalue of a square matrix $B$ is denoted by $\lambda_{\max}(B)$, the smallest  by $\lambda_{\min}(B)$,  while   $\|B\|_2^2=\lambda_{\max}(BB^\top)$.
For the $p$-dimensional identity matrix $I_p$,  
$I_{p,-J}\in \R^{p\times (p-|J|)}$ denotes the  matrix obtained after deleting  the  columns $j \in J$ of $I_p$, where $J\neq\emptyset$ and  $J \subset \{1,2, \ldots, p\}$. 
$A_{u,\cdot}$ denotes the  $u$th row of  a matrix $A$ while  $A_{\cdot,u}$ the $u$th column.   $ x_v$ stays for the $v$th element of a vector $x$ while for  a matrix  $A \in \C^{r\times s}$,  $\Re(A)$ denotes its   real,  $\Im(A)$ its  imaginary part,  $A^{(C)}$   the complex conjugate  and  $A^{H}=(A^{(C)})^\top$  the complex conjugate and transpose. Finally, for a complex random vector  $X$, $Var(X)=Cov(X,X^{H})$, where $X^{H}$ denotes the conjugate transpose of $X$.  

\vspace*{-0.5cm}
\section{Estimation of Frequency Domain Parameters}
\label{sec.2}
Consider  a $p$-dimensional, zero mean  stochastic process  $ \{X_t,t\in\Z\}$,  where for each $t\in \Z$,  the random vector $X_t=(X_{j,t}, j=1,2, \ldots, p)^{\top}$, is generated as
\begin{equation}\label{eq.X}
X_t=R(\eps_t,\eps_{t-1}, \ldots).
\end{equation}
Here $ R :\R^{\tilde p\times\infty} \rightarrow \R^p$ is some measurable function and $\{\eps_t,t\in\Z\}$ a $\tilde  p$-dimensional sequence of independent and identically distributed (i.i.d.) random vectors with mean zero and covariance matrix $ \Sigma_\eps$.  $\tilde p=p$ is possible but not necessary. Denote by  $ \Gamma(u)=E(X_{u}X_{0}^\top)$, $ u \in \Z$,  the lag $u$ autocovariance matrix  and assume that $\{X_t,t\in\Z \} $ possesses a 
spectral density matrix denoted by $ f(\omega)$ with inverse 
 $ f^{-1}(\omega)$ which exists for all frequencies $\omega\in \R$.  

For an observed stretch 
$ X_1, X_2, \ldots, X_n $ stemming from $\{X_t,t\in \Z\}$,   let  $\widehat{\Gamma}(u)$ be   the  estimator of $ \Gamma(u)$ given by 
$ \widehat{\Gamma}(u)=n^{-1}\sum_{t=\max\{1,1-u\}}^{min\{n,n-u\}}X_{t+u}X_t^\top$ and      $ Z_n(\omega)$, $\omega \in [0,2\pi]$,   the discrete finite Fourier transform, 
$ Z_n(\omega) = (2\pi n)^{-1/2}\sum_{t=1}^n X_t \exp\{-i \omega t\}$.
 Let    $\Sigma_n(\omega) = {\rm Var}(Z_n(\omega))$ which equals  
\begin{equation} \label{eq.fFt} \Sigma_n(\omega)= \frac{1}{2\pi} \sum_{k=-n+1}^{n-1}\big( 1-|k|/n\big)\Gamma(k) \exp\{-i\omega k\}. 
\end{equation}

Using the estimators $ \widehat{\Gamma}(u)$,  a   lag window estimator of $f(\omega)$ is obtained as
\begin{align} \label{eq.lag.window}
\flw(\omega)= \frac{1}{2\pi}  \sum_{u=-n+1}^{n-1} K(u/M) \hat \Gamma(u) \exp(-iu \omega),    
\end{align}
where $M$ is  a truncation lag that  determines the number of sample autocovariances effectively taken into account in  estimating $f(\omega)$.  $ K(\cdot)$ is a lag-window kernel which satisfies certain conditions to be specified later on. Recall that the lag-window estimator $ \flw(\omega)$ also  has  a (discrete) smoothed periodogram analogue given by 
\begin{equation}\label{eq.f-smooth-per}
 \flw(\omega)=\frac{M}{n}  \sum_{k=1}^{n} \kappa_{M}(\omega-\omega_k)  Z(\omega_k)Z^H(\omega_k),
 \end{equation}
  where  $\kappa_M(\cdot)=1/M \sum_{u=-n+1}^{n-1} K(u/M) \exp(-i  u \cdot )$ is  the  discrete Fourier transform  of $K$ and $ \omega_k=2\pi k/n$, $ k=1,2, \ldots, n$ are  Fourier frequencies.


\subsection{Estimation of  Coherence and  Partial Coherence} \label{sec.coh.partial.coh}
Recall the  coherence $ \sigma_{u,v}(\omega)$  between two components $u$ and $v$ of the vector process $\{X_t,t\in\Z\}$ which for any frequency $\omega \in [0, 2\pi]$,  is  given by
\begin{align} \label{eq.coherence}
\sigma_{u,v}(\omega)=|s_{u,v}(\omega)|,  \ \  \mbox{where} \ \ s_{u,v}(\omega)= f_{u,v}(\omega)/\sqrt{f_{u,u}(\omega) f_{v,v}(\omega)}. \end{align}
Here,  $f_{r,s}(\omega)$ denotes the $(r,s)$th element of the spectral density matrix $f(\omega)$. Analogous  to the partial correlation in the i.i.d. context, the partial coherence $R_{u,v}(\omega) $  at any frequency $\omega$ between  any two processes  $ \{X_{u,t}\}$ and $ \{X_{v,t}\}$, describes the direct linear relation between these components, that is, their cross-correlation  structure, after taking into account  the  linear effects due to all   other  components of the process.  Using the   inverse  spectral density matrix $f^{-1}$, a useful expression of the  partial coherence 
is given by
\begin{align} \label{eq.partial.coherence}
    R_{u,v}(\omega)=|\rho_{u,v}(\omega)|, \ \ \mbox{where} \ \ \rho_{u,v}(\omega)=-f_{u,v}^{-1}(\omega)/\sqrt{f_{u,u}^{-1}(\omega)f_{v,v}^{-1}(\omega)};
\end{align}
 see \cite{dahlhaus2000graphical}, where  $ f^{-1}_{r,s}(\omega)$ denotes the $(r,s)$th element of $f^{-1}(\omega)$. 
 
Complex coherence,  $ s_{u,v}$, and complex partial coherence, $\rho_{u,v}$, respectively,  can be approximated  using the correlation and partial correlation of the discrete Fourier transform $Z_n$.
More specifically, we have 
\begin{align} \label{eq.coherence.approx}
\cor(Z_{n,u}(\omega),Z_{n,v}(\omega))& =
\Sigma_{n;u,v}(\omega)/\sqrt{\Sigma_{n;v,v}(\omega)\Sigma_{n;u,u}(\omega)}=s_{u,v}(\omega)+O(1/n), 
\end{align}
where $ \Sigma_{n;r,s}(\omega)$ denotes the $(r,s)$th element of the matrix  $ \Sigma_n(\omega)$; see (\ref{eq.fFt}). 
An analogous   expression for  $ \rho_{u,v}(\omega)$ also can be derived. 
For this, consider for $ v\in\{1,2, \ldots, p\}$ the   regression of the component $ Z_{n,v}(\omega)$ of the vector $Z_n(\omega)$ on all other components of the same vector, that is,  the regression,    
\begin{equation} \label{eq.Regr}
Z_{n,v}(\omega)=\beta_v^H(\omega) Z_{n,-v}(\omega)+E_{u,v}(\omega).
\end{equation} 
Notice that 
$ E_{u,v}(\omega)=\Sigma_{n;v,\cdot}^{-1}(\omega) Z_n(\omega)/\Sigma_{n;v,v}^{-1}(\omega)$, while  
$
\beta_v(\omega)= -\Sigma_{n;-v,v}^{-1}(\omega)/\Sigma_{n;v,v}^{-1}(\omega),$
where $ \Sigma^{-1}_{n;r,s}(\omega)$ denotes the $(r,s)$th element of the inverse matrix  $ \Sigma_n^{-1}(\omega)$. 
For $u\not =v$, we get, 
\begin{align} \label{eq.partial.coherence.approx}
e_u^\top I_{p,-v} \beta_v(\omega) \sqrt{\frac{\Sigma_{n;v,v}^{-1}(\omega)}{\Sigma_{n;u,u}^{-1}(\omega)}}=-\frac{\Sigma_{n;u,v}^{-1}(\omega)}{\sqrt{\Sigma_{n;v,v}^{-1}(\omega)\Sigma_{n;u,u}^{-1}(\omega)}}=\rho_{u,v}(\omega)+O(1/n).
\end{align}
Expressions  (\ref{eq.coherence.approx}) and (\ref{eq.partial.coherence.approx}) suggest that estimators of the functions  $s_{u,v}(\cdot) $ and $\rho_{u,v}(\cdot) $  can be obtained by  
replacing the  matrices  $ \Sigma_n$ and  $\Sigma_n^{-1}$ appearing in these expressions,   by  appropriate sample estimators. Moreover,   to ensure consistency, smoothing over  frequencies is necessary. 

Estimation of the complex coherence is straightforward. To elaborate,  from  (\ref{eq.coherence.approx}) and using the estimator $ \fsp$ given in (\ref{eq.f-smooth-per}),   the  following, commonly used, estimator,  \ 
$
    \hat s_{u,v}(\omega)=
     \fracd{\fspi{u,v}(\omega)}{\sqrt{\fspi{v,v}(\omega) \fspi{u,u}(\omega)}}
$
of $s_{u,v}(\omega)$, is obtained. 

In contrast to coherence,  the deviation of an estimator for the complex  partial coherence $\rho_{u,v}(\omega)$,  
 is   more involved. This is due to the fact  that   some form of regularization is required in order to obtain a consistent  estimator of the inverse spectral density matrix   $ f^{-1}$, respectively, of the inverse matrix $ \Sigma_n^{-1}$ appearing   in (\ref{eq.partial.coherence.approx}). However,  
regularization-based estimators have largely unknown distributional properties which makes the derivation of distributional results  for   plug-in type estimators of  partial coherences a difficult task. Moreover, the use of  regularization  introduces  some  additional  bias  in estimating the  coefficients $\beta_v(\omega), v=1,\dots,p$, which  affects the distributional properties of the corresponding estimator and calls for  a bias correction. We stress the fact that for our purposes, it  
is important for  the estimators 
used  to       possess    manageable limiting distributions which  are suitable  for inference  and more specifically,  for properly implementing statistical tests for the corresponding population parameters.  Taking  the aforementioned   distributional requirements   into account,  we introduce  in the following an  estimator of  $\rho_{u,v}$ which   builds upon a bias correction of a regularized estimator of the component $\beta_{v,\tilde u}(\omega)
=e_u^\top I_{p,-v} \beta_v(\omega)$ appearing in (\ref{eq.partial.coherence.approx}) and which posses a Gaussian limiting distribution. 

To introduce this estimator,   we rewrite \eqref{eq.Regr} as  \begin{equation} \label{eq.Z}
Z_{n,v}(\omega)=\beta_{v,\tilde u}^{(C)}(\omega)Z_{n,u}(\omega)+\beta_{v,-\tilde u}^H(\omega) Z_{n,-(v,u)}(\omega)+E_{n,v}(\omega),
\end{equation}
and  we  focus on estimation of the parameter $\beta_{v,\tilde u}(\omega)$. The idea now is to obtain a  rotated version of $Z_{n,u}(\omega)$, say $\widetilde Z_{n,u}(\omega)=\gamma_{-v,\tilde u}^H(\omega) Z_{n,-v}(\omega)$,  so that $\cov(\widetilde Z_{n,u}(\omega),Z_{n,-(v,u)}(\omega))=0$ and $\cov(\widetilde Z_{n,u}(\omega),Z_{n,u}(\omega)) \neq 0$. If such a rotation is possible, then    the parameter $\beta_{v,\tilde u}(\omega)$ of interest  can  be expressed  as   $\beta_{v,\tilde u}(\omega)=\cov(\tilde Z_{n,u}(\omega),Z_{n,v}(\omega))/\cov(\tilde Z_{n,u}(\omega),Z_{n,u}(\omega))$. Notice that from  expression (\ref{eq.Z}) we get,  
\begin{align} \label{eq.cov-expr}
    \cov(\tilde Z_{n,u}(\omega),Z_{n,v}(\omega)) &= 
    \beta_{v,\tilde u}(\omega)\cov(\widetilde Z_{n,u}(\omega),Z_{n,u}(\omega)) \nonumber \\
    & \ \ +\beta_{v,-\tilde u}(\omega) \cov(\widetilde Z_{n,u}(\omega),Z_{n,-(v,u)}(\omega))+\cov(\widetilde Z_{n,u}(\omega),E_{n,v}(\omega)).
\end{align}
To implement  (\ref{eq.cov-expr}), an  estimator of the rotation $ \widetilde{Z}_{n,u}(\omega)$  is required. Given such an estimator,  the covariance $\cov(\widetilde Z_{n,u}(\omega),Z_{n,u}(\omega)) $  can then be  replaced by a sample version, which can be  obtained via   kernel smoothing over the frequencies using a  kernel $K$ and a bandwidth $M$. However, in high dimensions an additional problem appears. This is due to the fact that  the sample version of the covariance 
$\cov(\tilde Z_{n,u}(\omega),Z_{n,-(v,u)}(\omega))=\gamma_{-v,\tilde u}^H(\omega)\cov(Z_{n,-v}(\omega),Z_{n,-(v,u)}(\omega))$  is with probability one not exactly zero as this is the case for its theoretical counterpart. Notice that  the latter  covariance 
is obtained as  a linear combination of a high-dimensional vector and a matrix, and 
the sample version of this covariance will be of low-rank in the high-dimensional set-up.
This means that  in the high dimensional set-up it is not possible to come up with an estimated  rotation $\widetilde{Z}_{n,u}(\omega)$ which exactly  fulfills the desired orthogonality requirements. Therefore, 
the sample version of the term $\beta_{v,-\tilde u}(\omega) \cov(\tilde Z_{n,u}(\omega),Z_{n,-(v,u)}(\omega))$ appearing in (\ref{eq.cov-expr}) cannot be considered as zero and will be  treated, in what follows,  as a bias term. Fortunately, this bias term can  be (consistently) estimated 
by  using a (consistent)  estimator of $ \beta_{v,-\tilde u}(\omega)$
and can, therefore,  be  appropriately  taken into account in constructing the final estimator of  $\beta_{v,\tilde u}(\omega) $.

Regarding the estimations of the inverse $ f^{-1}$, notice that if sparsity assumptions are imposed,  known procedures for covariance matrix estimation such as graphical lasso or CLIME (see \cite{cai2016estimating_overview} for an overview) 
can  be applied to obtain  consistent estimators of $f^{-1}$ 
 that  satisfy  Assumption \ref{asp.spectral.density.estimator} stated in Section~\ref{sec.2.properties}. Using as starting point   the lag-window estimator $\fsp$, such estimators  have 
 been investigated by several authors;  see among others, \cite{sun2018large,fiecas2019spectral,zhang2020convergence,jung2015graphical} and \cite{tugnait2021sparse}. The last two papers  also  consider regularization procedures applied not only locally, i.e., to   each frequency,   but also globally and  group-wise, that is, applied  to  all frequencies in the interval $[0,2\pi]$. Such a global regularization approach makes use of smoothness  assumptions imposed on the  spectral density matrix and  can be beneficial.   In  our approach we use a   CLIME type  estimator as in \cite{fiecas2019spectral} which is  discussed  in great detail in section~\ref{sec.est.inv.f} of the supplementary material, where we also show   
 how such an estimator fulfills the assumptions required in our paper. Notice that another alternative to graphical lasso/CLIME will be one which uses a semiparametric  approach to  estimate  $f^{-1}$ by  using  sparse vector autoregressive models;  we refer here to  \cite{krampe2020Est} and \cite{krampe2021dynamic} for such an approach. In the following, we denote by $\hat f^{-1}$ an estimator of $f^{-1}$ which uses some form of regularization.

Taking  the considerations made  so far into account and using the  estimators
$
\hat \gamma_{-v,\tilde u}^H (\omega)=\hat f_{v,v}^{-1}(\omega)\hat f_{u,-v}^{-1}(\omega)- \hat f_{u,v}^{-1}(\omega) \hat f_{v,-v}^{-1}(\omega) 
$
and 
$
\hat \beta_{v}^H(\omega)=\hat f^{-1}_{v,-v}(\omega)/\hat f_{v,v}^{-1}(\omega),
$
for $\gamma_{-v,\tilde u} (\omega)$ and $\beta_{v}(\omega) $, respectively, the following  de-biased estimator 
$\hat \beta_{v,\tilde u}^{(de)}(\omega)$ of $ \beta_{v,\tilde u}(\omega)$, is proposed.   
\begin{align} \label{eq.beta.debiased}
    \hat \beta_{v,\tilde u}^{(de)}(\omega)=\hat \beta_{v,\tilde u}(\omega)+\frac{\sum_{k=1}^n \kappa_M(\omega-\omega_k)[Z_{n,v}(\omega_k)-\hat \beta_v^H(\omega) Z_{n,-v}(\omega_k)]Z_{n,-v}^{H}(\omega_k) \hat \gamma_{-v,\tilde u}(\omega)}{\sum_{k=1}^n \kappa_M(\omega-\omega_k) Z_{n,u}(\omega_k) Z_{n,-v}^{H}(\omega_k) \hat \gamma_{-v,\tilde u}(\omega)}.
\end{align}
Observe that $\hat \beta_{v,\tilde u}^{(de)}(\omega) $  in 
\eqref{eq.beta.debiased} is constructed in such a way that only an estimator of $f^{-1}$ at frequency $\omega$ is used. 
To proceed with  the construction of a de-biased  estimator for $\rho_{u,v}(\omega) $,   we use   the  estimators $\widehat \beta_{v,\tilde u}^{(de)}(\omega)$ and $\widehat \beta_{\tilde v, u}^{(de)}(\omega) $, where the latter estimator is obtained  as $\hat \beta_{v,\tilde u}^{(de)}(\omega) $ in (\ref{eq.beta.debiased}) but by replacing $v $ and $\widetilde{u} $ by $\widetilde{v} $ and $u $, respectively. Given these two estimators,  the estimator  of the complex partial coherence  proposed,  is finally given by  
\begin{align}
   \hat \rho^{(de)}_{u,v}(\omega):=\frac{1}{2}\Big(\hat \beta_{u,\tilde v}^{(de)}(\omega) \sqrt{\frac{\hat f_{u,u}^{-1}(\omega)}{\hat f_{v,v}^{-1}(\omega)}}+\hat \beta_{v,\tilde u}^{(de)}(\omega)^{(C)} \sqrt{\frac{\hat f_{v,v}^{-1}(\omega)}{\hat f_{u,u}^{-1}(\omega)}}\Big).
\end{align}
Note  the appropriate rescaling of  $\widehat \beta_{v,\tilde u}^{(de)}(\omega)$  and $\widehat \beta_{u,\tilde v}^{(de)}(\omega) $  as well as 
the fact that  $\widehat \rho^{(de)}_{u,v}(\omega) $ fulfills the property 
$\widehat{\rho}^{(de)}_{u,v}(\omega)=\big(\widehat{\rho}^{(de)}_{v,u}(\omega)\big)^{(C)}$,  which also  is fulfilled    by 
its population counterpart  $\rho_{u,v}(\omega) $. We postpone the discussion of the asymptotic properties   of the estimators derived to the supplementary material and their use for testing hypotheses to  Section~\ref{sec.2.properties} of the paper.

\section{Testing} \label{sec.testing}
We now focus on    the problem of testing hypotheses about the
 frequency domain parameters considered so far and which  build upon the estimators  derived in  Section~\ref{sec.2}. For brevity of presentation, we only concentrate on the more involved case, that is on the case of testing hypotheses about the partial coherences for  frequencies belonging to a frequency  band   $\mathcal{W}\subset [0,\pi]$ of interest. Applications of the procedure presented in this section to testing hypothesis for  a particular frequency $\omega$ only, or  for  the  population  coherences,  are straightforward and will be omitted.

\subsection{Testing Single Hypothesis}
We begin our discussion  with  the problem of testing a single hypothesis, that is of testing that, within a frequency band $\mathcal{W}\subset[0,\pi]$ of interest,  the values of  a particular partial coherence  are 
below some desired  level for all frequencies belonging to ${\mathcal W}$.
 More specifically,  let ${\mathcal W}$ be a subset of $ [0,\pi]$ having positive Lebesgue measure, $ \delta \in [0,1)$ be a user specified threshold and     $ (u,v)\in \{ (u,v)| \, 1 \leq u <v \leq p\}$ be a pair of indices. Consider the testing problem 
\begin{enumerate}
    \item[$H_0^{(u,v)}$:]   $ 
    \sup_{\omega  \in \mathcal{W}} |\rho_{u,v}(\omega)|\leq \delta$,
    \vspace*{-0.2cm}
   \item[vs.] 
    \vspace{0.2cm}
    \item[$H_1^{(u,v)}$:]   $  |\rho_{u,v}(\omega)|> \delta$ for all $ \omega \in A$, where the set  $ A \subset \mathcal{W}$ has   positive Lebesgue measure.
\end{enumerate}

Our  aim is to obtain a test statistic which is pivotal that is, its  distribution is independent of the parameters of the underlying process. For this  it  is  convenient  if the estimators  $\widehat{\rho}^{(de)}_{u,v}(\omega)$ involved in the construction of the test statistic, are  asymptotically  independent at nearby frequencies.  To achieve this,   we consider a  grid of frequencies $\omega_l^\prime=\pi l N/M$, $l =1,2, \ldots, M/N-1 $, where  neighborhood  frequencies are  $\pi N/M$ apart from each other and where the value of  $N$ depends on the particular kernel used. For instance, for the uniform kernel,  $N=1$ can be chosen; see also the discussion preceding Theorem~\ref{thm.partial.coherence.max} of Section~\ref{sec.2.properties}. 
Let $\mathcal{L}=\{ l =1,\dots,M/N-1 : \omega_l^\prime \in \mathcal{W}\}$ and set $d:=|\mathcal{L}|$.   The following test statistic is then proposed,  
\begin{align}
     T_n^{(u,v)}=&\ind\big(\max_{l \in \mathcal{L}}  |\hat \rho^{(de)}_{u,v}(\omega_l^\prime)|> \delta\big) \times  \label{eq.test.statistic}\\
     & \max_{l \in \mathcal{L}}\Big\{ \frac{n}{M} \begin{pmatrix}
    \Re(\hat \rho^{(de)}_{u,v}(\omega_l^\prime)-\delta \exp(i \tilde \omega_l)) \\
    \Im(\hat \rho^{(de)}_{u,v}(\omega_l^\prime)-\delta \exp(i \tilde \omega_l)) \\
    \end{pmatrix}^\top
    \hat V^{-1}_{(u,v)}(\omega_l^\prime)
    \begin{pmatrix}
    \Re(\hat \rho^{(de)}_{u,v}(\omega_l^\prime)-\delta \exp(i \tilde \omega_l)) \\
    \Im(\hat \rho^{(de)}_{u,v}(\omega_l^\prime)-\delta \exp(i \tilde \omega_l)) \\
    \end{pmatrix}\Big\}, \nonumber
\end{align}
where 
$\tilde \omega_l=\arg(\hat \rho^{(de)}_{u,v}(\omega_l^\prime))$ 
and
\begin{align*}
    \hat V^{-1}_{(u,v)}(\omega)=\frac{2}{C_{K_2}(1-|\hat\rho_{u,v}(\omega)|^2)^2}\begin{pmatrix}
    1-\Im(\hat\rho_{u,v}(\omega))^2 & \Re(\hat\rho_{u,v}(\omega))\Im(\hat\rho_{u,v}(\omega)) \\
    \Re(\hat \rho_{u,v}(\omega))\Im(\hat\rho_{u,v}(\omega)) & 1-\Re(\hat\rho_{u,v}(\omega))^2 \\
    \end{pmatrix}.
\end{align*}

Observe  that for a given frequency $\omega$, and as Theorem~\ref{thm.partial.coherence} of Section~\ref{sec.2.properties} shows, the real and imaginary parts of $ \hat{\rho}^{(de)}_{u,v}(\omega)$ are asymptotically normal which implies that the above transformation leads to an asymptotic $\chi^2_2$ distributed random variable. The  frequency $\omega=0$ is excluded from the calculation of the maximum above   for two reasons. First,  for  $\omega=0$, the imaginary part is zero and 
therefore  a  $\chi^2_2$-approximation is not appropriate. Second, deterministic trends in the time domain lead to peaks of the spectral density for frequencies close to zero. Hence, excluding  $\omega=0$   makes the  test  more robust with respect to such  distortions.

Given the test statistic $T_n^{(u,v)}$, for any desired  level $ \alpha \in (0,1)$, the  null hypothesis    $H_0^{(u,v)}$   is rejected  if 
\begin{align}
T_n^{(u,v)}\geq \mathcal{G}(1-\alpha), \label{eq_test}   
\end{align}
where $\mathcal{G}(1-\alpha)$ is the upper  $\alpha$ quantile of the  distribution function of the maximum of $d$ independent $\chi_2^2$ random variables, i.e., $G_d(t):=1-(1-\exp(-t/2))^d$ and $\mathcal{G}(1-\alpha)=\inf_t \{G_d(t)=\alpha\}$. 
The following theorem ensures  that the   test  (\ref{eq_test}) has   asymptotically the desired  level $\alpha$. 

\begin{thm} \label{thm.test-pc}
Suppose that $ H_0^{(u,v)}$ is true. Then, under the conditions  of Theorem~\ref{thm.partial.coherence} of Section~\ref{sec.2.properties}, the test  \eqref{eq_test} satisfies
$\lim_{n\rightarrow\infty} P\Big(T_n^{(u,v)}\geq \mathcal{G}(1-\alpha)\Big)\leq \alpha$.
\end{thm}

\subsection{Testing Multiple Hypotheses}
\label{sec.MultTest}
Let $\mathcal{Q}$ be   a  set of pairs of indices  for which  the hypothesis should be tested that the corresponding partial coherences do not exceed a desired level $\delta \geq 0$. Let $q=|\mathcal{Q}|$ and observe  that if all possible pairs of partial coherences are considered, then   $\mathcal{Q}=\{ (u,v): u,v=1,\dots,p, u<v\}$ and   $q=(p^2-p)/2$. For any $ (u,v)\in {\mathcal Q}$, let   $H_0^{(u,v)}$ and $H_1^{(u,v)}$ be the corresponding  null and alternative hypothesis specified  as in Section 3.1. 

The main problem to be solved in  implementing such a     multiple testing procedure is that of controlling the  false discovery rate (FDR). That is, to  ensure  that (at least asymptotically),  the expected ratio of false rejections to the total number of rejections,  does not exceed a  desired level $ \alpha$. Controlling the FDR in multiple testing problems has a long-standing history in statistics, and different approaches have been proposed  under a variety of assumptions; see among others  \cite{benjamini1995controlling,benjamini2001control} and \cite{barber2015controlling}. In  our setting, we use  a  thresholding based  approach which is applied to 
the test statistic $ T_n^{(u,v)}$
(see \eqref{eq.test.statistic}),  
for  all  pairs $(u,v)\in{\mathcal Q}$. In this context,  each    null hypothesis $ H_0^{(u,v)}$ is  rejected if and only if  the corresponding  test statistic $T_n^{(u,v)} $  exceeds a  specified threshold value, say $\widehat{t}$. An approach  for FDR control along these lines  has been introduced  in the i.i.d. context  by  \cite{liu2013gaussian}. Clearly, the key issue  here  is how to select  the threshold value $\widehat{t}$ so   that  the desired FDR control is  achieved. For our multiple testing procedure, the  threshold 
\begin{align}
    \label{eq.that}
    \widehat t = \operatorname{inf} \Big\{ 0 \leq t \leq 2 \log(dq) : \frac{G_d(t) q}{\max(1,\sum_{ (u,v) \in \mathcal{Q}} \ind(T_n^{(u,v)} \geq t))} \leq \alpha \Big\},
\end{align}
is used.
Using this threshold, the  multiple testing procedure  rejects  for every $(u,v) \in \mathcal{Q}$ the corresponding
 null hypothesis if and only if $T_n^{(u,v)}>\widehat t$. 
 Notice that the  main idea behind the particular  construction of $\hat t$ is that $G_d(t) q$ roughly approximates the expected number of falsely rejected nulls for any  given threshold $t$. Since a sparse signal setting is considered, the total number of nulls can be approximated by $q$, i.e., the total number of tests conducted.  At the same time and as Theorem~\ref{thm.partial.coherence.max} of Section~\ref{sec.2.properties} shows,   $G_d(t)$ approximates well enough the upper tail of the distribution of a single test statistic under the null. 

To state our next result which deals with the theoretical properties of the described thresholding procedure, we need to fix some additional  notation.  
Let  $\mathcal{H}_0=\{(u,v) \in \mathcal{Q} : \sup_{\omega\in[0,2\pi]}|\rho_{u,v}(\omega)|\leq \delta \}$ be  the set of true null hypotheses and let,  for $\mu >0$, $\mathcal{H}(\mu)=\{ (u,v)\in \mathcal{Q} :  \sup_{\omega} |\rho_{u,v}(\omega)|>\delta+\mu\}$, be the set of alternative hypotheses  for which the corresponding partial coherences exceed the value of $\delta$ by  $\mu$. Recall  the definition of the  false discovery rate   given by 
$$
FDR=E \Big( \frac{\sum_{(u,v) \in \mathcal{H}_0} \ind(T_n^{(u,v)} \geq \widehat t\,)}{\max(\sum_{(u,v) \in \mathcal{Q}} \ind(T_n^{(u,v)} \geq \widehat t\,),1)}\Big).
$$
Theorem~\ref{th.FDR_1} bellow 
shows that the proposed,  thresholding-based, multiple testing procedure,  succeeds in properly controlling the FDR.


\begin{thm} \label{th.FDR_1}
Suppose that the assumptions of Theorem~\ref{thm.partial.coherence.max} of Section~\ref{sec.2.properties} are satisfied, that 
 $q/|\mathcal{H}_0|=1+o(1)$, as $q\rightarrow \infty$, and that $\Big|\mathcal{H}\Big(2\sqrt{M/n \log(dq)}\Big)\Big|\geq \log(\log(n))$. It then holds true  for any $ \alpha \in (0,1)$, that, 
$
FDR\leq \alpha + o(1)$, as $ n,q \rightarrow \infty$.

\end{thm}

\section{Implementation Issues and Simulations}

\subsection{Implementation Issues} \label{subsec4.1}

Before summarizing the main steps involved in the practical implementation of our procedure, we first elaborate on bias reduction in spectral density matrix estimation. Tapering, using of  flat top kernels and prewhitening are some known procedures used for the purpose of bias reduction.  We elaborate here on prewhitening.

Prewhitening or prefiltering  is mainly used to  improve the  finite sample performance   of nonparametric spectral density estimators;  see   Section~5.8 in \cite{brillinger2001time}. The basic  idea is to apply  an appropriate filter  to the time series at hand,  so that the filtered time series has   a flatter spectral density which allows for  a less biased  nonparametric estimator.
In the multivariate  set-up,  an additional benefit of such an approach  is that prefiltering  homogenizes (to a certain extend)  the spectral densities of the   component processes,  allowing, therefore,  
for the application of the same bandwidth 
in order to estimate the spectral density matrix. 
 Using individual bandwidths in a multivariate context has the disadvantage that the estimated  spectral density matrix   is not equivariant   and  that semi-positive definiteness is not  guaranteed. 
 
Let $\Phi(z) = \sum_{j \in \Z} \Phi_j z^j$ be a linear filter 
and define for $\omega \in [0,2\pi]$,   $\Phi(\omega):=\Phi(\exp(-i \omega))$. Think of $\Phi(z)$ as a autoregressive filter
which can be implemented in the high-dimensional context considered here, by   fitting  a sparse vector autoregressive model to the vector   time series  at hand. Note  that prefiltering also can be used   in the construction of the de-biased estimator $\hat \beta_{v,\tilde u}^{(de)}(\omega)$ proposed. For this observe that  $\hat \beta_{v,\tilde u}^{(de)}(\omega)$ also  can be written as
\begin{align*} 
    \hat \beta_{v,\tilde u}^{(de)}(\omega)=\hat \beta_{v,\tilde u}(\omega)+\frac{
    (e_v-I_{p,-v}\hat \beta_v)^H
    (\sum_{k=1}^n \kappa_M(\omega-\omega_k)Z_n(\omega_k) Z_n(\omega_k)^H) I_{p,-v}
     \hat \gamma_{-v,\tilde u}(\omega)}{ e_u^\top(\sum_{k=1}^n \kappa_M(\omega-\omega_k) Z_{n}(\omega_k) Z_{n}^{H}(\omega_k))I_{p,-v}  \hat \gamma_{-v,\tilde u}(\omega)}.
\end{align*}
Applying an autoregressive filter $\Phi(z)=1-\sum_{j=1}^m \Phi_j z$ in the time domain leads to the 
filtered time   $Y_t=X_t-\sum_{j=1}^m \Phi_j X_{t-j}, t=m+1, m+2, \ldots, n$.
Let   
$\breve Z_n(\omega)=(2 \pi (n-m))^{-1/2} \sum_{t=1}^{n-m} Y_{t+m} \exp(-i \omega t)$. 
Then, the  term $\sum_{k=1}^n \kappa_M(\omega-\omega_k)Z_n(\omega_k) Z_n(\omega_k)^H$
 appearing  in the above expression for  $\hat \beta_{v,\tilde u}^{(de)}(\omega)$,  can  be replaced by 
$$
\Phi(\omega)^{-1}
    (\sum_{k=1}^{n-m} \kappa_M(\omega-\tilde \omega_k) \Breve Z_n(\omega_k) \Breve Z_n(\tilde \omega_k)^H) (\Phi(\omega)^H)^{-1},
$$
where $\tilde \omega_k=2\pi (k-1)/(n-m),k=1,\dots,n-m$, are the Fourier frequencies corresponding to  the $n-m$ observations of the filtered time series $Y_t$.

We can now summarize the main  steps involved in the practical implementation of our inference procedure as follows:

\begin{enumerate}
    \item[Step 1:] \ Select a filter $\Phi(z)$ to prewhiten  the vector time series at hand.
    \item[Step 2:] \ Select a global bandwidth (truncation lag)  $M$ and a kernel $K$. 
    \item[Step 3:] For $N_1=\lfloor M/N \rfloor$, use the grid of frequencies $\mathcal{L}=\{\omega_l^\prime=l \pi /N_1  \in \mathcal{W} :l \in {1,\dots,N_1-1} \}$ to cover the frequency band $ {\mathcal W}$ of interest, where  $N=\log^{2/r}(M)$ and  $r$ is determined by the decay behavior of the Fourier coefficients of the kernel $K$ used; see Assumption 2 in Section 6.
    \item[Step 4:] \  Estimate the inverse spectral density matrix $f^{-1}$ at the  frequencies $\omega_l^\prime$ for every $ l \in \mathcal{L}$. 
    \item[Step 5] \  Compute for all $(u,v) \in Q$ and for all $l \in \mathcal{L}$, the de-biased estimator 
    $\hat \rho_{u,v}^{(de)}(\omega_l^\prime)$ and  the test statistic $T_n^{(u,v)}$.
    \item[Step 6:] \  Set $d=|\mathcal{L}|$,  $G_d(t)=1-(1-\exp(-t/2))^d$ and calculate  the threshold   
    $$\widehat t = \operatorname{inf} \{ 0 \leq t \leq 2 \log(dq) : \frac{G_d(t) q}{\max(1,\sum_{ (u,v) \in \mathcal{Q}} \ind(T_n^{(u,v)} \geq t))} \leq \alpha \}.$$
    \item[Step 7:] \  For each  $(u,v) \in \mathcal{Q}$ reject $H_0^{(u,v)}$ if $T_n^{(u,v)} \geq \widehat t$.
\end{enumerate}

 Some remarks regarding  the above steps are in order. In  Step 1, the use of a vector autoregressive filter, implemented by fitting a sparse vector autoregressive model, is recommended.
We present here the approach using a global bandwidth since prewhitening can already homogenize the spectrum such that there is less need for different bandwidths, and a global bandwidth with a positive definite kernel ensures a positive semi-definite estimate which stabilizes  numerical results. The algorithm using different bandwidths for each component time series is described in the supplementary material in Section~\ref{sec.multiple.bandwidth}. For Step 2, we adapt to the high-dimensional context, the adaptive rule proposed by \cite{politis2003adaptive}. For this, the Frobenius norm of the matrices of the sample autocorrelations  of the filtered time series obtained  in Step 1 is used.  
Regarding the kernel $K$, we suggest the use of  kernels that ensure (semi-)positive definite results, like, for instance, the modified Bartlett kernel $K(u)=\ind(|u|\leq 1)(1-|u|)$. 
For  Step 4 and  depending on the time series at hand, 
 different estimation procedures can be used.
For instance, a vector autoregressive filter together with a lag-window estimator, that is a prewhitening type 
  estimator, can be applied. In this case, the inverse spectral density matrix can be estimated using, for instance, graphical lasso. We implemented the algorithm in the R-package \emph{HDSpectralAnalysis} which is available to download at \url{https://github.com/JKrampe/HDSpectralAnalysis}.

\subsection{Simulations} \label{sec.simulations}  We investigate the finite sample performance of the frequency domain inference procedures developed in this paper. For this we consider the following  six  data generating processes  where  $S_1$ denotes the percentage  of non-zero partial coherences and $S_2$ the percentage  of partial coherences that  exceed the value $0.2$.  The particular parameterization of the coefficient  matrices of the corresponding vector autoregressive moving-average  (VARMA) and vector moving-average (VMA) processes  are reported  as  \emph{R-Data} files in the supplementary material. 
\begin{enumerate}[(a)]
\item[] {\bf Models (a)-(c)}, where $\{X_t,t\in\Z\}$ is a  $p$-dimensional, $p\in\{50, 100, 200\}$, sparse VARMA$(1,1)$ model,  with $S_1\in\{7.5\%, 4.1\%, 1.9\%\}$  and  $S_2\in\{6.3\%, 3.5\%, 1.5\%\}$, respectively.
\item[] {\bf Models (d)-(f)},  where $\{X_t,t\in\Z\}$ is a  $p$-dimensional, $p\in\{50, 100, 200\}$, sparse VMA$(5)$ model,  with  $S_1\in\{3.9\%, 4\%, 4\%\}$ and   $  S_2\in\{3.8\%, 3.6\%, 2.8\%\}$, respectively.
\end{enumerate}
    
Three sample sizes $n$, namely $512$, $2048$ and $4096$, are considered and the results presented are based on $ B=1000$ replications and on implementations in \emph{R} \citep{R}.
In these  implementations we follow the steps described in  Subsection~\ref{subsec4.1}. The frequency band  ${\mathcal W}$ of interest is  set equal to $[0,\pi]$ and  the set of  hypotheses tested in given by    $\mathcal{Q}=\{(u,v) | 1 \leq u < v \leq p\}$. Two different threshold values $\delta$ for the null hypothesis are considered, $\delta=0$ and $\delta=0.2$. For the prefiltering step, we use a sparse vector autoregressive model of order $\lceil log_{10}(n) \rceil$ estimated by a row-wise adaptive lasso with tuning parameter selected by BIC; see for instance, \cite{krampe2020Est} for details. The BIC is adapted to handle a diverging number of parameters using the approach in \cite{wang2009shrinkage} and the parameter $C_n$ therein is set equal to  $C_n=\log(p)$. To determine the bandwidth of the lag-window estimator, we use the rule described in Section~\ref{subsec4.1} adapted from \cite{politis2003adaptive} with tuning parameters $K_n=5$ and $ c_{thres}=1.5$. Additionally, we apply Jenkin's kernel bandwidth correction, see Section 7.3.2 (2) in \cite{priestley1988spectral}. We use the modified Bartlett kernel, i.e. $K(u)=\ind( u \in (-1,1))(1-|u|)$, which has Fourier representation $k_{1/M}(\omega)=(2 \pi M)^{-1} \sin^2(M \omega/2)/\sin^2(\omega/2)$ and is positive definite. Furthermore,  we  set $N=log(M)$. To estimate the inverse spectral density matrix, the discussed prefiltering procedure  is used in combination with the lag-window estimator \eqref{eq.lag.window} applied to the filtered vector time series. The inverse spectral density matrix is  estimated by  applying a frequency-by-frequency graphical lasso procedure with  tuning parameter $\lambda_n$  chosen by BIC with the modification of \cite{wang2009shrinkage} and parameter $C_n=\log(\log(p))$.

Table~\ref{table.1} presents the empirical false discovery rates and powers for the case $\delta=0$ and for the different models, levels, and sample sizes considered. 
 We compare the results of our inference procedure with those obtained  using a thresholding and  regularizing-based  estimator of the inverse spectral density matrix. For this, we estimate the inverse spectral density matrix $f^{-1}$ using the frequency-by-frequency graphical lasso. The tuning parameter $\lambda_n$ is selected as above and  the outcome is thresholded at $\lambda_n$ again  to get a sparse result. This approach is called in the following  \methodb, while the procedure proposed in this paper and which uses statistical testing with FDR control based on  de-biased estimators of partial coherence,  is called  \methoda. 

\begin{table}[t]
\spacingset{1}
\centering
\begin{tabular}{|cc|cc|cc|cc|}
\hline
  & &\multicolumn{4}{c|}{\methoda} & \multicolumn{2}{c|}{\methodb}\\ 
 & &\multicolumn{2}{c}{$\alpha=0.05$} & \multicolumn{2}{c|}{$\alpha=0.1$} & \multicolumn{2}{c|}{} \\
 $p$& $n$ &  FDR & Power & FDR & Power & FDR & Power  \\ 
 \hline
 \multicolumn{8}{|c|}{DGP: VARMA $(1,1)$}\\ \hline
   \multirow{4}{1cm}{50}&190 & 0.04(0.04) & 0.57(0.04) & 0.07(0.06) & 0.59(0.06) & 0.00(0.01) & 0.29(0.01) \\ 
  &512 & 0.05(0.04) & 0.75(0.03) & 0.10(0.05) & 0.77(0.03) & 0.08(0.04) & 0.52(0.03) \\ 
  &2048 & 0.07(0.03) & 0.90(0.02) & 0.13(0.05) & 0.91(0.02) & 0.32(0.04) & 0.74(0.02) \\ 
  &4096 & 0.07(0.04) & 0.95(0.01) & 0.13(0.05) & 0.95(0.01) & 0.41(0.05) & 0.84(0.02) \\ 
   \hline
\multirow{4}{1cm}{100}&190   & 0.02(0.02) & 0.24(0.03) & 0.04(0.04) & 0.26(0.03) & 0.00(0.01) & 0.05(0.01) \\ 
  &512   & 0.03(0.02) & 0.52(0.02) & 0.07(0.03) & 0.55(0.02) & 0.04(0.02) & 0.32(0.02) \\ 
  &2048  & 0.05(0.02) & 0.77(0.02) & 0.10(0.03) & 0.79(0.02) & 0.09(0.02) & 0.60(0.02) \\ 
  &4096  & 0.05(0.02) & 0.85(0.01) & 0.11(0.03) & 0.86(0.01) & 0.11(0.01) & 0.70(0.01) \\ 
   \hline
\multirow{4}{1cm}{200}&190  & 0.22(0.16) & 0.53(0.02) & 0.27(0.16) & 0.54(0.02) & 0.00(0.00) & 0.18(0.01) \\ 
  &512    & 0.09(0.07) & 0.69(0.01) & 0.14(0.07) & 0.70(0.01) & 0.12(0.02) & 0.55(0.01) \\ 
  &2048   & 0.06(0.02) & 0.78(0.01) & 0.12(0.03) & 0.79(0.01) & 0.29(0.01) & 0.66(0.01) \\ 
  &4096   & 0.06(0.02) & 0.81(0.01) & 0.12(0.02) & 0.82(0.01) & 0.35(0.01) & 0.71(0.01) \\ 
   \hline
\multicolumn{8}{|c|}{DGP: VMA $(5)$}\\ \hline
\multirow{4}{1cm}{50} &190  & 0.04(0.06) & 0.45(0.04) & 0.05(0.06) & 0.46(0.05) & 0.00(0.01) & 0.25(0.05) \\ 
   &512  & 0.04(0.05) & 0.69(0.04) & 0.07(0.07) & 0.71(0.04) & 0.00(0.01) & 0.48(0.04) \\ 
   &2048 & 0.03(0.04) & 0.88(0.02) & 0.06(0.06) & 0.88(0.02) & 0.01(0.02) & 0.78(0.05) \\ 
   &4096 & 0.03(0.04) & 0.94(0.03) & 0.06(0.06) & 0.95(0.02) & 0.04(0.04) & 0.93(0.04) \\ 
   \hline
\multirow{4}{1cm}{100} &190   & 0.04(0.09) & 0.15(0.02) & 0.07(0.10) & 0.16(0.02) & 0.00(0.00) & 0.03(0.01) \\ 
   &512   & 0.04(0.03) & 0.40(0.03) & 0.07(0.04) & 0.43(0.03) & 0.00(0.01) & 0.17(0.02) \\ 
   &2048  & 0.03(0.02) & 0.74(0.02) & 0.07(0.03) & 0.77(0.02) & 0.00(0.00) & 0.48(0.03) \\ 
   &4096  & 0.03(0.02) & 0.87(0.01) & 0.06(0.02) & 0.88(0.01) & 0.00(0.01) & 0.65(0.02) \\ 
   \hline
\multirow{4}{1cm}{200} &190   & 0.01(0.02) & 0.07(0.01) & 0.03(0.03) & 0.08(0.01) & 0.00(0.00) & 0.02(0.00) \\ 
   &512   & 0.02(0.01) & 0.22(0.01) & 0.04(0.02) & 0.24(0.01) & 0.00(0.00) & 0.09(0.01) \\ 
   &2048  & 0.02(0.01) & 0.53(0.01) & 0.05(0.01) & 0.56(0.01) & 0.00(0.00) & 0.27(0.02) \\ 
   &4096  & 0.02(0.01) & 0.72(0.01) & 0.04(0.01) & 0.74(0.01) & 0.00(0.00) & 0.41(0.01) \\ 
  \hline
\end{tabular}
\caption{Empirical false discovery rate and power for the case $\delta=0$ and for different models, different levels, and different sample sizes.} \label{table.1}
\end{table}
\spacingset{1.9}

For the case $\delta=0$, the empirical FDR of {\methoda} is for all processes, sample sizes, and dimensions (except the case VARMA$(1,1)$, $p=200, n=190$ case) close to the nominal level. In terms of power, {\methoda} outperforms {\methodb} in all situations. The dimension affects the power only slightly for the case of the VARMA(1,1) model  whereas the same effect is larger for the case of the VMA(5) model and 
 the power 
 improves as the sample size increases.
  The corresponding results for the case $\delta=0.2$ are presented in the Supplementary File. 
  We additionally, visualize in Figure~\ref{fig.Example1} in the Supplementary File the results for the VARMA$(1,1)$ process with parameters as specified  in (a) and the VMA$(5)$ process with parameters as in (d)  and  for  $n=512$ and $n=4096$ observations.

\begin{figure}[!h]
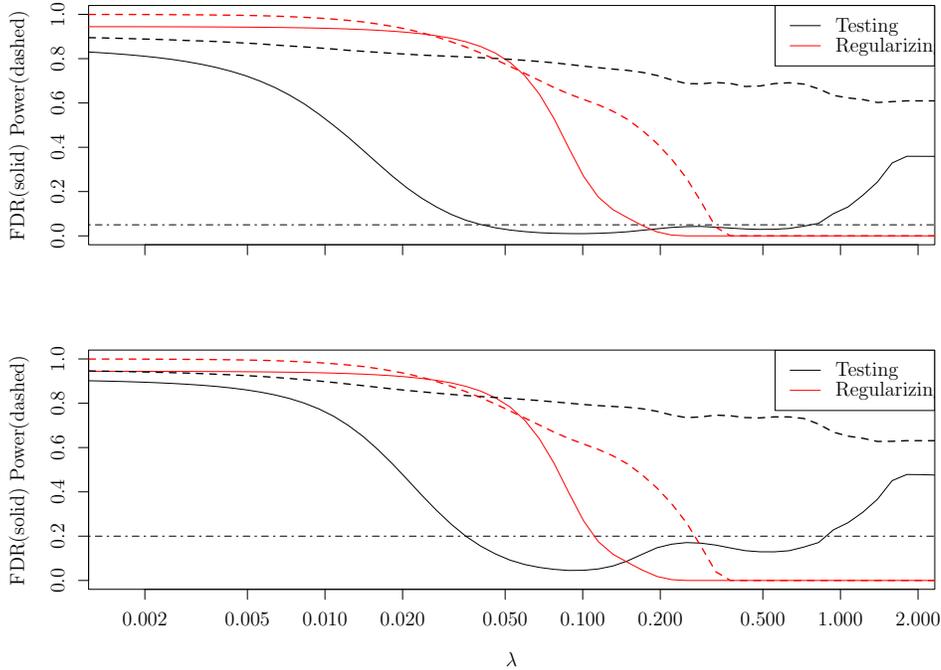

\spacingset{1}
\centering
   \vspace*{-0.5cm}\resizebox{0.8\textwidth}{!}{\input{FDR_Power_lambda_005}}\vspace*{-1.1cm}\\
   \resizebox{0.8\textwidth}{!}{\input{FDR_Power_lambda_020}}\vspace*{-0.6cm}\\
   \caption{Results for detecting  non-zero partial coherences for the $50$-dimensional VARMA$(1,1)$ process with parameters as  in (a)   and for the sample size $n=512$ as well as  for several choices of the tuning parameter $\lambda$ used in the  implementation of graphical lasso. The solid lines refer to   FDR and the dashed lines  to power. The behavior of  {\methoda}  is described by the solid  and the dashed black lines while that of  {\methodb}  by the  solid and the dashed red lines. 
    The horizontal line in each figure denotes the target FDR-level $\alpha$ (which is $0.05$ in the top figure and $0.2$ in the bottom figure).}
    \label{fig.FDR.Power}
    \spacingset{1.9}
\end{figure}

In Figure~\ref{fig.FDR.Power} we visualize the sensitivity of both approaches with respect to the choice of the tuning parameter $\lambda$ used in the implementation of   the graphical lasso estimator. Both approaches rely on a good estimator of the inverse spectral density which has been achieved by  using  a proper choice of the tuning parameters involved. As Figure~\ref{fig.FDR.Power} shows,   {\methoda} is more robust with respect to the choice of the tuning parameter $\lambda$  than  {\methodb}. The first approach gives good results for a broader range of choices of this tuning parameter. In particular, we observe  that {\methoda} is able to achieve a FDR which is close to the target value $\alpha$ for a wide range of values of $\lambda$.  {\methoda} also outperforms {\methodb} in terms of power for all values  of $\lambda$ for which the corresponding FDR is close or smaller than the desired target level.

\section{A Graphical Model for Brain Connectivity}
In this section, we study brain connectivity based on electroencephalography (EEG) data and  use the coherence and partial coherence to measure  connectivity as well as  conditional connectivity of several regions of the brain. We display non-zero coherences and partial coherences in terms of graphs keeping in mind that a graph based on  partial coherences is the time series context, can be considered as an  analogue of a graphical model based on partial correlations for i.i.d. data. 

We use the data set provided by \cite{trujillo2017effect}. This data set consists of EEG recordings of $22$ undergraduate students, where for each student,   $72$ channels with a sampling rate of $256$Hz are used. The students were recorded in a resting state with $4$ minutes of eyes open and $4$ minutes of eyes closed. \cite{trujillo2017effect} preprocess the data. Among other steps, they split the data into epochs of one second, i.e., each epoch consists of $256$ observations. The epochs are constructed with $50\%$ overlap. Epochs with artifacts caused, for instance, by blinks or muscles were removed automatically and by hand. Additionally, they apply band-pass filters to remove linear trends and other noise effects. {These preproessing steps aim to obtain stationary time series for each epoch.} Finally, they focus on the alpha band  ($4-13$Hz) and the beta band ($14-25$Hz). In our  application, we use the preprocessed data and  focus on the beta band, which is available for download from \url{https://doi.org/10.18738/T8/CNVLAM}. We consider two students (students with number 19 and  20). For these two students, no channels were interpolated and $209$ to $387$ epochs are available. Then, for each epoch (i.e. $n=256, p=72$), we compute the brain connectivity for the beta band. That is we use the multiple testing procedure described in Section~\ref{sec.testing} to test whether coherences and partial coherences having indices belonging to the set $Q=\{(u,v) : u,v=1,\dots,72,u<v\}$ are   zero ($\delta=0$) in the frequency band $\mathcal{W}=[14\text{Hz},25\text{Hz}]$. We use the same implementation with prewhiteting, automatic bandwidth selection,  and the modified Bartlett kernel as the one used in the simulation study presented in the previous section. For the FDR control, we set $\alpha=0.1$. The  results  obtained are averaged for each student and each state (eyes open and eyes closed). To display the results, we only keep those edges which  are present in more than $50\%$ of the epochs. 

\begin{figure}[!h]
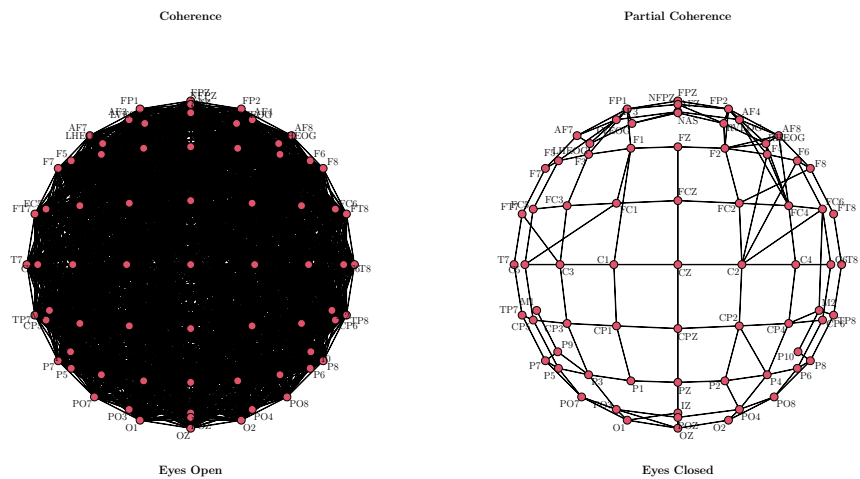

\spacingset{1}
        Student 19:\\
    \vspace*{-1cm}
    \begin{center}
   \resizebox{0.8\textwidth}{!}{\input{PC_Network_Pat_Coh_19}}         
    \end{center}   
   \vspace*{-2cm}Student 20:\vspace*{-1cm}\\
   \begin{center}
      \resizebox{0.8\textwidth}{!}{\input{PC_Network_Pat_20}} 
   \end{center}
        \vspace{-2cm}
    \caption{
    Graphical model representing the brain connectivity for students 19 and 20. The vertex labels denote the channel labels of the EEG recording and the patient's nose is located at the top. The figures in the top row representing the brain connectivity for student 19 in the state of eyes open. In this row, the left  figure displays the unconditional connectivity based on  coherence and the right  figure, the conditional connectivity based on  partial coherence. The two figures in the second  row show  the conditional brain  connectivity of student 20 in the two states considered, eyes open and eyes closed. }
    \label{fig.realdata1}
    \spacingset{1.9}
\end{figure}

In the top row of Figure~\ref{fig.realdata1}, we display the graphs obtained using  coherences and  partial coherences  for student $19$ in the same state of eyes open. As it is  seen, the brain connectivity based on  coherences is quite dense with $75\%$ of all possible connections identified as non-zero. For other states and students, we can observe values up to  $95\%$. This is in stark contrast to the results obtained using   partial coherences where the focus is on the direct effects only. Here, only $4\%$ to $6\%$ of all possible connections are identified as non-zero. 
Furthermore, connections based on  partial coherences are mainly between  neighboring brain regions. This is not the case for the results of brain connectivity analysis  based on  coherences, where connections are  strong not only between  neighboring regions but also between  regions which are  far apart. 

Concerning  the differences in brain connectivity   between the two states, eyes open and eyes closed, we focus on the partial coherence results only. The bottom row of Figure~\ref{fig.realdata1} presents the 
corresponding  graphical model in the two aforementioned states. We observe that in the state of eyes open, brain  connectivity is slightly higher. In particular, measuring brain connectivity  by the number of edges divided by the number of all possible edges,  we have for student $19$ a connectivity of $ 5.9\%$ in the state of open eyes and of $4.4\%$ in the state of closed eyes. The corresponding percentages for student $20$, are  $4.3\%$ and $3.7\%$, respectively.  Note that the inference presented  in only within a condition and  between-condition statements are exploratory.

\section{Asymptotic Considerations} \label{sec.2.properties}
To derive the limiting distribution of 
$\widehat{s}_{u,v}(\omega)$ and $\widehat{\rho}^{(de)}_{u,v}(\omega)$,  
some assumptions  have to be imposed on  the stochastic properties of the underlying high-dimensional process $ \{X_t,t\in\Z\}$,  the inverse spectral density matrix $ f^{-1}$ and  the lag-window kernel $K$. Furthermore, and since we do not want to restrict our considerations to a particular estimator
of $f^{-1}$, we only state the required consistency properties of the regularized estimator $\widehat{f}^{-1}$ used, covering in this way  a wide range of possible estimators.  

 To control the temporal dependence of the high-dimensional process $ \{X_t,t\in\Z\}$,   we make  use of the concept of functional dependence,    \cite{wu2005nonlinear}; also see  \cite{wu2016performance} and \cite{zhang2020convergence}. For this let for $ k \leq t$,  $X_{t,\{k\}}=R(\eps_t,\dots,\eps_{k+1},\eps_k^+,\eps_{k-1},\dots)$, be a coupled processes, where $\eps_k^+$ is an i.i.d.~copy of $\eps_k$. Let $\tau \in \N$ be  the number of finite moments of $ X_t$. Define the so-called functional dependence measures  $\delta_{t,\tau}^{[i]}$ and $ w_{t,\tau}$,  as
$$\delta_{t,\tau}^{[i]}=\big\{E|e_i^\top(X_t-X_{t,\{0\}})|^\tau\big\}^{1/\tau} \ \ \mbox{and}\ \  
w_{t,\tau}=\big\{E( \max_i |e_i^\top(X_t-X_{t,\{0\}})|^\tau)\big\}^{1/\tau}
.$$
Let $\delta_{t,\tau}^{[{\max}]}=\max_{\|v\|_2=1} \sum_{i=1}^p v_i\delta_{t,\tau}^{[i]} $
and for $\alpha \geq 0$, 
let 
$\| |X_{\cdot}|_\infty\|_{\tau,\alpha}=\sup_{m\geq 0} (m+1)^\alpha \sum_{t=m}^\infty w_{t,\tau}$. Note that $w_{t,\tau}\leq p^{1/\tau}  \max_i \delta_{t,\tau}^{[i]} \leq p^{1/\tau} \delta_{t,\tau}^{[{\max}]} $. The following assumption is then  imposed.




\begin{asp} \label{asp.moments}
For some $\tau\geq 8$ and for all $p\in \N$,  it holds true that $\delta_{t,\tau}^{[\max]}\leq C \rho^t$ and  $\sup_{\|v\|_2=1}  (E |v^\top X_t|^\tau)^{1/\tau}\leq C<\infty$, where   $\rho\in (0,1)$ is a  constant. Furthermore,  $\| |X_{\cdot}|_\infty\|_{\tau,\alpha}\leq Cp^{r(\tau)}$ for some
 $\alpha>1/2-1/\tau$, where   $ r(\tau)$  is a positive number that depends on the number of finite moments $ \tau$.
\end{asp}

Stationary Markov chains as well as  stationary linear processes, see Example~2.1 and 2.2 in \cite{chen2013covariance}, are examples of  processes fulfilling the  conditions of Assumption~\ref{asp.moments}. Note that by this assumption,  $\max_{\|v\|_2=1} (E(v^\top(X_t-X_{t,\{0\}}))^\tau)^{1/\tau}\leq C \delta_{t,\tau}^{[\max]}$
and $\sum_{j=m}^\infty \delta_{t,q}^{[\max]}\leq C \rho^m$ for some  $q\leq \tau$; see also \cite{krampe2022inverse,yi-22} for the Spectral-norm/$\ell_2$ physical dependence $\delta_{t,\tau}^{[{\max}]}$.  Furthermore, since $w_{t,\tau}\leq p^{1/\tau} \delta_{t,\tau}^{[{\max}]}$, we have $\| |X_{\cdot}|_\infty\|_{\tau,\alpha}\leq C p^{1/\tau}$, i.e.,  $r(\tau)\leq 1/\tau$ holds true. We mention here that the geometric decay of $ \delta_{t,\tau}^{[\max]}$ given in Assumption~\ref{asp.moments}  can be relaxed to a polynomial decay; see  condition~(12) in \cite{wu2018asymptotic}. Such a polynomial decay would, however,  require  additional restrictions on the allowed increase  of the dimension  $p$ of the process  compared to those  stated in Assumption~\ref{asp.rates} below.

The following assumption summarizes our requirements regarding  the kernel function $K$ used in  the   non-parametric estimators considered in this paper.
\begin{asp} \label{asp.kernel}
$K$ is  an even and bounded kernel function with compact  support  $[-1,1]$ satisfying $K(0)=1$ and $C_{K_2}=\int_{-1}^1 K^2(u)du<1$. Furthermore,  $K$ is Lipschitz continuous in $[-1,1]$ with Lipschitz constant $L_{K}$ and
$1/M\sum_{s=-M}^M K^2(ks\pi/M) \exp(-i \pi k s/M)\leq C [(1/k)^r+1/M]$ for all $k\geq 1$ and some $r\geq 1$.
\end{asp}
The last condition in Assumption 2, specifies the  decay behavior of the Fourier coefficients of the kernel $K$ and it affects the covariance of the lag-window estimator at different frequencies. Note that for the uniform kernel $ K(u)=\ind(|u|\leq1)  $,  $(1/k)^r$ vanishes for every $r\in\N$,  while  for the modified Bartlett kernel, $K(u)=(1-|u|)\ind(|u|\leq1)$, we have $r=2$.
Under Assumption~\ref{asp.moments} and \ref{asp.kernel}, we have by Proposition 4.3 in \cite{zhang2020convergence} that  
$
    P(\sup_\omega \| \flw(\omega)) -f(\omega)\|_{\max}\geq x) \leq \gps x,
$
where  the function $g$ is defined as 
\begin{equation} \label{eq.g2n}
\gps x=C_{\tau} n \left(\frac{M\max(1,\log(p)^{C_1})p^{2r(\tau)}}{nx }\right)^{\tau/2}+
CMp^2\exp\left(-C nx^2/M\right),
\end{equation}
with  $C_\tau,  C_1 $ and $ C$  generic constants. \cite{zhang2020convergence} obtained the above  result for  $C_1=5/2$. Observe that $r(\tau)$ in (\ref{eq.g2n}) is in some sense the price paid for allowing the dimension $p$ to  increase to infinity  and  $r(\tau)$ is bounded by $1/\tau$.

  Our  next two assumptions deal with the conditions  imposed on the inverse spectral density matrix $f^{-1}$  and  on its estimator $ \widehat{f}^{-1}$.
\begin{asp} \label{asp.spectral}
The inverse spectral density matrix $f^{-1}$  exists for all $p\in \N$ and satisfies for some positive constant $C<\infty$,
$\sup_\omega \|f^{-1}(\omega)\|_2<C \text{\  and } \sup_\omega \|f^{-1}(\omega)\|_1\leq C\cdot \spar,$
where $ \spar$ is an  increasing  sequence of the dimension $p\in \N$ of the process.
Furthermore,   
 $\sup_\omega \|\Sigma_n(\omega)-f(\omega)\|_{\max}\leq C/n$, $\sup_\omega \|\Sigma_n^{-1}(\omega)-f^{-1}(\omega)\|_{1}\leq C\spar/n$ and 
\begin{align} \label{asp.bias.spectral.estimator}
    \sup_\omega\|\sqrt{M/n} \sum_{k=1}^n \kappa_M(\omega-\omega_k) (\Sigma_n(\omega_k)-f(\omega))\|_{\max}\leq C\cdot \gpbias, 
\end{align}
with  $\gpbias=o(\log(n))$.
 \end{asp}
The condition $\sup_\omega \|f^{-1}(\omega)\|_2<C$ can  be interpreted  as a lower bound condition  for  the eigenvalues of the spectral density matrix $f$. The parameter $\spar$ restricts in fact  the growth of the cross-sectional (conditional) dependence structure of the process and if $f^{-1}$ is (weakly) sparse,
then $\spar$ also depends on  the (weakly) sparsity parameters. Expression \eqref{asp.bias.spectral.estimator} restricts  the order of the bias of the lag-window estimator $\fsp$ to be  $\gpbias$, where 
the condition $\gpbias=o(\log(n))$ is in fact a  lower bound condition  on  the truncation lag $M$. For instance, for   $K(u)=\ind(|u|\leq 1)$ the uniform kernel, we have $\gpbias\leq C \sqrt{n/M} \sum_{|h|\geq M} \|\Gamma(u)\|_{\max}$. Hence, $\sum_{h\in \Z} |h| \|\Gamma(h)\|_{\max} <\infty$   leads  to $M=n^a$ for some  $ a \geq 1/5$. Under Assumption~\ref{asp.moments}, we have $C \sqrt{n/M} \sum_{|h|\geq M} \|\Gamma(u)\|_{\max} \leq C \sqrt{n/M} \rho ^{M}$ and $a\geq \eps>0$ will  suffice. Similar results can be obtained  for other kernels.




As already mentioned,  we do not want to restrict ourselves to a particular estimator $ \widehat{f}^{-1}$ of  $ f^{-1}$. Therefore,   the following  assumption summarizes the conditions we impose on the estimator of $ f^{-1}$;  see Section~\ref{sec.coh.partial.coh} for a discussion of estimators fulfilling this rate.

 \begin{asp} \label{asp.spectral.density.estimator}
 A (regularized) estimator $\widehat f^{-1}$ of the inverse  spectral density matrix  $ f^{-1}$ exists which satisfies for $\gps x$ as given in \eqref{eq.g2n}, that,
 $
 P(\sup_\omega \|\widehat f^{-1}(\omega)-f^{-1}(\omega)\|_{\max} \geq x ) \leq \gps x
 $
 and
 $
 P( \sup_\omega \|\widehat f^{-1}(\omega)-f^{-1}(\omega)\|_1 \geq x) \leq \gps {x/\spar}.
  $
 \end{asp}

Our last  assumption deals with the rates at which the different parameters involved in our estimation procedure are allowed to increase to infinity, also  depending on the number of existing finite moments, so that their interplay delivers the desired behavior of the estimators involved. 
 
 \begin{asp}
 \label{asp.rates}
   $M=C n^a$ \  for some   $ a \in [1/5,4/7]$, \ $ \spar=n^s_1$ \ for  some $ s_1\leq 3/8(1-a)$ and  $ p\leq n^b$ for $b\geq 0$, such that, as $n\rightarrow\infty$, 
   \[ p^{\tau\cdot r(\tau)} s(p)^{\tau/2} n ^{-7\tau(1-a)/16+1+2b+a} =o(1).\]
   
 \end{asp}


Observe that the condition  imposed on  $a$, which determines the rate of increase of   the truncation lag  $M$, also allows for this parameter to increase  at the  rate $n^{1/5}$, which is the standard  optimal rate with respect to the mean square error of nonparametric estimators. Furthermore, the rate of increase  of  the  sparsity parameter $s(p)$, also  depends on the rate of increase of $M$ (i.e., the parameter $a$), which is due to the fact that  the truncation lag $M$,  affects the effective  number of observations used in the construction of the nonparametric estimators involved in our inference procedure. Finally,  if $b>0$,  the dimension $p$ of the multivariate system  is allowed to increase to infinity  at a polynomial rate as $n$ increases. This rate, however,    also depends  on the number of existing finite moments $\tau$. The larger $\tau$ is, the larger this rate can  be. If  $\{X_t,t\in\Z\}$ is a  Gaussian process, then even a geometric rate of increase of the dimension $p$ can be established.

We now consider  the limiting distribution of the estimators $\widehat{s}_{u,v} $ and $\widehat{\rho}_{u,v}^{(de)} $ introduced in the previous sections. 
 For the coherence estimator,   asymptotic normality can be established   under validity of Assumption  \ref{asp.moments} and Assumption  \ref{asp.kernel}. For instance,  \cite{koopmans1995spectral},
Section 8.4,  establishes such a result under slightly different assumptions than those used in this paper. More precisely,  we have for $u,v \in \{1,\dots,p\}$ and $\omega\not= \pi \Z$, that, as $n\rightarrow\infty$,
 $   \sqrt{n/M}(\hat s_{u,v}(\omega)-s_{u,v}(\omega))
    \overset{d}{\to}\psi_1^{(u,v)}+i \psi_2^{(u,v)}$,
where 
$$
\begin{pmatrix}
\psi^{(u,v)}_1 \\
\psi^{(u,v)}_2 
\end{pmatrix} \sim
\mathcal{N}\left(0,
\frac{C_{K_2}(1-|s_{u,v}(\omega)|^2)}{2}
\begin{pmatrix}
1-\Re(s_{u,v}(\omega))^2 & -\Re(s_{u,v}(\omega))\Im(s_{u,v}(\omega)) \\
-\Re(s_{u,v}(\omega))\Im(s_{u,v}(\omega))& 1-\Im(s_{u,v}(\omega))^2
\end{pmatrix}\right).
$$
Moreover, for $\omega \in \pi \Z$ we have,
$
    \sqrt{n/M}(\hat s_{u,v}(\omega)-s_{u,v}(\omega))\overset{d}{\to}\mathcal{N}(0,C_{K_2}(1-|s_{u,v}(\omega)|^2)^2).
$

Deriving the limiting distributional properties of the de-biased estimator
$\widehat{\rho}_{u,v}^{(de)}(\omega)$ introduced in Section~\ref{sec.coh.partial.coh} is  more involved.
The following theorem  establishes asymptotic normality and shows
that the limiting Gaussian  distribution  has 
a  structure which is similar to that of the coherence estimator   $\hat s_{u,v}(\omega)$. 

\begin{thm}\label{thm.partial.coherence}
Under validity of Assumptions 1-5
and  for $u,v\in \{1,\dots,p\}$,   we have  for $\omega\not= \pi \Z$, that, as $n\rightarrow \infty$, 
$
    \sqrt{n/M}(\hat \rho_{u,v}^{(de)}(\omega)-\rho_{u,v}(\omega))\overset{d}{\to} \xi_1^{(u,v)}+i \xi_2^{(u,v)},
$
where
$\big(
    \xi_1^{(u,v)}, \
    \xi_2^{(u,v)}\big)^{\top}
     \sim \mathcal{N} \left(0, V_{(u,v)}(\omega)\right)
$
and
$$
V_{(u,v)}(\omega)=\frac{C_{K_2}(1-|\rho_{u,v}(\omega)|^2)}{2}\begin{pmatrix}
1-\Re(\rho_{u,v}(\omega))^2 & -\Re(\rho_{u,v}(\omega))\Im(\rho_{u,v}(\omega)) \\
-\Re(\rho_{u,v}(\omega))\Im(\rho_{u,v}(\omega)) & 1-\Im(\rho_{u,v}(\omega))^2 
\end{pmatrix}.
$$
Furthermore, for  $\omega \in \pi \Z$, 
$
    \sqrt{n/M}(\hat \rho_{u,v}^{(de)}(\omega)-\rho_{u,v}(\omega))\overset{d}{\to}\mathcal{N}(0,C_{K_2}(1-|\rho_{u,v}(\omega)|^2)^2).
$


\end{thm}

The following corollary   deals with     the limiting covariance between the  estimators 
$\hat\rho^{(de)}_{u_i,v_j}(\omega)$ 
for  different pairs of indices  $ (u_i,v_j)$, $i,j=1,2$.

\begin{corollary} \label{cor.coherence.cov} Let  $u_1,u_2,v_1,v_2\in \{1,\dots,p\}$ and $\omega \in [0,2\pi]$. Under the assumptions of Theorem~\ref{thm.partial.coherence},  we have,  as $n \to \infty$,   
\begin{align*}
    \frac{n/M}{C_{K_2}}&\cov(\hat \rho_{u_1,v_1}^{(de)}(\omega),\hat \rho_{u_2,v_2}^{(de)}(\omega))\to \rho_{u_1,u_2}(\omega)\rho_{v_2,v_1}(\omega)
    +\ind(\omega \in \pi \Z) \rho_{u_1,v_2}(\omega)\rho_{u_2,v_1}(\omega)
    \\&
    \frac{1+\ind(\omega \in \pi \Z)}{2}\Big[-
    \rho_{v_2,u_2}(\omega)[\rho_{u_1,v_2}(\omega)\rho_{v_2,v_1}(\omega)+\rho_{u_1,u_2}(\omega)\rho_{u_2,v_1}(\omega)]\\
    &-\rho_{u_1,v_1}(\omega)[\rho_{v_2,v_1}(\omega)\rho_{v_1,u_2}(\omega)+\rho_{v_2,u_1}(\omega)\rho_{u_1,u_2}(\omega)]\\
    &+1/2\rho_{u_1,v_1}(\omega)\rho_{v_2,u_2}(\omega)[
    |\rho_{v_1,v_2}(\omega)|^2+|\rho_{v_1,u_2}(\omega)|^2+|\rho_{u_1,v_2}(\omega)|^2+ |\rho_{u_1,u_2}(\omega)|^2]\Big].
\end{align*}
\end{corollary}

In our applications
the focus is not only on the behavior of $ \rho_{u,v}$ at a particular  frequency $\omega$ but also  on a set of  frequencies belonging to a frequency  band,  $\mathcal{W}\subset [0,\pi]$, of interest.
 In such a case and in order to handle the behavior of the de-biased coherence $ \widehat{\rho}^{(de)}_{u,v}(\omega)$  for all  $\omega\in {\mathcal W}$,  distributional  results are needed which simultaneously  hold true  for   the entire  set of frequencies  considered. The following theorem
 deals with the asymptotic distribution of the maximum of the standardized real and imaginary parts of the estimator $\hat \rho_{u,v}^{(de)} $ evaluated over a growing grid of frequencies belonging  to the frequency band  ${\mathcal W} $.    \cite{liu2010asymptotics} and \cite{wu2018asymptotic} derived an asymptotic,  Gumbel type  approximation for the maximum deviation of the spectral density over  growing sets of frequencies. 
The following theorem   extends this result to the case of the  de-biased estimator of the partial coherence proposed in this paper. Furthermore, it   gives a uniform convergence result for  the ratio of  the upper tails of the distribution of the maximum deviation of the standardized,
de-biased partial coherence and its  asymptotic  $\chi^2$ approximation. Such a  result  is of great  importance  if  interest is focused on the behavior of   the upper quantiles of the aforementioned distributions, which  is the case if  one deals with multiple testing problems.
This result,  stated in Theorem~\ref{thm.partial.coherence.max} below,  is  
the main ingredient in proving the false discovery rate control associated with the thresholding-based,  multiple testing procedure proposed in this paper.

\begin{thm}\label{thm.partial.coherence.max}
Let $\omega_l^\prime=\pi l N/M
, M=C n^a$ and $\mathcal{L}=\{ l =1,\dots,M/N-1 : \omega_l^\prime \in \mathcal{W}\}$ with $d=|\mathcal{L}|$. Then, 
  under Assumptions 1 to 5, we have for $u,v\in \{1,\dots,p\}$, that, as $n\rightarrow\infty$, 
$$
\sup_{0\leq t \leq 2(a+2b)\log(n)} \Big| \frac{P(n/M \max_l \chi_{(u,v)}(\omega_l^\prime)  \geq t)}{G_d(t)} -1 \Big|=o(1), 
$$
where $G_d(t)=1-(1-\exp(-t/2))^d=P(\max_{1\leq l \leq d} Z_l \geq t), Z_l\sim \chi_2^2, iid$,
$$
\chi_{(u,v)}(\omega)=\begin{pmatrix}
\Re(\hat \rho_{u,v}^{(de)}(\omega)-\rho_{u,v}(\omega)) \\
\Im(\hat \rho_{u,v}^{(de)}(\omega)-\rho_{u,v}(\omega))
\end{pmatrix}^\top
\hat V^{-1}_{(u,v)} (\omega)
\begin{pmatrix}
\Re(\hat \rho_{u,v}^{(de)}(\omega)-\rho_{u,v}(\omega)) \\
\Im(\hat \rho_{u,v}^{(de)}(\omega)-\rho_{u,v}(\omega))
\end{pmatrix}.
$$
\end{thm}
Notice that  $\sup_t |G_d(t)/\widetilde{G}_d(t)-1|\leq \exp(-1)/d$ holds true. Hence,  instead of $G_d(t)=1-(1-\exp(-t/2))^d$ we could also use $\widetilde{G}_d (t)=1-\exp(-\exp(-(t+2\log(d))/2))$, with   $\widetilde{G}_d$ the Gumbel distribution with scaling factor $2$ and shifted by $2\log(d)$.

  \begin{center}
    {\LARGE\bf Supplementary Material to \\Frequency Domain Statistical Inference for High-Dimensional Time Series}
\end{center}
\appendix

\section{Auxiliary Results}
We split this section  into four subsections. The first  is devoted to error bounds related to  the construction of the de-biased partial coherence estimator, the second  contains  useful lemmas for  the Gaussian approximation, the third deals with  the covariance structure of the lag-window estimator
and the last subsection presents   some additional  useful lemmas.

\subsection{Error bounds in the construction of de-biased partial coherences} \label{subsection.lemmas.partial}

\begin{lem} \label{lem.beta.gamma}
Under Assumption~\ref{asp.spectral} and \ref{asp.spectral.density.estimator} we have $\max_v \sup_w \|\beta_v(\omega)\|_1=O(\spar)$, 
$$
\max_{v} \sup_\omega \|\hat \beta_v-\beta_v\|_1=\bp(\gps {(x-\gpbias)/\spar}),
$$
and
$$
\max_{v} \sup_\omega \|\hat \gamma_{-v}(\omega)-\gamma_{-v}(\omega)\|_1=\bp(\gps {(x-\gpbias)/\spar}).
$$
\end{lem}
\begin{proof}[Proof]
The first assertion follows immediately by Assumption~\ref{asp.spectral}. Note that  
\begin{align*}
\hat \beta_v(\omega)-\beta_v(\omega)=&-I_{p,-v}^\top(\hat f^{-1}(\omega)-\Sigma_n^{-1}(\omega) )e_v (\Sigma_{n,v,v}^{-1}(\omega)^{-1})\\
&-I_{p,-v}^\top\Sigma_n^{-1}(\omega) e_v [(\hat f_{n,v,v}^{-1}(\omega))^{-1}-(\Sigma_{n,v,v}^{-1}(\omega))^{-1}]\\
&-I_{p,-v}^\top(\hat f^{-1}(\omega)-\Sigma_n^{-1}(\omega)) e_v [(\hat f_{n,v,v}^{-1}(\omega))^{-1}-(\Sigma_{n,v,v}^{-1}(\omega))^{-1}]    
\end{align*}
from which  the second assertion follows by \eqref{eq.g2n}
and since $\|\cdot\|_1$ is sub-multiplicative. The third assertion follows by similar arguments.
\end{proof}

\begin{lem}\label{lem.beta.bias}
Under Assumption \ref{asp.moments},\ref{asp.kernel}, \ref{asp.spectral}, and \ref{asp.spectral.density.estimator}, we have for  $u,v=1\dots,p$ and $\omega\in[0,2\pi]$
\begin{align*}
&\sqrt{n/M}(\hat \beta_{v,\tilde u}^{(de)}(\omega)-\beta_{v,\tilde u}(\omega))=\delta(\omega)+\\
&\frac{\sum_{k=1}^n \kappa_M(\omega-\omega_k) e_v^\top f^{-1}(\omega) [Z(\omega_k) Z^H(\omega_k)-f(\omega)][f^{-1}(\omega) e_u f_{v,v}^{-1}(\omega)-f^{-1}(\omega) e_v f_{v,u}^{-1}(\omega)]}{\sqrt{nM}(f_{v,v}^{-1}(\omega))^2},
\end{align*}
where $=\sup_\omega |\delta(\omega)|=\bp(\gps {((x-\gpbias)/(\spar (n/M)^{1/4}))^{1/2}}+ \gps 1)=:\bp(\gpbetade x)$. 
\end{lem}
\begin{proof}[Proof]
To simplify notation let 
$$
DN=1/n\sum_{k=1}^n \kappa_M(\omega-\omega_k) Z_{u}(\omega_k) Z_{-v}^{H}(\omega_k) \hat \gamma_{-v,\tilde u}(\omega).
$$
By Lemma~\ref{lem.beta.gamma} we have  $\sup_\omega |DN-f^{-1}_{v,v}(\omega)|=\bp(\gps {(x-\gpbias)/\spar}).$
Furthermore, note that for some vector $c \in \R^p$,
$e_v^\top \Sigma_n^{-1}(\omega) \Sigma_n(\omega) I_{p;-v} c =0$.
Then, using $e_v^\top - \beta_v^H(\omega) I_{p,-v}= e_v^\top \Sigma_n^{-1}(\omega)^{-1}/\Sigma_{n,v,v}^{-1}(\omega)$  we have
\begin{align*}
    \sqrt{n/M}&(\hat \beta_{v,\tilde u}^{(de)}(\omega)-\beta_{v,\tilde u}(\omega))=\frac{\sum_{k=1}^n \kappa_M(\omega-\omega_k)[Z_v(\omega_k)-\hat \beta_{v-u}^H(\omega) Z_{-(v,u)}(\omega_k)]Z_{-v}^{H}(\omega_k) \hat \gamma_{-v,\tilde u}(\omega)}
    {\sqrt{Mn} DN}\\
    &-\frac{ \sqrt{n/M}\beta_{v,\tilde u}(\omega) DN}{DN}\\
    &=\frac{\sum_{k=1}^n \kappa_M(\omega-\omega_k) [Z_v(\omega_k)-\beta_v^H(\omega) Z_{-v}(\omega_k)] Z_{-v}^H(\omega_k)\hat \gamma_{-v,u}(\omega)}{\sqrt{Mn} DN} \\
    &+\frac{\sum_{k=1}^n \kappa_M(\omega-\omega_k) [\beta_{v,-u}(\omega)-\hat \beta_{v,-u}(\omega)]^H Z_{-(v,u)}(\omega_k) Z_{-v}^H(\omega_k)\hat \gamma_{-v,u}(\omega)}{\sqrt{Mn} DN}\\
    &=\frac{\sum_{k=1}^n \kappa_M(\omega-\omega_k) e_v^\top \Sigma_n^{-1}(\omega) [Z(\omega_k) Z^H(\omega_k)-\Sigma_n(\omega)]
    I_{p;-v} \gamma_{-v,u}(\omega)}{\sqrt{Mn} DN \Sigma_{n,v,v}^{-1}(\omega)} \\
    &+\frac{\sum_{k=1}^n \kappa_M(\omega-\omega_k) e_v^\top \Sigma_n^{-1}(\omega) [Z(\omega_k) Z^H(\omega_k)-\Sigma_n(\omega)]
    I_{p;-v} [\hat \gamma_{-v,u}(\omega)- \gamma_{-v,u}(\omega)]}{\sqrt{Mn} DN \Sigma_{n,v,v}^{-1}(\omega)} \\
&+\frac{\sum_{k=1}^n \kappa_M(\omega-\omega_k) [\beta_{v,-u}(\omega)-\hat \beta_{v,-u}(\omega)]^H [Z_{-(v,u)}(\omega_k) Z_{-v}^H(\omega_k)-\Sigma_n(\omega)]\gamma_{-v,u}(\omega)}{\sqrt{Mn} DN}\\
&+\frac{\sum_{k=1}^n \kappa_M(\omega-\omega_k) [\beta_{v,-u}(\omega)-\hat \beta_{v,-u}(\omega)]^H [Z_{-(v,u)}(\omega_k) Z_{-v}^H(\omega_k)][\hat \gamma_{-v,u}(\omega)-\gamma_{-v,u}(\omega)]}{\sqrt{Mn} DN}\\
&=\frac{\sum_{k=1}^n \kappa_M(\omega-\omega_k) e_v^\top f^{-1}(\omega) [Z(\omega_k) Z^H(\omega_k)-f(\omega)]
    I_{p;-v} \gamma_{-v,u}(\omega)}{\sqrt{Mn} (f_{v,v}^{-1}(\omega))^2}
    +\delta(\omega).
\end{align*}
We investigate the remainder  $\delta(\omega)$. This remainder  consists of the errors caused by  replacing $\hat \beta_{v,-u}(\omega)$ by $\beta_{v,-u}(\omega)$, $\hat \gamma_{-v,u}(\omega)$ by $ \gamma_{-v,u}(\omega)$, $DN$ by $f_{v,v}^{-1}(\omega)$, and $\Sigma_n(\omega)$ by $f(\omega)$. Denote these errors by $I$ to $IV$. By Lemma~\ref{lem.beta.gamma} that $I, II, III$ are of order $\bp(\gps {(x-\gpbias)/\spar})$
in $\|\cdot\|_1$-norm and  uniformly in $\omega$. Furthermore, we have by Assumption~\ref{asp.spectral} $\|\Sigma_n(\omega)-f(\omega)\|_{\max}=O(1/n)$ and $\|\Sigma_n^{-1}(\omega)-f^{-1}(\omega)\|_{1}=O(\spar/n)$. Additionally, note that we have by \eqref{eq.g2n} $\sup_\omega\|1/(n h) \sum_{k=1}^n \kappa_M(\omega-\omega_k) [Z(\omega_k) Z^H(\omega_k)-\Sigma_n(\omega)]\|_{\max}=\sup_\omega\|\fsp(\omega)-E \fsp(\omega)\|_{\max}=\bp (\gps{x})$ and $\sup_\omega \|\Sigma_n^{-1}(\omega)\|_1=O(\spar)$. From this we have
\begin{align*}
&\frac{\sum_{k=1}^n \kappa_M(\omega-\omega_k) e_v^\top \Sigma_n^{-1}(\omega) [Z(\omega_k) Z^H(\omega_k)-\Sigma_n(\omega)]
    I_{p;-v} [\hat \gamma_{-v,u}(\omega)- \gamma_{-v,u}(\omega)]}{\sqrt{Mn} DN \Sigma_{n,v,v}^{-1}(\omega)}  \\
    &\leq \sqrt{n/M}\|\Sigma_n^{-1}(\omega)\| \|\fsp(\omega)-E \fsp(\omega)\|_{\max} \|[\hat \gamma_{-v,u}(\omega)- \gamma_{-v,u}(\omega)]\|_1 |DN \Sigma_{n,v,v}^{-1}(\omega)|\\
    &\leq (n/M)^{1/4}\spar \|\fsp(\omega)-E \fsp(\omega)\|_{\max} (n/M)^{1/4}\|[\hat \gamma_{-v,u}(\omega)- \gamma_{-v,u}(\omega)]\|_1 |DN \Sigma_{n,v,v}^{-1}(\omega)|\\
    &= \bp(\gps {((x-\gpbias)/(\spar (n/M)^{1/4}))^{1/2}}+ \gps 1),
\end{align*}
where  the above equality  holds uniformly in $\omega$. The other  terms  of $\delta(\omega)$ are of the same order and, therefore,
$$\sup_\omega |\delta(\omega)|=\bp(\gps {((x-\gpbias)/(\spar (n/M)^{1/4}))^{1/2}}+ \gps 1).$$

\end{proof}

\subsection{Error bounds for Gaussian approximations} \label{subsection.lemmas.Gaussian.approximaiton}
In all Lemmas of  this subsection,  Assumption~\ref{asp.moments} to \ref{asp.rates} hold true.

\begin{lem} \label{lem.approximation.g.m}
For $l=1,\dots,M/N-1$ and if $\Theta_{0,\tau}<\infty$ and for some sequence $g_n$ $d_{m,\tau}=\sum_{t=0}^\infty \min (\delta_{t,\tau},\Psi_{m+1,\tau})=o(g_n)$ , we have for some $q \leq \tau $ 
$$
\max_l \frac{1}{\sqrt{Mn}} |g_{n}^{[u,v]}(\omega_l^\prime)-g_{n,m}^{[u,v]}(\omega_l^\prime)|=\bp( (\rho^m/x)^{q/2})
$$
and 
$E(\frac{1}{\sqrt{Mn}} |g_{n}^{[u,v]}(\omega_l^\prime)-g_{n,m}^{[u,v]}(\omega_l^\prime)|)^{q/2}=O(d_{m,q}^{q/2})$.
\end{lem}
\begin{proof}
Following the proof of Lemma A.1 in \cite{liu2010asymptotics}, we have by Lemma A.2 in \cite{wu2018asymptotic} and some $q \leq \tau $ 
$$E(\frac{1}{\sqrt{Mn}} |g_{n}^{[u,v]}(\omega_l^\prime)-g_{n,m}^{[u,v]}(\omega_l^\prime)|)^{q/2}=O(d_{m,q}^{q/2}).$$
Under Assumption~\ref{asp.moments} we have $d_{m,q}=\sum_{t=0}^\infty \min(\delta_{t,q}^{[\max]},(\sum_{j=m}^\infty (\delta_{j,q}^{[max]})^2)^{1/2})\leq C \rho^m$. Hence,  by Markov's inequality
\begin{align*}
    P(\max_l \frac{1}{\sqrt{Mn}} |g_{n}^{[u,v]}(\omega_l^\prime)-g_{n,m}^{[u,v]}(\omega_l^\prime)| \geq x)\leq C (\rho^m/x)^{q/2}.
\end{align*}
\end{proof}

\begin{lem} \label{lem.approximation.g.m.bar}
For $q\leq \tau/2$ it holds true that 
\begin{align*}
E( \frac{1}{\sqrt{Mn}}\max_l |\bar g_{n,m}(l)^{[u,v]} & -g_{n,m}(l)^{[u,v]}|)^q  \leq C_\tau m^{2q} \big[(Mn)^{-1+q/\tau}\\
&  +M^{-q/2}(Mn)^{-1+(2q-1)/\tau+(q^2-q)/\tau^2}\big]
\end{align*}
and 
$$
\frac{1}{\sqrt{Mn}}\max_l |\bar g_{n,m}(l)^{[u,v]}-g_{n,m}(l)^{[u,v]}|=\bp(C_\tau m^{3/2\tau-2} ((x \sqrt{Mn})^{-\tau/2} n+m^{\tau/2}x^{-\tau} (Mn)^{-\tau/2+1})).
$$
\end{lem}
\begin{proof}
First note that since the kernel is bounded, $\bar g_{n,m}(l)$ can be written as 
$\bar g_{n,m}(l)=\sum_{t=2}^n \bar U_{t,l,m} \sum_{s=1}^{t-1} a_{n,t-s} \bar V_{s,l,m}$ for some bounded coefficients $a_{n,i} \in \C$. A similar expression for  $g_{n,m}(l)$ holds true. We then  obtain 
for some constant $C>0$, that, 
\begin{align*}
    \max_l | \bar g_{n,m}(l)^{[u,v]}-g_{n,m}(l)^{[u,v]}|\leq& C \max_l \sum_{t=2}^n |\bar U_{t,l,m}-U_{t,l,m}| \sum_{s=\max(1,t-M)}^{t-1} |  V_{s,l,m} | \\
    &+
    C \max_l \sum_{t=2}^n |U_{t,l,m}| \sum_{s=\max(1,t-M)}^{t-1} |V_{s,l,m}-\bar V_{s,l,m}|\\
    &+C \max_l \sum_{t=2}^n |\bar V_{t,l,m}-V_{t,l,m}| \sum_{s=\max(1,t-M)}^{t-1} |  U_{s,l,m} | \\
    &+
    C \max_l \sum_{t=2}^n |V_{t,l,m}| \sum_{s=\max(1,t-M)}^{t-1} |U_{s,l,m}-\bar U_{s,l,m}|.
\end{align*}
The four terms  of the above inequality, can be treated by the same arguments. We focus here on the first only. Note that for all $t$ it holds true that $E |\max_l U_{t,l,m}|^\tau<\infty $, $E |\max_l V_{t,l,m}|^\tau<\infty$. Hence, by Hölder's inequality and Markov's inequality
\begin{align*}
E (\max_l |\bar U_{t,l,m}-U_{t,l,m}|)^q &= E \max_l |U_{t,l,m}|^q \ind(|U_{t,l,m}|> (Mn)^\iota)\\
&\leq  E |U_{t,l,m}|^\tau P(|U_{t,l,m}|\geq (Mn)^\iota)^{1-q/\tau}=O((Mn)^{-1+q/\tau}).
\end{align*}

Using  $m$-dependency and Hölder's inequality, we obtain  

\begin{align*}
    &(1/(Mn))^{q/2}E \left(\max_l \sum_{t=2}^n |\bar U_{t,l,m}-U_{t,l,m}| \sum_{s=\max(1,t-M)}^{t-1} |  V_{s,l,m} |)\right)^q \\
    \leq& C_\tau (1/(Mn))^{q/2} E \left(\max_l \sum_{t=m+1}^n |\bar U_{t,l,m}-U_{t,l,m}| \sum_{s=\max(1,t-M)}^{t-m} |  V_{s,l,m} |\right)^q \\
    &+C_\tau E \left( \max_l \sum_{t=2}^n |\bar U_{t,l,m}-U_{t,l,m}| \sum_{s=\max(1,t-m+1)}^{t-1} |  V_{s,l,m}|\right)^q \\
    \leq& C_\tau (1/(Mn))^{q/2} \Bigg[(Mn)^{q/2} m^{2(q-1)} (Mn)^{-1+q/\tau}+\\
    &+n^{q/2}m^{2q-1}(Mn)^{-1+(\tau/q-1)^{-1}+q/\tau-(q/\tau)/(\tau/q-1)}\Bigg] \\
        &\leq C_\tau m^{2q} [(Mn)^{-1+q/\tau}+M^{-q/2}(Mn)^{-1+(\tau/q-1)^{-1}+q/\tau-(q/\tau)/(\tau/q-1)}].
\end{align*}
This proves  the first assertion of the lemma. For the second assertion, we  use the same splitting arguments as above and get 
\begin{align*}
    \max_l& \sum_{t=2}^n |\bar U_{t,l,m}-U_{t,l,m}| \sum_{s=\max(1,t-M)}^{t-1} |  V_{s,l,m} |\\
    =& \sum_{t=m+1}^n \max_l |\bar U_{t,l,m}-U_{t,l,m}| \sum_{s=\max(1,t-M)}^{t-m} |  V_{s,l,m} |
    \\
    &+ \sum_{t=2}^n \max_l |\bar U_{t,l,m}-U_{t,l,m}| \sum_{s=\max(1,t-m+1)}^{t-1} |  V_{s,l,m}|\\
    =&: \sum_{t=m+1}^n W_t+\sum_{t=2}^n \tilde W_t,
\end{align*}
with an obvious notation for $W_t$ and $\widetilde W_t$. We obtain by Lemma~2 in \cite{liu2010asymptotics} and  for $Q=\tau/2$ and the arguments used above, that 
\begin{align*}
    P(\sum_{t=m+1}^n W_t \geq x) &\leq C_Q (m/x^2 E (\sum_{t=m+1}^n W_t)^2)^Q+C_Q m^{\tau-1}/x^\tau \sum_{t=m+1}^n E W_t^\tau \\
    & \leq C_Q (m/x^2 m^2 (Mn)^{2/\tau})^Q+C_Q m^{\tau-1}x^{-\tau} n M m^{\tau-1}\\
    & \leq C_\tau m^{2\tau-2}x^{-\tau} n M.
\end{align*}
Furthermore, we have for $\widetilde W_t$
\begin{align*}
    P(\sum_{t=m+1}^n \widetilde W_t \geq x) &\leq C_Q (m/x^2 E (\sum_{t=m+1}^n \widetilde W_t)^2)^Q+C_Q m^{\tau/2-1}/x^{\tau/2} \sum_{t=m+1}^n E \widetilde W_t^{\tau/2} \\
    & \leq C_Q (m/x^2 m^2 M^{-1} (Mn)^{(2\tau-1)^{-1}})^Q+C_Q m^{\tau/2-1}x^{-\tau/2} n m^{\tau-1}\\
    & \leq C_\tau m^{3/2\tau-2}x^{-\tau/2} n.
\end{align*}
Putting this together with $x=x\sqrt{Mn}$ gives the second assertion.

\end{proof}





\begin{lem} \label{lem.variance.u.v}
We have 
$$
 E (\max_l \sum_{j=1}^{k_n+1} v_j(l)^{[u,v]})^2\leq C n(M)^{1-\beta} m^{4}
$$

and
$$
  E (\max_l \sum_{j=1}^{k_n} u_j(l)^{[u,v]})^2\leq C Mn m^{4}.
$$
\end{lem}
\begin{proof}
First note that by independence $$
E (\sum_{j=1}^{k_n+1} \max_l v_j(l)^{[u,v]})^2=\sum_{j=1}^{k_n+1} E (\max_l v_j(l)^{[u,v]})^2.
$$ 
For $j=1,\dots,k_n$ and similarly to the proof of Lemma~\ref{lem.approximation.g.m.bar},   the properties of the  kernel $K$ and the   $m$-dependence of $\bar U_{t,l,m},V_{t,l,m}$, we have 
\begin{align*}
E (\max_l v_j(l)^{[u,v]})^2&\leq C E(\sum_{t\in I_j} \max_l \bar U_{t,l,m} )^2 E (\sum_{s=\max(1,t-M)}^{t-4m} K((t-s)/M) \exp(-i(t-s)\omega_l^\prime)\bar V_{s,l,m})^2\\
&+C_\tau  E( \sum_{t\in I_j} \max_l \bar U_{t,l,m} \sum_{s=t-4m+1}^{t-1} K((t-s)/M) \exp(-i(t-s)\omega_l^\prime) \bar V_{s,l,m})^{2} \\
&\leq C |I_j|M m^{3}+|I_j|m^{2q-1} \leq C_q |I_j| m^{3}(M+m).
\end{align*}
Similar arguments apply to $u_j(l)$ with $I_j$ replaced by $H_j$. We have  $|I_j|=CM+m, j=1,\dots,k_n$, $|I_{k_n+1}|< C(M)^{1+\beta}+M+m, \beta>0$, $|H_j|=C(M)^{1+\beta}$, and $k_n=\floor{n/(|I_1|+|H_1|)}$. 
Thus, 
\begin{align*}
\max_l E (\sum_{j=1}^{k_n+1} v_j(l)^{[u,v]})^2&\leq C (n/(1+|H_1|/|I_1|) m^{3}(M+m))
\leq C_q n(M)^{1-\beta} m^{4}
\end{align*}
and
\begin{align*}
\max_l E (\sum_{j=1}^{k_n+1} u_j(l)^{[u,v]})^2&\leq C Mn m^{4}.    
\end{align*}

\end{proof}

\begin{lem}\label{lem.approximation.gamma.i}
It holds true  that
\begin{align*}
\sqrt{1/(Mn)} \max_l & |\Gamma^{[u,v]}(l)|\\
& =\bp(C \log(n)^{C_\tau} (x n^{1/2-(a+2)/\tau-2(1+a)/\tau^2})^{-\tau}+x^{-C\tau} (Mn)^{-\tau}),
\end{align*}
where  $\Gamma^{[u,v]}(l)=\sum_{j=1}^{k_n+1} v_j^{[u,v]}(l).$
\end{lem}
\begin{proof}
Note first that 
\begin{align*}
    P(\max_l \sqrt{1/(Mn)} |\Gamma^{[u,v]}(l)|>x) & \leq \sum_{l} P( |\Gamma^{[u,v]}(l)|> \sqrt{Mn}x)\\
    & \leq M P( |\Gamma^{[u,v]}(l)|> \sqrt{Mn}x).
    \end{align*}
We focus on the term on the RHS of the last inequality. For this, we use  Lemma~2 in \cite{liu2010asymptotics} and obtain for any $Q$ and some positive constants $C_Q$
$$
P(|\Gamma(l)^{[u,v]}|\geq \sqrt{Mn} x) \leq C (\frac{\sum_{j=1}^{k_n+1} E |v_j^{[u,v]}(l)|^2}{Mn x})^Q+C \sum_{j=1}^{k_n+1} P(|v_j^{[u,v]}(l)|\geq C_q \sqrt{Mn} x). 
$$
For the first term on the RHS we have by Lemma~\ref{lem.variance.u.v}
$$\frac{\sum_{j=1}^{k_n+1} E |v_j^{[u,v]}(l)|^2}{Mn x }\leq C (n^{-\beta a}m^2 x).$$ 
Hence, for some $Q=C \tau$ we have that this term is of order $O( x^{-C\tau} (Mn)^{-\tau}).$

Furthermore, by Lemma A.5 in \cite{wu2018asymptotic}, see also Proposition~3 in \cite{liu2010asymptotics} (note that we do not use the replacement of $m$ by $n$ for the second and third term on the RHS of Proposition~3), we have with  $M=(Mn)^{\iota}, k=M+m$ and  $ x=\sqrt{Mn} x, y=4 \log((nxM)^{\tau})$, that  
\begin{align*}
P(|v_j^{[u,v]}(l)|\geq& C_q \sqrt{Mn}x)\leq 2 (Mnx)^{-\tau}\\
&+C nm^2 (Mn)^{2\iota} \Big( (Mn)^{-1} x^{-2} \log((nxM)^{\tau})^{-2} m^3 [(Mn)^{2\iota}+M+m]M\Big)^Q\\
&+C n^2m^2 (Mn)^{2\iota} \max_{Z \in \{U,V\}}P\left(|\bar Z_{0,l,m}|\geq \frac{C  \log((nxM)^{\tau})^{-1} x\sqrt{ Mn }}{m^2 ((Mn)^{\iota}+(M+m))^{1/2}}\right)\\
&=I+II+III.
\end{align*}
Consider  the second term on the RHS and recall that $M=n^{a}$.
\begin{align*}
  \frac{m^3 [(Mn)^{2\iota}+M+m]M}{Mn x^{2} \log((Mn)^{\tau})^{2}}\leq x^{-2} \log(n)^{C_\tau} (n^{-1+a}+n^{(1-a)(-1+2/\tau)+a(4/\tau-1)}).
\end{align*}
Hence, for $Q=C \tau$ we obtain $II=O( x^{-C\tau} (Mn)^{-\tau}) $. 

For the third term we use Markov's inequality and obtain for $\tau\geq 8$
\begin{align*}
    III&\leq C n^2m^2 (Mn)^{2\iota} \max_{Z \in \{U,V\}} E |\bar Z_{0,l,m}|^\tau \left(\frac{m^2 ((Mn)^{\iota}+(M+m))^{1/2}}{ \log((Mn/x)^{\tau}) x\sqrt{Mn}}\right)^\tau \\
    & \leq C \log(n)^{C_\tau}  n^{-\tau/2+2+(1+a)2/\tau}x^{-\tau}=C \log(n)^{C_\tau} (x n^{1/2-2/\tau-2(1+a)/\tau^2})^{-\tau}.
    \end{align*}


\end{proof}

\begin{lem} \label{lem.approximation.u.hat}
Let $\xi$ be defined as in  Remark~\ref{rmk.definition.u.v} of the main paper  and let $$\hat u_j(l)^{[u,v]}=u_j(l)^{[u,v]} \ind(|u_j(l)^{[u,v]}|\leq \sqrt{Mn} \xi)- E (u_j(l)^{[u,v]} \ind(|u_j(l)^{[u,v]}|\leq \sqrt{Mn} \xi)$$ be a truncated version of $u_j(l)^{[u,v]}$, $j=1,\dots,k_n, l=1,\dots,M/N-1$. 
We have  
\begin{align*}
\frac{1}{\sqrt{Mn}}\max_l |&  \sum_{j=1}^{k_n} \hat u_j(l)^{[u,v]}- u_j(l)^{[u,v]} |\\
& = \bp \Big( \frac{\log(n)^{C_\tau} }{(x n^{1/2(1-1/4a)-1/\tau(2-a)-2(1+a)/\tau^2})^{\tau}}+x^{-C\tau} \frac{1}{(Mn)^{\tau}}\Big)
\end{align*}
and
\begin{align*}
\frac{1}{Mn} \max_l  &  \sum_{j=1}^{k_n} E|\hat u_j(l)^{[u,v]}- u_j(l)^{[u,v]}|^2\\
& =O\Big(\frac{\log(n)^{C_\tau}}{ (\xi n^{1/2(1-1/4a)-1/\tau(2-a)-2(1+a)/\tau^2})^{\tau}}+\xi^{-C\tau} \frac{1}{(Mn)^{\tau}}\Big)
\end{align*}

\end{lem}
\begin{proof}
Let $\tilde Z_j(l)=u_j(l)^{[u,v]} \ind(|u_j(l)^{[u,v]}|> \sqrt{Mn} \xi)$. We  follow the lines of proof of Lemma~\ref{lem.approximation.gamma.i}. The main difference is that $u_j$ consist of $p_n=(M)^{1+\beta}$ elements whereas in Lemma~\ref{lem.approximation.gamma.i},  $v_j$ consist of only $M$ elements. We obtain
$$
P(|\sum_{j=1}^{k_n} \tilde Z_j(l) |\geq \sqrt{Mn} x) \leq C \Big(\frac{\sum_{j=1}^{k_n+1} E |\tilde Z_j(l)|^2}{Mn x }\Big)^Q+C \sum_{j=1}^{k_n+1} P(|\tilde Z_j(l)|\geq C_q \sqrt{Mn} x). $$
We have by Cauchy-Schwarz's  inequality, that $E |\tilde Z_j(l)|^2 \leq (E(u_j(l)^{[u,v]})^4 P(|u_j(l)^{[u,v]}|\geq C_q \sqrt{Mn} \xi))^{1/2}$.
Furthermore, we have with similar arguments as in Lemma~\ref{lem.variance.u.v} $E(u_j(l)^{[u,v]})^4\leq C (|H_j|M)^2 m^8$. This  means that  
$
(\frac{\sum_{j=1}^{k_n+1} E |\tilde Z_j(l)|^2}{Mn x })^Q \leq (P(|u_j(l)^{[u,v]}|\geq C_q \sqrt{Mn} \xi))^{Q/2}.
$
In the above derivation eight  moments are required. Note that since $Q$ can be chosen large, it is also possible to use Hölder's inequality so that only $4+\delta$ for some $\delta>0$ moments are required. 

We  first investigate  $P(|\tilde Z_j(l)|\geq C_q \sqrt{Mn} x)$. Similar to the arguments used for $P(|v_j^{[u,v]}(l)|\geq C_q \sqrt{Mn} x)$ in the proof of Lemma~\ref{lem.approximation.gamma.i},  we have by Lemma A.5 in \cite{wu2018asymptotic}, see also Proposition~3 in \cite{liu2010asymptotics}, 
with $M=(Mn)^{\iota}, k=(M)^{1+\beta}, x=\sqrt{Mn} x, y=4 \log((nxM)^{\tau})$ 
\begin{align*}
P(|u_j^{[u,v]}(l)|\geq& C_q \sqrt{Mn}x)\leq 2 (nx/(h))^{-\tau}\\
&+C nm^2 (Mn)^{2\iota} \Big( (Mn)^{-1} x^{-2} \log((nxM)^{\tau})^{-2} m^3 [(Mn)^{2\iota}+(M)^{1+\beta}]M\Big)^Q\\
&+C n^2m^2 (Mn)^{2\iota} \max_{Z \in \{U,V\}}P\left(|\bar Z_{0,l,m}|\geq \frac{C  \log((nxM)^{\tau})^{-1} x\sqrt{ Mn }}{m^2 ((Mn)^{\iota}+(M)^{1+\beta})^{1/2}}\right)\\
&=I+II+III.
\end{align*}

Consider  the second term on the RHS above. Recall  that $M=n^{a}$ and $a\leq 2/3$.
\begin{align*}
  (Mn)^{-1}& x^{-2} \log((Mn)^{\tau})^{-2} m^3 [(Mn)^{2\iota}+(M)^{1+\beta}]M\\
  &\leq x^{-2} \log(n)^{C_\tau} (n^{-1+a(1+\beta)}+n^{(1-a)(-1+2/\tau)+a(4/\tau-1)}).
\end{align*}
Note that $n^{-1+a(1+\beta)}=O(k_n)=n^{8/(5\tau)}$. Hence, for $Q=C_\tau$ we obtain $II=O( x^{-C_\tau} (Mn)^{-\tau})$. 

For the third term we use Markov's inequality and obtain for $\tau\geq 8$
\begin{align*}
    III&\leq C n^2m^2 (Mn)^{2\iota} \max_{Z \in \{U,V\}} E |\bar Z_{0,l,m}|^\tau \left(\frac{m^2 ((Mn)^{\iota}+(M)^{1+\beta})^{1/2}}{ \log((n/(hx))^{\tau}) x\sqrt{Mn}}\right)^\tau \\
    & \leq C \log(n)^{C_\tau}  n^{-\tau/2+\beta a/2 \tau+2+(1+a)2/\tau}x^{-\tau}=C \log(n)^{C_\tau} (x n^{1/2(1-1/4a)-1/\tau(2-a)-2(1+a)/\tau^2})^{-\tau}
    \end{align*}

Since $1/\xi=\log(n)^{C_\tau}$,  $Q$ can chosen as $Q= C_\tau$ such that $C (\frac{\sum_{j=1}^{k_n+1} E |\tilde Z_j(l)|^2}{Mn x })^Q=O( x^{-C\tau} (Mn)^{-\tau})$.

\end{proof}

\subsection{Covariance structure of lag-window estimators} \label{subsection.covariance}
\begin{lem} \label{lem.spec.var.rate}
Let $\fspc(\omega)=\frac{M}{n}  \sum_{k=1}^{n} \kappa_M(\omega-\omega_k)  (Z(\omega_k)Z^{H}(\omega_k)-\Sigma_n(\omega_k))$ and $u,v \in \{1,\dots,p\}$. Under Assumption~\ref{asp.moments},\ref{asp.kernel},\ref{asp.spectral}, we have
$$
\sup_\omega |n/M\var(e_{v}^\top \fspc(\omega) e_{u})-C_{K_2}(f_{v,v}(\omega)f_{u,u}(\omega)+\ind(\omega \in \pi \Z)f_{v,u}(\omega)^2)|=O(1/M).
$$
\end{lem}
\begin{proof}
We express the occurring fourth order moments in terms of covariances and cumulants,
see among others Section~5.1 in \cite{rosenblatt2012stationary}. To elaborate,  we have
\begin{align*}
    n/M\var(e_{v}^\top \fsp(\omega) e_{u})=I_1+I_2+I_3,
\end{align*}
\begin{eqnarray*}
I_{1}&=&  \frac{M}{n} \sum_{k_1,k_2=1}^n \kappa_M(\omega-\omega_{k_1}) \kappa_M(\omega-\omega_{k_2})
\cov(e_v^\top Z_n(\omega_{k_1}), e_v^\top Z_n(\omega_{k_2})) \cov(Z_n(\omega_{k_1})^H e_u, Z_n(\omega_{k_2})^H e_u)\\
  I_{2}  &=&\frac{M}{n} \sum_{k_1,k_2=1}^n \kappa_M(\omega-\omega_{k_1}) \kappa_M(\omega-\omega_{k_2})
\cov(e_v^\top Z_n(\omega_{k_1}),Z_n(\omega_{k_2})^H e_u)   \cov(Z_n(\omega_{k_1})^H e_u, e_v^\top Z_n(\omega_{k_2}) ) \\
  I_{3} &=& \frac{M}{n} \sum_{k_1,k_2=1}^n \kappa_M(\omega-\omega_{k_1}) \kappa_M(\omega-\omega_{k_2}) \cum(
      e_{u}^\top Z_n(\omega_{k_1}),
    e_{v}^\top Z_n(\omega_{k_1}),e_{u}^\top Z_n(\omega_{k_2}),e_{v}^\top Z_n(\omega_{k_2})).
\end{eqnarray*}

By Theorem~4.1 in \cite{shao2007local} and the summability of fourth order cumulants, i.e., $$\max_{a_1,\dots,a_4}\sum_{\tau_1,\tau_2,\tau_3\in
    \mathbb{Z}}|\cum( e_{a_1}^\top X_{0},e_{a_2}^\top X_{\tau_1},e_{a_3}^\top X_{\tau_2}, e_{a_4}^\top X_{\tau_3})|<\infty.$$ This implies uniformly for all $\omega_{1},\dots,\omega_{4} \in [0,2\pi]$
    $$\max_{a_1,\dots,a_4}\left|\cum[e_{a_1}^\top Z_n(\omega_1),
    e_{a_2}^\top Z_n(\omega_2),e_{a_3}^\top Z_n(\omega_3),e_{a_4}^\top Z_n(\omega_4)] \right|  = O(n^{-1}),$$
   due to  Assumption~\ref{asp.moments}, 
 we have that $I_3=O(1/M)$. Following the arguments of \cite{rosenblatt2012stationary},  Section 5.1, see also Section 5.4 in \cite{brillinger2001time}, we obtain uniformly for all $\omega \in [0,2\pi]$

\begin{align*}
I_{1}=&\frac{M}{n} \sum_{k_1,k_2=1}^n \kappa_M(\omega-\omega_{k_1}) \kappa_M(\omega-\omega_{k_2}) \frac{1}{(2 \pi n)^2} \sum_{t_1,t_2=1}^n \Gamma_{v,v}(t_1-t_2) \exp( i(t_1 \omega_{k_1} -t_2 \omega_{k_2})\\
&\times \sum_{\tau_1,\tau_2=1}^n \Gamma_{u,u}(\tau_1-\tau_2) \exp( -i(\tau_1 \omega_{k_1} -\tau_2 \omega_{k_2}))\\
=&\frac{M}{n} \sum_{k_1}^n \kappa_M(\omega-\omega_{k_1})^2  (f_{u,u}(\omega_{k_1})+O(1/n))(f_{u,u}(\omega_{k_1})+O(1/n))\\
=& C_{K_2}f_{u,u}(\omega)f_{v,v}(\omega)+O(1/M).
\end{align*}

\begin{align*}
I_{2}=&\frac{M}{n} \sum_{k_1,k_2=1}^n \kappa_M(\omega-\omega_{k_1}) \kappa_M(\omega-\omega_{k_2}) \frac{1}{(2 \pi n)^2} \sum_{t_1,t_2=1}^n \Gamma_{v,u}(t_1-t_2) \exp( i(t_1 \omega_{k_1} +t_2 \omega_{k_2})\\
&\times \sum_{\tau_1,\tau_2=1}^n \Gamma_{u,v}(\tau_1-\tau_2) \exp( i(\tau_1 \omega_{k_1} +\tau_2 \omega_{k_2}))\\
=& \ind(\omega \in \pi \Z)C_{K_2}f_{v,u}(\omega)f_{v,u}(\omega)+O(1/M).
\end{align*}

\end{proof}


\begin{lem}
\label{lem.cov.uhat}
Under Assumption ~\ref{asp.moments} to \ref{asp.rates} the following assertions hold true. If $K$ is the uniform kernel, then 
$$
\frac{1}{Mn} \max_{l_1,l_2=1,\dots,M,l_1\not=l_2} | \cov(\sum_{j=1}^{k_n} \hat u_j(l_1)^{[u,v]},
\sum_{j=1}^{k_n} \hat u_j(l_2)^{[u,v]})| =O(1/M+1/k_n).
$$
For other kernels, we have 
$$
\frac{1}{Mn} | \cov(\sum_{j=1}^{k_n} \hat u_j(l_1)^{[u,v]},
\sum_{j=1}^{k_n} \hat u_j(l_2)^{[u,v]})| =O(1/M+1/k_n+(M(\omega_{l_1}^\prime+\omega_{l_2}^\prime))^{-r}+(M(\omega_{l_1}^\prime-\omega_{l_2}^\prime))^{-r}).
$$

\end{lem}
\begin{proof}
First note that  by independence 
$
    \cov(\sum_{j=1}^{k_n} \hat u_j(l_1)^{[u,v]},
\sum_{j=1}^{k_n} \hat u_j(l_2)^{[u,v]})= \sum_{j=1}^{k_n}\cov( \hat u_j(l_1)^{[u,v]}, \hat u_j(l_2)^{[u,v]}).
$
Furthermore, 
\begin{align*}
    \sum_{j=1}^{k_n}&\cov( \hat u_j(l_1)^{[u,v]}, \hat u_j(l_2)^{[u,v]})=\\
    &\sum_{j=1}^{k_n} [\cov( u_j(l_1)^{[u,v]}, u_j(l_2)^{[u,v]})
    -\cov(u_j(l_1)^{[u,v]},\hat u_j(l_2)^{[u,v]}- u_j(l_2)^{[u,v]}) \\
    &-\cov(\hat u_j(l_1)^{[u,v]}- u_j(l_1)^{[u,v]},u_j(l_2)^{[u,v]})
    +\cov(\hat u_j(l_1)^{[u,v]}- u_j(l_1)^{[u,v]},\hat u_j(l_2)^{[u,v]}- u_j(l_2)^{[u,v]})\\
    =& I+II+III+IV
\end{align*}
For $II,III$ and $IV$ note that by Lemma~\ref{lem.approximation.u.hat} and $\tau \geq 8$, 
$
Mn^{-1} \max_l \sum_{j=1}^{k_n} \var(\hat u_j(l)^{[u,v]}- u_j(l)^{[u,v]})=O(M^{-2}).
$
Hence, we obtain by Cauchy-Schwarz's and Hölder's inequality, that 
\begin{align*}
II&=\frac{1}{Mn} \sum_{j=1}^{k_n} \cov(\hat u_j(l_1)^{[u,v]}- u_j(l_1)^{[u,v]},u_j(l_2)^{[u,v]}) \\
&\leq (\sum_{j=1}^{k_n} \frac{1}{Mn} \var(\hat u_j(l_1)^{[u,v]}- u_j(l_1)^{[u,v]}))^{1/2}
(\sum_{j=1}^{k_n} \frac{1}{Mn} \var (u_j(l_2)^{[u,v]}))^{1/2}=O(M^{-1}).    
\end{align*}

Similar arguments apply to $III$ and $IV$. 

For $I$ note that 
\begin{align*}
u_j(l)^{[u,v]}=& \sum_{t \in H_j} (\sum_{s=1}^{t-1} K((t-s)/M)\exp(-i (t-s) \omega_l^\prime) [\bar U_{t,l,m}\bar V_{s,l,m}- E U_{t,l,m}\bar V_{s,l,m}]\\
    &+\sum_{s=1}^{t-1} K((t-s)/M)\exp(i (t-s) \omega_l^\prime) [\bar V_{t,l,m}\bar U_{s,l,m}-E\bar V_{t,l,m}\bar U_{s,l,m}])
\end{align*}
Evaluating  $\cov( u_j(l_1)^{[u,v]}, u_j(l_2)^{[u,v]})$ using the the above decomposition leads to  four terms, i.e., 
\begin{align*}
    &\cov( u_j(l_1)^{[u,v]}, u_j(l_2)^{[u,v]})=\\
    &\sum_{t\in \Z} \sum_{s_1,s_2=-M}^M K(\frac{s_1}{M}) K(\frac{s_2}{M}) \cov(\bar U_{0,l_1,m}\bar V_{-s_1,l_1,m}, \bar U_{t,l_2,m}\bar V_{t-s_2,l_2,m})\exp(-i s_1 \omega_{l_1}^\prime+s_2 \omega_{l_2}^\prime) +O(\frac{1}{k_n})\\
    &=\sum_{t\in \Z} \sum_{s_1,s_2=-M}^M K(\frac{s_1}{M}) K(\frac{s_2}{M}) \exp(-i s_1 \omega_{l_1}^\prime+s_2 \omega_{l_2}^\prime)\big[\cov(\bar U_{0,l_1,m},\bar U_{t,l_2,m}) \cov(\bar V_{-s_1,l_1,m},\bar V_{t-s_2,l_2,m})\\
     &+\cov(\bar U_{0,l_1,m},\bar V_{t-s_2,l_2,m})\cov(\bar V_{-s_1,l_1,m}, \bar U_{t,l_2,m})
     +\cum(\bar U_{0,l_1,m},\bar V_{-s_1,l_1,m}, \bar U_{t,l_2,m},\bar V_{t-s_2,l_2,m})\big] +O(\frac{1}{k_n}).
\end{align*}

By the  summability of the  fourth order cumulants, see also Theorem 4.1 in \cite{shao2007local}, we have that the cumulant term is of order $O(1/M)$. 
Furthermore, we have with the decay conditions of the autocovariance, smoothness of the kernel and the decay condition of the Fourier coefficients of the kernel
\begin{align*}
    I.1=& f_{\bar U_{l_1,m},\bar U_{l_2,m}}(\omega_{l_1}^\prime)f_{\bar V_{l_1,m},\bar V_{l_2,m}}(\omega_{l_1}^\prime) \frac{1}{M} \sum_{s=-M}^M K^2(s/M) \exp(-is (\omega_{l_1}^\prime-\omega_{l_2}^\prime)+O(1/M)\\
    & \leq C (M(\omega_{l_1}^\prime-\omega_{l_2}^\prime))^{-r}+O(1/M)
\end{align*}
and 
\begin{align*}
    I.2=& f_{\bar U_{l_1,m},\bar V_{l_2,m}}(\omega_{l_1}^\prime)f_{\bar V_{l_1,m},\bar U_{l_2,m}}(\omega_{l_1}^\prime) \frac{1}{M} \sum_{s=-M}^M K^2(s/M) \exp(-is (\omega_{l_1}^\prime+\omega_{l_2}^\prime)+O(1/M)\\
    & \leq C (M(\omega_{l_1}^\prime+\omega_{l_2}^\prime))^{-r}+O(1/M)
\end{align*}
For $K$ being the uniform kernel and $\omega_l^\prime=\pi l/M$, we have 
$\frac{1}{M} \sum_{s=-M}^M K^2(s/M) \exp(-is (\omega_{l_1}^\prime-\omega_{l_2}^\prime)=\ind(\omega_{l_1}^\prime=\omega_{l_2}^\prime)+O(1/M)$ and 
$\frac{1}{M} \sum_{s=-M}^M K^2(s/M) \exp(-is (\omega_{l_1}^\prime+\omega_{l_2}^\prime)=\ind(\omega_{l_1}^\prime+\omega_{l_2}^\prime\in 2M \Z)+O(1/M)$. 
\end{proof}

\subsection{Additional  Lemmas}

\begin{lem}
\label{lem.gaussian.gt}
Let $Z_l, l=1,\dots,d,d \geq 1$ be independent Gaussian random vectors and $Z_l \sim \mathcal{N}(0,I_2)$.
Furthermore, let $\gamma_n,g_n,h_n^{-1}>0$, $\gamma_n \leq 2 \log(d)$ and $W_n$ be a random variable with $P(|W_n|>\gamma_n)=g_n$. 
Set $G_d(t)=(1-(1-\exp(-\max(0,t)/2))^d$. Then,
\begin{align}
\sup_{0\leq t \leq h_n} \left| \frac{P(\max_l \|Z_l\|^2+W_n \geq t)}{G_d(t)}-1\right|\leq(g_n \max(2,\exp(h_n/2)/d)+|\gamma_n| \max(2,\exp(|\gamma_n|/2)).
\label{eq.bound.gaussian.gt}
\end{align}

\end{lem}
\begin{proof}
We compute derivatives under the condition $d \geq 2$. The same arguments with simpler derivatives can also be applied in the case $d=1$.
We have $P(\max_l \|Z_l\|^2+W_n\geq t)\leq P(\max_l \|Z_l\|^2 \geq t -\gamma_n)+P(|W_n|>\gamma_n)$ and by Lemma~\ref{lem.prob.bound},  $P(\max_l \|Z_l\|^2+W_n\geq t)\geq P(\max_l \|Z_l\|^2 \geq t +\gamma_n)-P(|W_n|>\gamma_n)$. Thus,
\begin{align*}
    \left| \frac{P(\max_l \|Z_l\|^2+W_n \geq t)}{G_d(t)}-1\right|\leq g_n/G_d(t)+G_d(t)^{-1} | G_d(t \pm \gamma_n)-G_d(t)|.
\end{align*}
Note that $(G_d(t))^{-1}$ is monoton increasing in $t$, where $G_d(0)^{-1}=1$. Additionally, we have with 
\begin{align}
(1+x)^d\leq 1/(1-dx) \text{ for }x\in [-1,0]    \label{eq.inequality1}
\end{align}
\begin{align}
G_d(t)^{-1}\leq \exp(t/2)/d+1.   \label{eq.bound.Gt}
\end{align}
We consider two cases $t\leq 2 \log(d)$ and $ t > 2\log(d)$. In the first case, we further have 
$G_d(t)^{-1} \leq 2$ and in the second, $G_d(t)^{-1} \leq 2 \exp(t/2)/d$. With this and using the monotonicity of $(G_d(t))^{-1}$, we get  the first summand of the  bound given in \eqref{eq.bound.gaussian.gt}, that is,  $g_n \max(2,\exp(h_n/2)/d)$. Regarding  the second summand of the same bound, note first that we have a positive and a negative case. Only in the case $t \leq \gamma_n$ the two cases need to be treated separately. We begin with this case and consider $t \leq \gamma_n$. Then, we  have $G_d(t)^{-1} | G_d(t - \gamma_n)-G_d(t)|\leq 2(G_d(0)-G_d(\gamma_n))$. Further, we have by the mean value theorem for some $\eps \in [0,\gamma_n]$ and the inequality \eqref{eq.inequality1}
\begin{align*}
2(G_d(0)-G_d(\gamma_n)) &\leq |\gamma_n|  d (1-\exp(-\eps/2))^{d-1} \exp(-\eps/2)    \\
&\leq |\gamma_n| \frac{1}{\exp(\eps/2)/d+(d-1)/d} \leq \gamma_n d/(d-1) \leq 2 \gamma_n.
\end{align*}
With similar arguments, we obtain for the case $\gamma_n\leq t \leq 2\log(d)$ that for  $\eps \in [0,\gamma_n]$ (not necessarily the same $\eps$ for the positive and negative  case)
\begin{align*}
    G_d(t)^{-1} | G_d(t \pm \gamma_n)-G_d(t)| &\leq  2 |\gamma_n| 2 d (1-\exp(-(t \pm \eps)/2))^{d-1} \exp(-(t \pm \eps)/2) \\
    & \leq |\gamma_n| \frac{1}{\exp( (t \pm \eps)/2)/d +(d-1)/d} \leq |\gamma_n| d/(d-1) \leq 2 \gamma_n.
\end{align*}
The same arguments applied to the case $t \leq \gamma_n$ and we obtain  $G_d(t)^{-1} | G_d(t + \gamma_n)-G_d(t)$. 
Now, let $t > 2 \log(d)$. Then,  by the mean value theorem we have for some $\eps \in [0,\gamma_n]$ (not necessarily the same $\eps$ for the positive  and negative  case)
\begin{align*}
    G_d(t)^{-1} | G_d(t \pm \gamma_n)-G_d(t)| & \leq  |\gamma_n| \exp(t/2) (1-\exp(-(t \pm \eps)/2))^{d-1} \exp(-(t \pm \eps)/2) \\
    &\leq |\gamma_n| \exp(|\gamma_n|/2)
\end{align*}
Hence,  
$$
G_d(t)^{-1} | G_d(t \pm \gamma_n)-G_d(t)| \leq |\gamma_n| \max(2,\exp(|\gamma_n|/2)),
$$
where  this bound is independent of $t$ and holds for all $t\geq 0$.
\end{proof}

\begin{lem}\label{lem.sub.inv}
Let $A\in \R^{p\times p}$ be a  positive definite matrix and $v \in \{1,\dots,p\}$. Then, we have for the inverse of the sub-matrix $A_{-v,-v}=I_{p;-v}^\top A I_{p;-v}$ the following
\begin{align} \label{eq.lem.inv}
    (I_{p;-v}^\top A I_{p;-v})^{-1}=I_{p;-v}^\top A^{-1} I_{p;-v}-\frac{I_{p;-v}^\top A^{-1} e_v e_v^\top A^{-1} I_{p;-v}}{e_v^\top A^{-1} e_v}
\end{align}
\end{lem}
\begin{proof}
Note that $I_p=I_{p;-v}^\top I_{p;-v}+e_v e_v^\top$ and $I_{p;-v} e_v=0$. \eqref{eq.lem.inv} is equivalent to
\begin{align*}
    I_{p;-v}^\top I_{p;-v} (e_v^\top A^{-1} e_v)=& I_{p;-v}^\top A I_{p;-v} I_{p;-v}^\top A^{-1} I_{p;-v} e_v^\top A^{-1} e_v -I_{p;-v}^\top A I_{p;-v} I_{p;-v}^\top A^{-1} e_v e_v^\top A^{-1} I_{p;-v} \\ & \pm I_{p;-v}^\top A e_v e_v^\top A^{-1} I_{p;-v} e_v^\top
    A^{-1} e_v\\
    =&I_{p;-v}^\top I_{p;-v} e_v^\top A^{-1} e_v+ I_{p;-v}^\top A e_v e_v^\top A^{-1} e_v e_v^\top A^{-1} I_{p;-v}-I_{p;-v}^\top A e_V e_v^\top A^{-1} I_{p;-v} e_v^\top A^{-1} e_v \\
    =&I_{p;-v}^\top I_{p;-v} e_v^\top A^{-1} e_v.
\end{align*}
\end{proof}

\begin{lem} \label{lem.prob.bound}
For random variables $X,Y$ and $t \in \R, \eps>0$, we have
$$
P(X+Y \geq t) \geq P(X \geq t +\eps)-P(|Y| > \eps)
$$
\end{lem}
\begin{proof}
\begin{align*}
    P(X \geq t +\eps) &= P(X \geq t +\eps,|Y| >\eps) +P(X \geq t +\eps,|Y| \leq \eps) \\
    \leq P(|Y| >\eps)+P(X \geq t - Y)
\end{align*}
\end{proof}

\section{Auxiliary Lemmas} \label{sec.app}

In this section, we present some auxiliary lemmas which are used  to prove  the   main results of this paper. Lemma~\ref{lem.partial.coherence.first.approximation} is the key lemma used to express the de-biased partial coherence as a
 quadratic form and  uses results   given in Subsection~\ref{subsection.lemmas.partial} of the supplementary file. The expression as  quadratic form is the starting point to obtain a Gaussian approximation, where the steps  involved are described in more detail in Remark~\ref{rmk.definition.u.v} bellow. The corresponding lemmas 
 can be found  in Subsection~\ref{subsection.lemmas.Gaussian.approximaiton} of the supplementary file. Lemma~\ref{lem.gaussian.approximation}  deals with  the uniform convergence of the maximum of the de-biased partial coherence over a growing grid of frequencies. To establish this  result,   convergences rates of  the covariance structure of the lag-window estimator and of the quadratic forms involved are required. The  results used in this context are stated as auxiliary lemmas in Subsection~\ref{subsection.covariance} of the supplementary file. Lemma~\ref{lem.gaussian.approximation} is the key ingredient to prove Theorem~\ref{thm.partial.coherence.max}, Theorem~\ref{thm.test-pc} and Theorem~\ref{th.FDR_1}.
 

To proceed, we fix some additional notation  used in the proofs.  $\{Z_n\}$ denotes  a sequence of random variables and $\{f_n(\cdot)\}$ a sequence of functions. If $P(|Z_n|\geq x) \leq C f_n(x)$, for some constant $C>0$ and for all $n$, then the notation  $Z_n=\bp(f_n(x))$ is used. Note that for $A_n=\bp(f_n(x)), B_n=\bp(g_n(x))$, we have $A_n+B_n=\bp((f_n(x/2)+g_n(x/2)))$ since $P(|A_n+B_n|\geq x) \leq P(|A_n|\geq x/2)+P(|B_n|\geq x/2)$. Similarly, $A_n\cdot B_n=\bp((f_n(\sqrt{x})+ g_n(\sqrt{x})))$ and  for $Z_n=\bp(f_n(x))$ and  $x_n$ a sequence, we have 
$Z_n+x_n=\bp(f_n(x-x_n))$.


\begin{lem} \label{lem.partial.coherence.first.approximation}
Let $\omega_l^\prime=\pi l N/M, l=1,\dots,M/N-1$. Under Assumption \ref{asp.moments}, \ref{asp.kernel}, \ref{asp.spectral}, and \ref{asp.spectral.density.estimator}, we have
\begin{align*}
\sqrt{n/M}\sup_l&\left|(\hat \rho^{(de)}_{u,v}(\omega_l^\prime)-\rho_{u,v}(\omega_l^\prime))-
\Big[ \frac{1}{n}T_{n,u,v}(l)-\frac{1}{2n}\rho_{u,v}(\omega_l^\prime)\big[ T_{n,v,v}(l)+ T_{n,u,u}(l)\big]\Big]\right|\\
&=\bp(\gpbetade x),
\end{align*}
where $$\gpbetade x=(\gps {((x-\gpbias)/(\spar (n/M)^{1/4}))^{1/2}}+ \gps 1),$$
\begin{equation}
    \label{eq.Tn-app}
T_{n,u,v}(l)=\frac{1}{2\pi} \sum_{u=-n+1}^{n-1} K(u/M) \sum_{t=\max(1,1-u)}^{\min(n,n-u)} [U_{t+u,l} V_{t,l}-E U_{t+u,l} V_{t,l}]\exp(-i u \omega_l^\prime),
\end{equation} 
and, for $l=1,\dots,M/N-1$,
$$U_{t,l}=e_u^\top f^{-1}(\omega_l^\prime) X_t(f_{u,u}^{-1}(\omega_l^\prime))^{-1/2}
\ \ \mbox{and} 
\ \ V_{t,l}=e_v^\top f^{-1}(\omega_l^\prime) X_t(f_{v,v}^{-1}(\omega_l^\prime))^{-1/2}.$$
Under Assumption~\ref{asp.spectral}, 
the processes $\{U_{t,l}\}$ and $ \{V_{t,l}\}$ possesses  the same functional dependence as the process  $\{X_t\}$.
\end{lem}

\begin{proof}
Let $g(x_1,x_2,x_3,x_4,x_5,x_6)=1/2((x_1+i x_2)\sqrt{x_5/x_6}+(x_3-i x_4)\sqrt{x_6/x_5})$. Then, 
$$\hat \rho^{(de)}_{u,v}(\omega)=g(Re(\hat \beta_{u,\tilde v}^{(de)}(\omega)),Im(\hat \beta_{u,\tilde v}^{(de)}(\omega)),Re(\hat \beta_{v,\tilde u}^{(de)}(\omega)),Im(\hat \beta_{v,\tilde u}^{(de)}(\omega)),\hat f_{v,v}^{-1}(\omega),\hat f_{u,u}^{-1}(\omega)).$$
Using a Taylor expansion of $\hat \rho^{(de)}_{u,v}(\omega)$ around $\rho_{u,v}(\omega)$, we get  that the terms given by the derivatives of $x_5$ and $x_6$ are of order $O(1/n)$ due to $\beta_{v,\tilde u}(\omega)\sqrt{\Sigma_{n,v,v}^{-1}(\omega) \Sigma_{n,u,u}^{-1}(\omega)}=
(\beta_{u,\tilde v}(\omega))^{(C)}\sqrt{\Sigma_{n,u,u}^{-1}(\omega)/\Sigma_{n,v,v}^{-1}(\omega)}\Sigma_{n,u,u}^{-1}(\omega)$ and $\sup_\omega \|\Sigma_n(\omega)-f(\omega)\|_{\max}=O(1/n)$.  Therefore, 
\begin{align}
    \hat \rho^{(de)}_{u,v}(\omega)-\rho_{u,v}(\omega)=&\frac{1}{2}\Big[
    (\hat \beta_{u,\tilde v}^{(de)}(\omega)-\beta_{u,\tilde v}(\omega))\sqrt{\frac{f_{u,u}^{-1}(\omega)}{f_{v,v}^{-1}(\omega)}}+(\hat \beta_{v,\tilde u}^{(de)}(\omega)-\beta_{v,\tilde u}(\omega))^{(C)}\sqrt{\frac{f_{v,v}^{-1}(\omega)}{f_{u,u}^{-1}(\omega)}}\Big]
    \nonumber
    \\&+
    Error_1(\omega), \label{eq.proof.rho}
\end{align}
where $\sup_\omega Error_1(\omega)=O(\sup_\omega \|\fsp(\omega)- E\fsp(\omega)\|_{\max}^2)=\bp(\gps{\sqrt{x}})$. Furthermore, we have by Lemma~\ref{lem.beta.bias} for $u,v=1,\dots,p$,
\begin{align*}
\sqrt{n/M}&(\hat \beta_{v,\tilde u}^{(de)}(\omega)-\beta_{v,\tilde v}(\omega))=\delta(\omega)+\\
&\sum_{k=1}^n \frac{\kappa_M(\omega-\omega_k)}{\sqrt{nM }(f_{v,v}^{-1}(\omega))^2} e_v^\top f^{-1}(\omega) [Z(\omega_k) Z^H(\omega_k)-f(\omega)]f^{-1}(\omega)[e_u f_{v,v}^{-1}(\omega)-e_v f_{v,u}^{-1}(\omega)]\\
=&\delta(\omega)+
\frac{\sqrt{n/M}}{(f_{v,v}^{-1}(\omega))^2} e_v^\top f^{-1}(\omega) [\fsp(\omega)-f(\omega)]f^{-1}(\omega)[e_u f_{v,v}^{-1}(\omega)-e_v f_{v,u}^{-1}(\omega)],    
\end{align*}
where $\sup_w |\delta(\omega)|=\bp(\gpbetade x)$. Define $\hat h_{u,v}(\omega):=e_v^\top f^{-1}(\omega) [\flw(\omega)-f(\omega)]f^{-1}(\omega)e_u.$ Inserting this into \eqref{eq.proof.rho} we  obtain
\begin{align*}
    \sqrt{n/M}(\hat \rho^{(de)}_{u,v}(\omega)-\rho_{u,v}(\omega))=&
    \sqrt{n/M}\Big[ \frac{\hat h_{u,v}(\omega)}{\sqrt{f_{v,v}^{-1}(\omega)f_{u,u}^{-1}(\omega)}}+\frac{\rho_{u,v}(\omega)}{2}
    [\frac{\hat h_{u,u}(\omega)}{f_{u,u}^{-1}(\omega)}+\frac{\hat h_{v,v}(\omega)}{f_{v,v}^{-1}(\omega)}\big]\Big]
    \\&+
    \sqrt{n/M} Error_1(\omega)\\
    &+\frac{\delta(\omega)}{2}\Big(\sqrt{\frac{f_{v,v}^{-1}(\omega)}{f_{u,u}^{-1}(\omega)}}-\sqrt{\frac{f_{u,u}^{-1}(\omega)}{f_{v,v}^{-1}(\omega)}} \Big).
\end{align*}
We have $$\sup_\omega \sqrt{n/M} Error_1(\omega)=\bp(\gps{\sqrt{x/(n/M)^{1/4}}})=\bp(\gpbetade x).$$

Now, we take a closer look at $\hat h_{u,v}(\omega_l^\prime)$. For $l=1,\dots,M/N-1$ we define  $U_{t,l}=e_u^\top f^{-1}(\omega_l^\prime) X_t/(f_{u,u}^{-1}(\omega_l^\prime) )^{1/2}$ and $V_{t,l}=e_v^\top f^{-1}(\omega_l^\prime) X_t/(f_{v,v}^{-1}(\omega_l^\prime) )^{1/2}$.
Since by Assumption~\ref{asp.moments} $\sup_\omega \|f^{-1}\|_2$ is bounded and $\max_{\|v\|_2=1} E | v^\top X|^\tau<\infty$, we have that $E |U_{t,l}|^\tau<\infty$ and $E|V_{t,l}|^\tau<\infty$ and that $\{U_{t,l}\}$ and $ \{V_{t,l}\}$ posses the same functional dependence as $\{X_t\}$. Furthermore, recall the definition of $ T_{n,u,v}(l)$ in (\ref{eq.Tn-app}).
Then,  by \eqref{asp.bias.spectral.estimator} we have 
$\max_l \sqrt{n/M}|\hat h_{u,v}(\omega_l^\prime)-1/n T_{n,u,v}(l)|=o(\gpbias)$. Thus  
\begin{align*}
\sqrt{n/M}\sup_\omega\Big|&(\hat \rho^{(de)}_{u,v}(\omega_l^\prime)-\rho_{u,v}(\omega_l^\prime))-
\Big[1/n T_{n,u,v}(l)-1/2 \rho_{u,v}(\omega_l^\prime)[1/n T_{n,v,v}(l)+1/n T_{n,u,u}(l)]\Big]\Big|\\
&=\bp(\gpbetade x).
\end{align*}
\end{proof}

\begin{rmk} \label{rmk.definition.u.v}
The strategy  used to  proof the results presented in Lemma~\ref{lem.gaussian.approximation},  
is to first approximate the random variables $U_{t,l} $ and $ V_{t,l} $ defined in Lemma~\ref{lem.partial.coherence.first.approximation},  by $m$-dependent and bounded random variables. The corresponding sums are then split into chunks of big and small blocks so that the big blocks are independent from each other. For this, we follow ideas used in  \cite{liu2010asymptotics}.
We  can  approximate $ \{X_t\}$ by the  $m$-dependent process $\{X_{t,m}=E (X_t | \eps_{t-m}, \dots, \eps_t)\}$
and denote by  $d_{m,\tau}$ the approximation error. By  the assumptions  made, we have  $d_{m,\tau}=\sum_{t=0}^\infty \min (\delta_{t,\tau},\Psi_{m+1,\tau})=O(\rho^m)$ for some $0<\rho<1$;  see Lemma~\ref{lem.approximation.g.m}. Consequently, for  $\eps>0$ small and $d_\xi>0$,  we set  $m=\big[\log((d_\xi+\eps)\log(n))/(-\log(\rho))\big]$, which implies  that $d_{m,\tau}=o(\log(n)^{-d_\xi})$. 

Replacing  $X_t$ by  $X_{t,m}$ in the construction of $U_{t,l} $ and $V_{t,l} $ used in   Lemma~\ref{lem.partial.coherence.first.approximation}, 
leads to the random sequences  $\{U_{t,l,m}\}$,   $\{V_{t,l,m}\}$ and $T_{n,u,v,m}(l)$. We further define 
$$g_n^{[u,v]}(\omega_l^\prime)=T_{n,u,v}(l)-E(T_{n,u,v}(l))- \sum_{t=1}^n \big(U_{t,l} V_{t,l}-E U_{t,l}V_{t,l}\big)$$ and $$g_{n,m}^{[u,v]}(\omega_l^\prime)=T_{n,u,v,m}(l)-E(T_{n,u,v,m}(l))- \sum_{t=1}^n\big( U_{t,l,m} V_{t,l,m}-E U_{t,l,m}V_{t,l,m})\big.$$

The next step in our proof is to truncate $ U_{t,l,m}$ and $ V_{t,l,m}$ in order to obtain bounded random variables. For this  we consider 
$$U_{t,l,m}^\prime=U_{t,l,m} \ind(|U_{t,l,m}| \leq (Mn)^\iota) \text{ and } V_{t,l,m}^\prime=V_{t,l,m} \ind(|V_{t,l,m}| \leq (Mn)^\iota),$$
where $\iota=1/\tau$. We then focus on  the centered random variables  $$\bar{U}_{t,l,m}=U_{t,l,m}^\prime-E U_{t,l,m}^\prime \text{ and }\bar{V}_{t,l,m}=V_{t,l,m}^\prime-E V_{t,l,m}^\prime.$$ 
For the construction of the big and small blocks,  let $\rho_n=\floor{(M)^{1+\beta}}, q_n=M+m,$ and $k_n=\floor{n/(p_n+q_n)}$. Note that $k_n=O(n^{1-a(1+\beta)})$ and recall that $M=n^a$. Since $k_n=n^{\eps}$ for some $\eps>0$ is needed, we set $\beta=1/4-2\iota$ and $a\leq 4/7$.

To proceed we  split the interval $[1,n]$ into alternating big and small blocks  $H_j$ and $I_j$ of lengths $p_n$ and $q_n$,  respectively, where  for $ 1 \leq j\leq k_n$, 
$$H_j=[(j-1)(p_n+q_n)+1,jp_n+(j-1)q_n], \ \ I_j=[jp_n+(j-1)q_n+1, j(p_n+q_n)], $$ and  $I_{k_n+1}=[k_n(p_n+q_n)+1,n]$ is the remaining block. We set 
\begin{align} \label{eq.Ybar}
    \bar Y_{t,m}^{[u,v]}(l)=&\frac{1}{2\pi}\bar U_{t,l,m}\sum_{s=1}^{t-1} K((t-s)/M)\exp(-i (t-s) \omega_l^\prime) \overline V_{s,l,m} \nonumber \\
    &+\frac{1}{2\pi}\bar V_{t,l,m}\sum_{s=1}^{t-1} K((t-s)/M)\exp(i (t-s) \omega_l^\prime) \overline U_{s,l,m}.
\end{align}
The above  lag-window estimator uses the sample autocovariances between $\overline{U}_{t,l,m}$ and $\overline{V}_{t,l,m}$ for lags $u=-n+1,\dots,n+1$. The first term on the right hand side 
of the above equation  corresponds to such sample autocovariances for lags $u\geq 1$,  while the  second term on the right hand side of (\ref{eq.Ybar}), to lags  $u\leq -1$. Observe that 
the sample autocovariance at lag $u=0$ is removed; also  see the construction of $g_n^{[u,v]}(\omega_l^\prime)$. 
Using $     \bar Y_{t,m}^{[u,v]}(l)$ we further introduce the random variables 
  $$u_j(l)^{[u,v]}=\sum_{t \in H_j} (\bar Y_{t,m}^{[u,v]}-E\bar Y_{t,m}^{[u,v]})$$ 
  for $1 \leq j \leq k_n$, and 
  $$v_j(l)^{[u,v]}=\sum_{t \in I_j} (\bar Y_{t,m}^{[u,v]}-E\bar Y_{t,m}^{[u,v]}),$$
  for $1 \leq j \leq k_n+1 $.
We then have 
\begin{align*}
\frac{1}{2\pi}& \sum_{u=-n+1}^{n-1} K(u/M) \sum_{t=\max(1,1-u)}^{\min(n,n-u)} [\bar U_{t+u,l,m} \bar V_{t,l,m}-E \bar U_{t+u,l,m} \bar V_{t,l,m}]\exp(-i u \omega_l^\prime)\\
&-\sum_{t=1}^n (\bar U_{t,l,m} \bar V_{t,l,m}-E \bar U_{t,l,m} \bar V_{t,l,m})\\
&=\sum_{j=1}^{k_n} u_j(l)^{[u,v]}+\sum_{j=1}^{k_n+1} v_j(l)^{[u,v]}=:\bar g_{n,m}(l)^{[u,v]}.    
\end{align*}

Note that since $K((t-s)/M)=0$ for $|t-s|>M$ and the sequences $\{U_{t,l,m}\}$,  and $ \{V_{t,l,m}\}$ are $m$-dependent, the random variables $u_{j_1}(l)^{[u,v]}$ and $u_{j_2}(l)^{[u,v]}$ are mutually independent for $j_1 \not = j_2$.

In the lemmas of Subsection~\ref{subsection.lemmas.Gaussian.approximaiton}, we prove 
that the  introduced approximations by $m$-dependent and bounded random variables are valid. We then focus on the behavior of the sum 
$ \sum_{j=1}^{k_n} u_j(l)^{[u,v]}$ involving  the big blocks  $u_j(l)^{[u,v]}$.
We show for $\omega_l^\prime=\pi l M/N, l=1,\dots,M/N-1$, that, 
\begin{align*}
 \sqrt{n/M}(\hat \rho^{(de)}_{u,v}(\omega_l^\prime)-\rho_{u,v}(\omega_l^\prime))& =
\sqrt{1/(Mn)} \sum_{j=1}^{k_n} \left(u_j(l)^{[u,v]} -    \frac{\rho_{u,v}(\omega_l^\prime)}{2}[u_j(l)^{[u,u]}+u_j(l)^{[v,v]}]\right)\\
& \ \ \ +Remainder,
\end{align*}
where,  the Remainder is asymptotically negligible.
We then  approximate 
$$ \left(u_j(l)^{[u,v]} -    \frac{\rho_{u,v}(\omega_l^\prime)}{2}[u_j(l)^{[u,u]}+u_j(l)^{[v,v]}]\right)$$
by a Gaussian vector. Toward this goal, we  need to introduce another truncation. To elaborate, let  
 $$w_j(l)=u_j(l)^{[u,v]} -    1/2 \rho_{u,v}(\omega_l^\prime)[u_j(l)^{[u,u]}+u_j(l)^{[v,v]}], $$
 $\xi=(\log(n^{d_\xi}))^{-\zeta_\xi}$, $ \zeta_\xi \in [2,3/2+\Nkernel)$ and  $ d_\xi= 2(b+1)$. We then consider  the following truncated version of $w_j(l)^{[u,v]}$, $j=1,\dots,k_n, l=1,\dots,M/N-1$,
$$\hat w_j(l)^{[u,v]}=w_j(l)^{[u,v]} \ind(|w_j(l)^{[u,v]}|\leq \sqrt{Mn}\xi)- E (w_j(l)^{[u,v]} \ind(|w_j(l)^{[u,v]}|\leq \sqrt{Mn}\xi).$$ 
\end{rmk}

\begin{lem}
\label{lem.gaussian.approximation}
Let $\omega_l^\prime=\pi l N/M, l \in \mathcal{L}$ and
 $T_{n,u,v}(l)$ as defined in Remark~\ref{rmk.definition.u.v}, and $Z_l$ i.i.d. with $Z_l\sim \mathcal{N}(0,I_2)$. We have under Assumption~\ref{asp.moments} to \ref{asp.rates} and  for $h_n>0$, that, 
 \begin{align*} 
    \sup_{0\leq t \leq h_n}& \left|\frac{P(\max_l 1/(Mn)\chi(l)^{[u,v]}\geq t)}{P(\max_l \|Z_l\|_2 \geq t)}-1\right| \leq C((\log(n^{d_\xi}))^{-\zeta_\xi+3/2}+\max(2,\exp(h_n/2)) \\
    &\times \log(n)^{C_\tau}\Big[\gpbetade{\log(n)} + (n^{1/2(1-1/4a)-1/\tau(2-a)-2(1+a)/\tau^2})^{-\tau}\Big],
\end{align*}
 where $\chi(l)^{[u,v]}:=\chi_{(u,v)}(\omega_l^\prime)$.
\end{lem}
\begin{proof}
Without loss of generality, we set $\mathcal{L}=\{1,\dots,M/N-1\}$. We first establish  the following equality 
$$
P(\max_l \chi(l)^{[u,v]}\geq t)=P(\max_l \|\tilde Z_l\|_2^2+W_n \geq t),
$$
where $\tilde Z_l$ are $2$-dimensional standard Gaussians random variables and $W_n$ denotes an  error term  satisfying for some $\delta>0$,  $P(|W_n|>\delta)\leq \Delta(\delta)$. To elaborate, let $$R_l:=\sqrt{n/M}((\Re(\hat \rho^{(de)}_{u,v}(\omega_l^\prime)-\rho_{u,v}(\omega_l^\prime)),
    \Im(\hat \rho^{(de)}_{u,v}(\omega_l^\prime)-\rho_{u,v}(\omega_l^\prime)))^\top$$
    such that
$\chi(l)=: R_l^\top \hat V^{-1}(\omega_l^\prime) R_l.$
Also let $\bar Z_l=(V^{-1}(\omega_l^\prime))^{1/2} \tilde Z_l$. We can then  write 
\begin{align*}
\chi(l)^{[u,v]}=& \bar Z_l^\top V^{-1}(\omega_l^\prime) \bar Z_l + (\bar Z_l-R_l)^\top V^{-1}(\omega_l^\prime)(\bar Z_l-R_l)
-\bar Z_l V^{-1}(\omega_l^\prime)(\bar Z_l-R_l)\\
&-(\bar Z_l-R_l)^\top V^{-1}(\omega_l^\prime)\bar Z_l+
R_l^\top (\hat V^{-1}(\omega_l^\prime)-V^{-1}(\omega_l^\prime)) R_l\\
&=: \|\tilde Z_l\|_2^2+ W_n.    
\end{align*}
That is, we need to determine the order of $W_n$,  i.e. of $\Delta(\cdot)$. Note that by Lemma~\ref{lem.partial.coherence.first.approximation} we have the following 
$$\max_l \|R_l\|_2=\bp(\gpbetade x+ \gps{x/\sqrt{n/M}}),$$
\begin{align*} 
\|(\hat V(\omega_l^\prime)-V(\omega_l^\prime))\|_2 & \leq C \|\sqrt{M/n} R_l\|_2=\bp(\gpbetade {x\sqrt{n/M}}+ \gps{x\sqrt{n/M}}),
\end{align*}
and
$$\|\hat V^{-1}(\omega_l^\prime)-V^{-1}(\omega_l^\prime)\|_2\leq C \|\hat V(\omega_l^\prime)-V(\omega_l^\prime)\|_2/(1- C \|\hat V(\omega_l^\prime)-V(\omega_l^\prime)\|_2).$$ 
Furthermore, since $(V^{-1}(\omega_l^\prime))^{1/2} Z_l \sim \mathcal{N}(0,I_2)$ and $d\leq M$, we have $$P(\max_l \|(V^{-1}(\omega_l^\prime))^{1/2} Z_l\|_2 \geq x)\leq M P(\|(V^{-1}(\omega_l^\prime))^{1/2} Z_1\|_2 \geq x)=M \exp(-\sqrt{x}/2).$$ 
It remains to establish the order of $\max_l \|(V^{-1}(\omega_l^\prime))^{1/2}(Z_l-R_l)\|_2$. We will show that  $(\bar Z_l-R_l)^\top V^{-1}(\omega_l^\prime)(\bar Z_l-R_l)$ is the dominating term. 
To determine the order of $\max_l \|(V^{-1}(\omega_l^\prime))^{1/2}(Z_l-R_l)\|_2$ we split this term  into three  terms  and derive for each of them its  order which is denoted $\Delta_i(\cdot),i=1,2,3$. 

For this note first that as outlined in Remark~\ref{rmk.definition.u.v}, we can approximate our random variables by bounded, $m$-dependent variables. That is we approximate $R_l$ by $\sum_{j=1}^{k_n} (\Re(\hat w_j(l)^{[u,v]}), \Im(\hat w_j(l)^{[u,v]}))^\top$.

Putting the results of Lemma~\ref{lem.beta.bias} 
to Lemma~\ref{lem.approximation.u.hat} together gives, 
\begin{align*}
    \|&\max_l (V^{-1}(\omega_l^\prime))^{1/2} [R_l-(\sum_{j=1}^{k_n} (\Re(\hat w_j(l)^{[u,v]}), \Im(\hat w_j(l)^{[u,v]}))^\top]\|_2=\bp(\gpbetade x+\\
    &C \log(n)^{C_\tau} (x n^{1/2(1-1/4a)-1/\tau(2-a)-2(1+a)/\tau^2})^{-\tau}+x^{-C\tau} (Mn)^{-\tau}+\frac{n}{x^{\tau/2}(Mn)^{\tau/4}})=:\bp(\Delta_1(x))
\end{align*}

For  the second term on the right hand side of the previous equation, we use a Gaussian approximation. We rewrite the statistic by using bounded, $m$-dependent variables as follows 
\begin{align*}
\max_l& (V^{-1}(\omega_l^\prime))^{1/2} \sum_{j=1}^{k_n} (\Re(\hat w_j(l)^{[u,v]}), \Im(\hat w_j(l)^{[u,v]}))^\top= \max_l ( \sum_{j=1}^{k_n} \tilde e_l^\top \tilde A_j),     
\end{align*}
where $\tilde e_j = (e_j \otimes (1,1)^\top)\in \R^{2d}$, and $\tilde A_j$ is a $2d$-dimensional vector given by
$$
\tilde A_j=\begin{pmatrix}
[(\sigma_1(1) \Re(\hat w_j(1)^{[u,v]}) + \sigma_3(1) \Im(\hat w_j(1)^{[u,v]})] \\ 
[(\sigma_3(1) \Re(\hat w_j(1)^{[u,v]}) + \sigma_2(1) \Im(\hat w_j(1)^{[u,v]})] \\ \vdots \\ [(\sigma_1(d) \Re(\hat w_j(d)^{[u,v]}) + \sigma_3(d) \Im(\hat w_j(d)^{[u,v]})] \\
[(\sigma_3(d) \Re(\hat w_j(d)^{[u,v]}) + \sigma_2(d) \Im(\hat w_j(d)^{[u,v]})]
\end{pmatrix},
$$
and  
$$(V^{-1}(\omega_l^\prime))^{1/2}=\begin{pmatrix}
\sigma_1(l) & \sigma_3(l) \\
\sigma_3(l) & \sigma_2(l)
\end{pmatrix}.
$$

Note first that for all $l$ we have $\|\tilde e_l^\top \tilde A_j\|_2\leq \|(V^{-1}(\omega_l^\prime))\|_2 |\hat w_j^{[u,v]}(l)|\leq 2 C\sqrt{Mn} \xi$. This means that $\sum_{j=1}^{k_n} \tilde e_l^\top \tilde A_j$ is a sum of bounded independent random vectors of dimension $2$. We can apply Fact~2.2 in \cite{einmahl1997gaussian} and  define a $2d$-dimensional Gaussian vector $Z=(Z_l)_{l=1,\dots,d}$, where each $Z_l$ is $2$-dimensional, with mean zero and the variance of the entire $2d$-dimensional Gaussian vector equals  $\var(Z)=1/(Mn)\var(\sum_{j=1}^{k_n} \tilde A_j)$. Furthermore, due to the boundedness condition $|\hat w_j^{[u,v]}(l)|\leq C\sqrt{Mn} \xi$,  the Gaussian vector is a close approximation in the sense that we have for the 2-dimensional marginal distributions 
$$
P( \| \sqrt{1/(Mn)} \sum_{j=1}^{k_n} \tilde e_l^\top \tilde A_j-Z_l\|_2 \geq x) \leq c \exp( - x/( c \xi)),
$$
where $c$ is a  positive constant. 
With this, we have further for $x>0$, 
\begin{align*}
    P&(\max_l \| \sqrt{1/(Mn)}\sum_{j=1}^{k_n} \tilde e_l^\top \tilde A_j-Z_l\|_2 \geq x) \leq
    \sum_{l=1}^d P( \| \sqrt{1/(Mn)}\sum_{j=1}^{k_n} \tilde e_l^\top \tilde A_j-Z_l\|_2 
    \geq x) 
    \\
    &\leq C \exp( - x /( C \xi)+\log(d))=:\Delta_2(x).
\end{align*}

Regarding  the third term, we  have $\var(Z)=1/(Mn) \var(\sum_{j=1}^{k_n} \tilde A_j)$. 
Let $H=1/(Mn)\var(\sum_{j=1}^{k_n} \tilde A_j)$, i.e., $Z \sim \mathcal{N}(0,H)$ and  $\tilde Z=H^{-1/2} Z$, that is,   $\tilde Z \sim \mathcal{N}(0,I_{2d})$. Then, we have  for $x>0$
\begin{align*}
    P&(\max_l (\| \tilde e_l^\top (H^{1/2} - I_{2d}) \tilde Z \|_2 \geq x) \leq
    \sum_{l=1}^d P( \| (e_{2(l-1)+1}^\top+e_{2(l-1)+2}^\top) (H^{1/2} - I_{2d}) \tilde Z \|_2 \geq x)  \\
    &\leq  \sum_{l=1}^d (P( |e_{2(l-1)+1}^\top (H^{1/2} - I_{2d}) \tilde Z| \geq x/2)+
    P( |e_{2(l-1)+2}^\top (H^{1/2} - I_{2d}) \tilde Z| \geq x/2)).
\end{align*}
The two probabilities can be bounded by the same arguments. We focus here on the first. Note that $e_{2(l-1)+1}^\top (H^{1/2} - I_{2d}) \tilde Z  \sim N(0,\|e_{2(l-1)+1}^\top (H^{1/2} - I_{2d})\|_2^2)$. Additionally, we have 
$$
\|e_{2(l-1)+1}^\top (H^{1/2} - I_{2d})\|_2 \leq \|e_{2(l-1)+1}^\top (H - I_{2d})\|_2\|(H^{1/2}+  I_{2d})^{-1}\|_2. 
$$
Since $H$ is a variance matrix and by definition positive semi-definite, we have $\|(H^{1/2}+  I_{2d})^{-1}\|_2\leq 1$. Furthermore, 
$$\|e_{2(l-1)+1}^\top (H - I_{2d})\|_2^2=\sum_{r=1,r\not=2(l-1)+1}^{2d}
(e_{2(l-1)+1}^\top H e_r)^2+(e_{2(l-1)+1}^\top H e_{2(l-1)+1}^\top - 1)^2
$$
and 
$e_{2(l-1)+1}^\top H e_{2(l-1)+1}^\top=e_1^\top \var(\tilde e_l^\top \sum_{j=1}^{k_n} \tilde A_j) e_1 $. The term $\sum_{r=1,r\not=2(l-1)+1}^{2d}
(e_{2(l-1)+1}^\top H e_r)^2$ contains the covariance between different frequencies and we use Lemma~\ref{lem.cov.uhat} to determine its order. For  other kernels than 
the uniform one, the additional terms $1/[(r \pm (2l-1)+1)]N^{-1}]$ appear. Note that the terms  with $r$ are $\ell_2$-summable and that no periodicity in terms of the Fourier coefficients occur. 
$$\sum_{r=1,r\not=2(l-1)+1}^{2d}
(e_{2(l-1)+1}^\top H e_r)^2=O(M(M^{-2}+k_n^{-2})+N^{-2r})=O(n^{-a}+n^{(3+2\beta)a-2}+N^{-2r}).$$ For $a <4/7$ and $ \beta <1/4$ we have $O(n^{-a}+n^{(3+2\beta)a-2})=o(n^{-\tilde \upsilon})$ for some $\tilde \upsilon>0$. For the variance term, we have 
$$\tilde e_l^\top \sum_{j=1}^{k_n} \tilde A_j=(\Sigma^{-1}(\omega_l))^{1/2}
\sum_{j=1}^{k_n}
\begin{pmatrix}
\Re\Big(\hat u_j(l)^{[u,v]}- \rho_{u,v}(\omega_l^\prime)/2[\hat u_j(l)^{[u,u]}+\hat u_j(l)^{[v,v]}]\Big)\\
\Im\Big(\hat u_j(l)^{[u,v]}- \rho_{u,v}(\omega_l^\prime)/2[\hat u_j(l)^{[u,u]}+\hat u_j(l)^{[v,v]}]\Big)
\end{pmatrix}.
$$
Additionally, we have by Lemma~\ref{lem.approximation.g.m} to 
Lemma~\ref{lem.approximation.u.hat}, that, 
\begin{align*}
\max_{1\leq l \leq d}1/(Mn)\var\Big(&\sum_{j=1}^{k_n}\Big[\hat u_j(l)^{[u,v]}- \rho_{u,v}(\omega_l^\prime)/2[\hat u_j(l)^{[u,u]}+\hat u_j(l)^{[v,v]}]\Big]\\
&-\Big(1/n T_{n,u,v}(l)-1/2 \rho_{u,v}(\omega_l^\prime)[1/n T_{n,v,v}(l)+1/n T_{n,u,u}(l)\Big)\Big)=o(n^{-\upsilon^\prime}),    
\end{align*}
where $0<\upsilon^\prime<\beta a$. The same rate also is obtained for the real and imaginary part. Additionally, we have by Lemma~\ref{lem.var.rho} 
$$
\max_l \|\var
\begin{pmatrix}
\Re\Big(1/n T_{n,u,v}(l)-1/2 \rho_{u,v}(\omega_l^\prime)[1/n T_{n,v,v}(l)+1/n T_{n,u,u}(l)\Big)\\
\Im\Big(1/n T_{n,u,v}(l)-1/2 \rho_{u,v}(\omega_l^\prime)[1/n T_{n,v,v}(l)+1/n T_{n,u,u}(l)\Big)
\end{pmatrix}
-V(\omega_l^\prime)
\|_2=O(M^{-1}).
$$
Thus, we get 
$\max_l \|1/(Mn) \var(\tilde e_l^\top \sum_{j=1}^{k_n} \tilde A_j) -I_2\|_2=O(1/M)+o(n^{-\upsilon^\prime})=o(n^{-\upsilon^\prime})
$. Consequently, we have with $N=\log^{\Nkernel/r}(M)$ $\|e_{2(l-1)+1}^\top (H - I_{2d})\|_2=O(\log^{-\Nkernel}(M))$.

With this  bound and additional tail bounds for Gaussian random variables, see among others Appendix~A in \cite{chatterjee2014superconcentration}, we have 
\begin{align*}
    P&(\max_l (\| \tilde e_l^\top (H^{1/2} - I_{2d}) \tilde Z \|_2 \geq x))\leq C(
    \exp(- (x \log^\Nkernel(M))^2+\log(d)) \frac{1}{x \log^\Nkernel(M)})=:\Delta_3(x).
\end{align*}

Now we set $x=(\log(n^{d_\xi}))^{-\zeta_\xi+3/2}$ with $d_\xi, \zeta_\xi$ as in Remark~\ref{rmk.definition.u.v}, especially $\zeta_\xi\geq 2$ and $\zeta_\xi < 3/2+\Nkernel$. Then,
$\Delta_3=O(\exp(-C \log(n)^{1+\eps})$,  where $ \eps>0$ and especially, we get that $\Delta_1$ obeys  the slowest rate among the three terms  $\Delta_1,\Delta_2$ and $\Delta_3$. Hence,
$
P(\|(V^{-1}(\omega_l^\prime))^{1/2}(Z_l-R_l)\|_2\geq x)\leq C\Delta_1(x)
$
and we get
$$
P(\max_l \chi(l)^{[u,v]}\geq t)=P(\max_l \|\tilde Z_l\|_2^2+W_n \geq t),
$$
where $W_n$ denotes the error and 
$$P(|W_n|>x)\leq C P(\|(V^{-1}(\omega_l^\prime))^{1/2}(Z_l-R_l)\|_2^2>x) \leq C \Delta_1(\sqrt{x}).$$

Now we apply Lemma~\ref{lem.gaussian.gt} and  we obtain with $G_d(t)=(1-(1-\exp(-t/2))^d$,
\begin{align*}
    \sup_{0\leq t   \leq h_n} \big| &  \frac{P(\max_l\tilde \chi(l)^{[u,v]}\geq t)}{G_d(t)} -1 \big|  \leq C \Delta_1\big(\log(n^{d_\xi})^{-\zeta_\xi/2+3/4}\big)\\
    & \ \  \times \max(2,\exp(h_n/2))+ 2 \log(n^{d_\xi})^{-\zeta_\xi+3/2} \\
    &\leq C \max(2,\exp(h_n/2)) \log(n)^{C_\tau}\Big(\gpbetade{\log(n)} \\
    & \ \ +(n^{1/2(1-1/4a)-1/\tau(2-a)-2(1+a)/\tau^2})^{-\tau}+n^{-\tau/4(1+a)+1}\Big)
    + 2 \log(n^{d_\xi})^{-\zeta_\xi+3/2}.
\end{align*}
\end{proof}

\begin{lem} \label{lem.spec.est.cov}
Let $\fspc(\omega)=\frac{M}{n}  \sum_{k=1}^{n} \kappa_M(\omega-\omega_k)  (Z(\omega_k)Z^{H}(\omega_k)-\Sigma_n(\omega_k))$ and $u_1,u_2,v_1,v_2\in\{1,\dots,p\}$. Under Assumption~\ref{asp.moments},\ref{asp.kernel} and \ref{asp.spectral}, we have 
\begin{align*}
\sup_\omega|&n/M\cov(e_{v_1}^\top \fspc(\omega) e_{v_2}, e_{u_1}^\top \fspc(\omega) e_{u_2}) \\
&-C_{K_2} \Big(f_{v_1,u_1}(\omega)f_{u_2,v_2}(\omega)+ \ind(\omega \in \pi \Z)f_{v_1,u_2}(\omega)f_{v_2,u_1}(\omega) \Big)|=O(1/M)
\end{align*}
and
\begin{align*}
\sup_\omega |&n/M\cov(e_{v_1}^\top f^{-1}(\omega) \fspc(\omega) f^{-1}(\omega) e_{v_2}, e_{u_1}^\top f^{-1}(\omega) \fspc(\omega) f^{-1}(\omega) e_{u_2})\\
&- C_{K_2} \Big(f_{v_1,u_1}^{-1}(\omega)f_{u_2,v_2}^{-1}(\omega)+ \ind(\omega \in \pi \Z)f_{v_1,u_2}^{-1}(\omega)f_{v_2,u_1}^{-1}(\omega) \Big)|=O(1/M).     
\end{align*}
\end{lem}
\begin{proof}
Let $\tilde f_{ax_1+by_1,cx_2+dy_2}:=(a e_{v_1}^\top+be_{u_1}^\top) \fspc(\omega) (c^{(C)} e_{v_2}+d^{(C)} e_{u_2})$.
Note that $E \fspc(\omega)=0$ and $\cov(\tilde f_{x_1,x_2}, \tilde f_{y_1,y_2})=E e_{v_1}^\top \fspc(\omega) e_{v_2} e_{u_2}^\top \fspc(\omega) e_{u_1}.$ Furthermore, we have 
\begin{align*}
4&(\var \tilde f_{x_1,x_2}+\var \tilde f_{y_1,y_2}+\var \tilde f_{x_1,y_2}+\var \tilde f_{y_1,x_2})-\\&
\var \tilde f_{x_1-y_1,x_2+y_2}-\var \tilde f_{x_1+y_1,x_2-y_2}-\var \tilde f_{ix_1-y_1,ix_2+y_2}-\var \tilde f_{ix_1+y_1,ix_2-y_2}-\\&i(\var \tilde f_{x_1-y_1,x_2-iy_2}+
\var \tilde f_{x_1+y_1,x_2+iy_2}-\var \tilde f_{x_1-iy_1,x_2-y_2}-\var \tilde f_{x_1+iy_1,x_2+iy_2})
\\=&8Ee_{v}^\top \fspc(\omega) e_{v} e_{u}^\top \fspc(\omega) e_{u}=8\cov(\tilde f_{x_1,x_2}, \tilde f_{y_1,y_2}).
\end{align*}

The sesquilinearity of the spectral density implies $\var(\tilde f_{ax_1+by_1,cx_2+dy_2})=C_{K_2}[(|a|^2 \tilde f_{x_1,x_1}+|b|^2\tilde f_{y_1,y_1}+ab^{(C)}\tilde f_{x_1,y_1}+ba^{(C)}\tilde f_{y_1,x_1})(|c|^2 \tilde f_{x_2,x_2}+|d|^2\tilde f_{y_2,y_2}+cd^{(C)}\tilde f_{x_2,y_2}+dc^{(C)}\tilde f_{y_2,x_2})+\ind(\omega \in \pi \Z)(ac^{(C)} \tilde f_{x_1,x_2}+bd^{(C)}\tilde f_{y_1,y_2}+bc^{(C)}\tilde f_{x_1,y_2}+ad^{(C)}\tilde f_{y_1,x_2})^2]$. Then, by  Lemma~\ref{lem.spec.var.rate} and   some algebra, we get 
$\cov(\tilde f_{x_1,x_2}, \tilde f_{y_1,y_2})=\tilde f_{x_1,y_1} \tilde f_{y_2,x_2}+\ind(\omega \in \pi \Z) \tilde f_{x_1,y_2} \tilde f_{x_2,y_1}$ from which the first assertion of the lemma  follows. 
For the second assertion, note that 
\begin{align*}
\cov(e_{v_1}^\top f^{-1}(\omega) \fsp(\omega) f^{-1}(\omega) e_{v_2}, e_{u_1} f^{-1}(\omega) \fsp(\omega) f^{-1}(\omega) e_{u_2})
=\\
\sum_{s_1,s_2=1}^p \sum_{t_1,t_2=1}^p f_{u_1,s_1}^{-1}f_{s_2,u_2}^{-1}f_{t_1,v_1}^{-1}f_{v_2,t_2}^{-1}
\cov( e_{s_1}^\top \fsp(\omega) e_{s_2} , e_{t_1}^\top \fsp(\omega) e_{t_2}). \end{align*}
Inserting in the above expression the first assertion of the lemma  leads to its second assertion. 
\end{proof}

\begin{lem} \label{lem.var.rho}
Under the conditions of Lemma~\ref{lem.partial.coherence.first.approximation} we have 
$$
\max_l \|\frac{n}{M}\var
\begin{pmatrix}
 \Re(1/n T_{n,u,v}(l)-1/2 \rho_{u,v}(\omega_l^\prime)[1/n T_{n,v,v}(l)+1/n T_{n,u,u}(l)) \\
 \Im(1/n T_{n,u,v}(l)-1/2 \rho_{u,v}(\omega_l^\prime)[1/n T_{n,v,v}(l)+1/n T_{n,u,u}(l))
 \end{pmatrix}
 -\Sigma(l)\|_2=O(1/M),
$$
where
\begin{align*}
    V(\omega)=\frac{C_{K_2}(1-|\rho_{u,v}(\omega)|^2)}{2}\begin{pmatrix}
    1-\Re(\rho_{u,v}(\omega))^2 & -\Re(\rho_{u,v}(\omega))\Im(\rho_{u,v}(\omega)) \\
    -\Re(\rho_{u,v}(\omega))\Im(\rho_{u,v}(\omega)) & 1-\Im(\rho_{u,v}(\omega))^2 \\
    \end{pmatrix}.
\end{align*}
\end{lem}
\begin{proof}
By Lemma~\ref{lem.spec.est.cov} and the definition of $T_{n,u,v}(l)$ for $l,l_2=1,\dots,M/N-1$, we have that  
$$
|\frac{1}{Mn}\cov(T_{n,u_1,v_1}(l), T_{n,u_2,v_2}(l_2))-  \frac{\ind(l=l_2) C_{K_2}f^{-1}_{v_2,v_1}(\omega_l^\prime)f^{-1}_{u_1,u_2}(\omega_l^\prime)}{(f^{-1}_{u_1,u_1}(\omega_l^\prime)f^{-1}_{u_2,u_2}(\omega_l^\prime)f^{-1}_{v_1,v_1}(\omega_l^\prime)f^{-1}_{v_2,v_2}(\omega_l^\prime))^{1/2}}|=O(\frac{1}{M}). 
$$
Both assertions of the lemma  then  follow by sesquilinarity of the covariance. 
\end{proof}

\section{Proofs of Main Results} \label{sec.additional.proofs}

\begin{proof}[Proof of Theorem~\ref{thm.partial.coherence}]
The assertion follows by Lemma~\ref{lem.partial.coherence.first.approximation}, Lemma~\ref{lem.var.rho} and following the  arguments of Lemma~\ref{lem.gaussian.approximation} for $d=1$.
\end{proof}

\begin{proof}[Proof of Theorem~\ref{thm.partial.coherence.max}]
The assertion follows by Lemma~\ref{lem.gaussian.approximation} with $h_n=2(a+2b)\log(n)$ and Assumption~\ref{asp.rates}.
\end{proof}

\begin{proof}[Proof of Corollary~\ref{cor.coherence.cov}]
The assertions follows by Lemma~\ref{lem.partial.coherence.first.approximation} and Lemma~\ref{lem.spec.est.cov}. 
\end{proof}

\begin{proof}[Proof of Theorem~\ref{thm.test-pc}]
Under $H_0$, we have \begin{align}
    P(T_n^{[u,v]}\geq x) \leq P\Big(\max_l
\frac{n}{M} \begin{pmatrix}
    \Re(\hat \rho^{(de)}_{u,v}(\omega_l^\prime)-\rho_{u,v}(\omega_l^\prime)) \\
    \Im(\hat \rho^{(de)}_{u,v}(\omega_l^\prime)-\rho_{u,v}(\omega_l^\prime)) \\
    \end{pmatrix}
    \hat V^{-1}(\omega_l^\prime)
    \begin{pmatrix}
    \Re(\hat \rho^{(de)}_{u,v}(\omega_l^\prime)-\rho_{u,v}(\omega_l^\prime)) \\
    \Im(\hat \rho^{(de)}_{u,v}(\omega_l^\prime)-\rho_{u,v}(\omega_l^\prime)) \\
    \end{pmatrix}
 \geq x\Big). \label{eq.H0.Tn}
\end{align}
 To see this, let $\tilde \omega \in \{\omega_l^\prime,l\in \mathcal{L}\}$ such that $|\hat \rho^{(de)}_{u,v}(\tilde \omega)|>\delta$. If  such a $\tilde \omega$ does not exist, then $T_n^{[u,v]}=0$ and the above statement holds true. Since $\hat V^{-1}(\omega_l^\prime)$ is positive (semi)-definite for all $l$, it  suffices to show that $|\hat \rho^{(de)}_{u,v}(\tilde \omega)-\delta \exp(i\arg(\hat \rho^{(de)}_{u,v}(\tilde \omega)))|\leq |\hat \rho^{(de)}_{u,v}(\tilde \omega)-\rho_{u,v}(\tilde \omega)|$. Let $\hat \rho^{(de)}_{u,v}(\tilde \omega)=\delta^\prime \exp(i\lambda^\prime)$ and $\rho_{u_v}(\tilde \omega)=\tilde \delta\exp(i \tilde \lambda)$, where $\delta^\prime,\tilde \delta \in [0,1]$. Under $H_0$ and since $|\hat \rho^{(de)}_{u,v}(\tilde \omega)|>\delta$, we have $\delta^\prime>\delta$ and $\tilde \delta\leq \delta$. Then, 
$|\hat \rho^{(de)}_{u,v}(\tilde \omega)-\delta \exp(i\arg(\hat \rho^{(de)}_{u,v}(\tilde \omega)))|=|\delta^\prime \exp(i \lambda^\prime)-\delta \exp(i\lambda^\prime)|=\delta^\prime-\delta$. Furthermore, 
$|\hat \rho^{(de)}_{u,v}(\tilde \omega)-\rho_{u,v}(\tilde \omega)|\geq |\hat \rho^{(de)}_{u,v}(\tilde \omega)|-|\rho_{u,v}(\tilde \omega)|=\delta^\prime-\tilde \delta \geq \delta^\prime-\delta$. This implies \eqref{eq.H0.Tn}. The assertion follows then by Lemma~\ref{lem.gaussian.approximation}.
\end{proof}

\begin{proof}[Proof of Theorem~\ref{th.FDR_1}]
Let $H_0=|\mathcal{H}_0|\leq q$ and set $b_n=2 \log(qd)$. Note first that by the inequalities  $(1+x)^d\leq 1/(1-dx)$ for $x\in [-1,0]$ and $(1+x)^d \geq 1+dx$ for $x\geq -1$, we have  the following  bounds: $1\leq q G_d(b_n)=q (1-(1-1/(qd))^d)\geq q/(q+1)\geq 1/2$. Recall  that 
\begin{align*}
    \widehat t = \operatorname{inf} \Big\{ 0 \leq t \leq 2 \log(dq) : \frac{G_d(t) q}{\max(1,\sum_{ (u,v) \in \mathcal{Q}} \ind(T_n^{(u,v)} \geq t))} \leq \alpha \Big\}.
\end{align*}

Our goal is to show that 
$$E \left( \frac{\sum_{(u,v) \in \mathcal{H}_0} \ind(T_n^{[u,v]} \geq \hat t)}{\max(\sum_{(u,v) \in \mathcal{Q}} \ind(T_n^{[u,v]} \geq \hat t),1)}\right)=E\left( \frac{\sum_{(u,v) \in \mathcal{H}_0} \ind(T_n^{[u,v]} \geq \hat t) G_d(\hat t) q}{\max(\sum_{(u,v) \in \mathcal{Q}} \ind(T_n^{[u,v]} \geq \hat t),1) G_d(\hat t) q}\right)\leq \alpha.$$
Consider the expression  $\max(\sum_{(u,v) \in \mathcal{Q}} \ind(T_n^{[u,v]} \geq \hat t),1)$. 
Let $\mathcal{H} (2 \sqrt{M/n b_n})=\mathcal{H}_a$. We have by assumption that $|\mathcal{H}_a|\geq \log(\log(n))$. Furthermore, 
$$\sum_{(u,v) \in \mathcal{Q}} \ind(T_n^{(u,v)} \geq b_n)\geq \sum_{(u,v) \in \mathcal{H}_a} \ind(T_n^{(u,v)} \geq b_n).
$$ 
We then have, for $(u,v) \in \mathcal{H}_a$, that   
\begin{align*}
    P&(T_n^{(u,v)} \geq b_n) = P((T_n^{(u,v)})^{1/2} \geq \sqrt{b_n}) \\
    &\geq P(\sqrt{n/M}\max_l \inf_\lambda (\lambda_{\min} (\hat V^{-1}(\omega_l^\prime)))^{1/2} |\hat \rho^{(de)}_{u,v}(\omega_l^\prime)-\delta \exp(i \lambda)| \geq \sqrt{b_n})\\
    &\geq P(\sqrt{n/M}\max_l (\lambda_{\min} (\hat V^{-1}(\omega_l^\prime)))^{1/2} (|\rho_{u,v}(\omega_l^\prime)|-\delta)\\
    &\qquad \qquad 
    -\sqrt{n/M}\max_l (\lambda_{\min} (\hat V^{-1}(\omega_l^\prime)))^{1/2}|\hat \rho^{(de)}_{u,v}(\omega_l^\prime)-\rho_{u,v}(\omega_l^\prime)|
    \geq \sqrt{b_n})\\
    &\geq P(\sqrt{n/M}\max_l (\lambda_{\min} (\hat V^{-1}(\omega_l^\prime)))^{1/2} (|\rho_{u,v}(\omega_l^\prime)|-\delta)-
    (n/M \max_l  \chi(l)^{[u,v]})^{1/2}
    \geq \sqrt{b_n})\\
    &\geq 1-P(\max_l \chi_l^{[u,v]} \geq b_n) \\
    &\geq 1- G_d(b_n)+(P(\max_l \chi_l^{[u,v]} \geq b_n)-G_d(b_n))\geq 1-(1+o(1))/q,
\end{align*}
where the second to last inequality follows by Lemma~\ref{lem.gaussian.approximation}. Note that $V(\omega_l^\prime)$ has the eigenvalues $
C_{K_2}(1-|\rho_{u,v}(\omega_l^\prime)|^2)/2
$
and $C_{K_2}(1-|\rho_{u,v}(\omega_l^\prime)|^2)^2/2$, which imply that  $\lambda_{\min}(\hat V^{-1}(\omega_l^\prime)) = 2(C_{K_2}(1-|\hat \rho_{u,v}(\omega_l^\prime)|^2))^{-1}\geq 2$.

This means that for $n$ large enough,  we have for some generic constant $C>0$
\begin{align*}
    P\Big(\sum_{(u,v) \in {\mathcal H}_a} \ind(T_n^{[u,v]} \geq b_n) \geq \log(\log(n))\Big)\geq 1 - |H_a| C /q=1-C(1-|\mathcal{H}_0|/q).
\end{align*}
By assumption $1-|\mathcal{H}_0|/q=o(1)$ and therefore,  it yields 
 with high probability that $\sum_{(u,v) \in \mathcal{H}_a} \ind(T_n^{(u,v)} \geq b_n)\geq \log(\log(n))$. Together with $q G_d(b_n)\leq 1$ we get for $n$ large enough  with high probability, that, 
$$\frac{G_d(b_n) q}{\max(1,\sum_{(u,v) \in \mathcal{Q}} \ind(T_n^{(u,v)} \geq b_n))}\leq C \log(\log(n))^{-1} < \alpha.$$
  Consequently, $P(\hat t \in [0,b_n))\to 1$ for $n\to \infty$. 

If $\hat t$ is chosen so that $\hat t< b_n$, then $(G_d(\hat t) q)/(\sum_{(u,v) \in \mathcal{Q}} \ind(T_n^{[u,v]} \geq \hat t)) \leq \alpha$. Hence, we have by using the definition of $\hat t$ 
\begin{align*}
    E& \left( \frac{\sum_{(u,v) \in \mathcal{H}_0} \ind(T_n^{[u,v]} \geq \hat t)}{\max(\sum_{(u,v) \in \mathcal{Q}} \ind(T_n^{[u,v]} \geq \hat t),1)}\right)\\
    =&E \left(\ind(\hat t < b_n)  \frac{\sum_{(u,v) \in \mathcal{H}_0} \ind(T_n^{[u,v]} \geq \hat t)}{ G_d(\hat  t) q}
    \frac{G_d(\hat  t) q}{\max(\sum_{(u,v) \in \mathcal{Q}} \ind(T_n^{[u,v]} \geq \hat  t),1)}
        \right)\\
    &+ E \left( \ind(\hat t \geq b_n) \frac{\sum_{(u,v) \in \mathcal{H}_0} \ind(T_n^{[u,v]} \geq \hat t)}{\max(\sum_{(u,v) \in \mathcal{Q}} \ind(T_n^{[u,v]} \geq \hat t),1)}\right)\\
    &\leq \alpha/q\sum_{(u,v) \in \mathcal{H}_0} \sup_{0\leq t < b_n}  \frac{P(T_n^{[u,v]} \geq t)  }{G_d(t) }+ P(\hat t = b_n) \\
    &\leq \alpha/q\sum_{(u,v) \in \mathcal{H}_0} \sup_{0\leq t < b_n}  \frac{P(\max_l 1/(Mn)\chi(l)^{[u,v]}\geq t) }{G_d(t) }+ P(\hat t = b_n) \\
    &= \alpha/q|\mathcal{H}_0| \sup_{0\leq t < b_n}  \frac{P(\max_l \|Z_l\|_2+W\geq t)}{G_d(t) }+ o(1) = \alpha/q|\mathcal{H}_0|+o(1),
\end{align*}
where for the second to last inequality note that we have, by the arguments of the proof of Theorem~\ref{thm.test-pc}, that,  for $1\leq u < v \leq p$  
\begin{align*}
P_{H_0}(T_n^{[u,v]} \geq t) &\leq P\Big(\max_l
\frac{n}{M} \begin{pmatrix}
    \Re(\hat \rho^{(de)}_{u,v}(\omega_l^\prime)-\rho_{u,v}(\omega_l^\prime)) \\
    \Im(\hat \rho^{(de)}_{u,v}(\omega_l^\prime)-\rho_{u,v}(\omega_l^\prime)) \\
    \end{pmatrix}
    \hat V^{-1}(\omega_l^\prime)
    \begin{pmatrix}
    \Re(\hat \rho^{(de)}_{u,v}(\omega_l^\prime)-\rho_{u,v}(\omega_l^\prime)) \\
    \Im(\hat \rho^{(de)}_{u,v}(\omega_l^\prime)-\rho_{u,v}(\omega_l^\prime)) \\
    \end{pmatrix}
 \geq {t}\Big) \\
 &=P(\max_l \frac{1}{Mn}\chi(l)^{[u,v]}\geq t).
\end{align*}
Furthermore,  for the last equality note that  by Lemma~\ref{lem.gaussian.approximation}
\begin{align*} 
    \sup_{0\leq t \leq b_n}& \left|\frac{P(\max_l 1/(Mn)\chi(l)^{[u,v]}\geq t)}{G_d(t)}-1\right| \leq C((\log(n^{d_\xi}))^{-\zeta_\xi+3/2}+qd\log(n)^{C_\tau} \\
    &\times \Big[\gpbetade{\log(n)} + (n^{1/2(1-1/4a)-1/\tau(2-a)-2(1+a)/\tau^2})^{-\tau}\Big])
\end{align*}
because of Assumption~\ref{asp.rates}. The assertion of the theorem follows then because  $q/|\mathcal{H}_0|=1+o(1)$. 
\end{proof}

\section{An Estimator of $f^{-1}$}\label{sec.est.inv.f}
In this section we  elaborate on  an example of an estimator of $ f^{-1}$ which satisfies  the  requirements of Assumption~\ref{asp.spectral.density.estimator}, we  consider  a  CLIME type  estimator as in \cite{fiecas2019spectral}, see also \cite{cai2011constrained} for details about CLIME type estimators. Notice that similar results can be obtained  using estimators based on  graphical lasso or other types of regularization.
Let $\operatorname{CLIME}_\lambda(A)$ denote the CLIME estimator applied  to an  input matrix $A$ with  tuning parameter $\lambda$ and obtained as   the solution of the optimization problem $\min \|B\|_1$ such that  $\|AB-I\|_{\max}\leq \lambda$,  plus a possible correction for  symmetry. Then, an estimator of  the inverse  spectral density matrix  at frequency $\omega$  based on the lag-window estimator $\fsp$ and using the tuning parameter $\lambda$,  is   given by 
\begin{align} \label{eq.fhat.inverse}
   \hat f^{-1}(\omega):=(1,i)\left(
  \operatorname{CLIME}_\lambda\left(
    \begin{pmatrix}
   Re(\fsp(\omega)) & Im(\fsp(\omega)) \\
    -Im(\fsp(\omega)) & Re(\fsp(\omega))
   \end{pmatrix}\right)\right)\begin{pmatrix}
    1\\
   -i
   \end{pmatrix}/2;
\end{align}
with   $\fsp(\omega)$ as  given in  (\ref{eq.f-smooth-per}). 
\cite{zhang2020convergence},  Theorem 5.1 and Remark 4, showed that the above   estimator 
satisfies, 
\begin{align}
P(\sup_\omega \|\hat f^{-1}(\omega)-f^{-1}(\omega)\|_{\max} \geq x ) \leq \gps {x/\sup_\omega \|f^{-1}(\omega)\|_1^2}. \label{eq.rates.max.spec.inv}
\end{align}
For the class of  so-called weak sparse inverse spectral density matrices, the above result also can be extended to other matrix norms. To elaborate,   let  $\mathcal{G}_r(w(p))$ be  the set of  inverse spectral density matrices which are weakly sparse within a small $\ell_r$ ball, that is, 
\begin{align}
\mathcal{G}_r(w(p)):=\Bigg\{& f^{-1} : [0,2\pi] \to \C^{p\times p}\, \Bigg| \, \sup_\omega \max_j \sum_{i=1}^p | f^{-1}_{i,j}(\omega)|^r \leq w(p)  \Bigg\}. \label{eq.finv.sparse}
\end{align}
Then, $\sup_\omega \|f^{-1}(\omega)\|_1 \leq C w(p)$ and by the arguments of \cite{cai2011constrained}, 
it can be shown  that  for all $f^{-1}\in \mathcal{G}_r(w(p))$ and  any $l \in [1,\infty]$,
\begin{align}
   P(\sup_\omega \|\hat f^{-1}(\omega)-f^{-1}(\omega)\|_l 
\geq x ) & \leq \gps {\frac{x}{w(p)\sup_\omega \|f^{-1}(\omega)\|_1^2}} \nonumber \\
   & \leq \gps {x/w(p)^3} . \label{eq.rates.spec.inv.freq}
\end{align}

\section{Additional Simulation Results}
\FloatBarrier

\begin{figure}[!h]
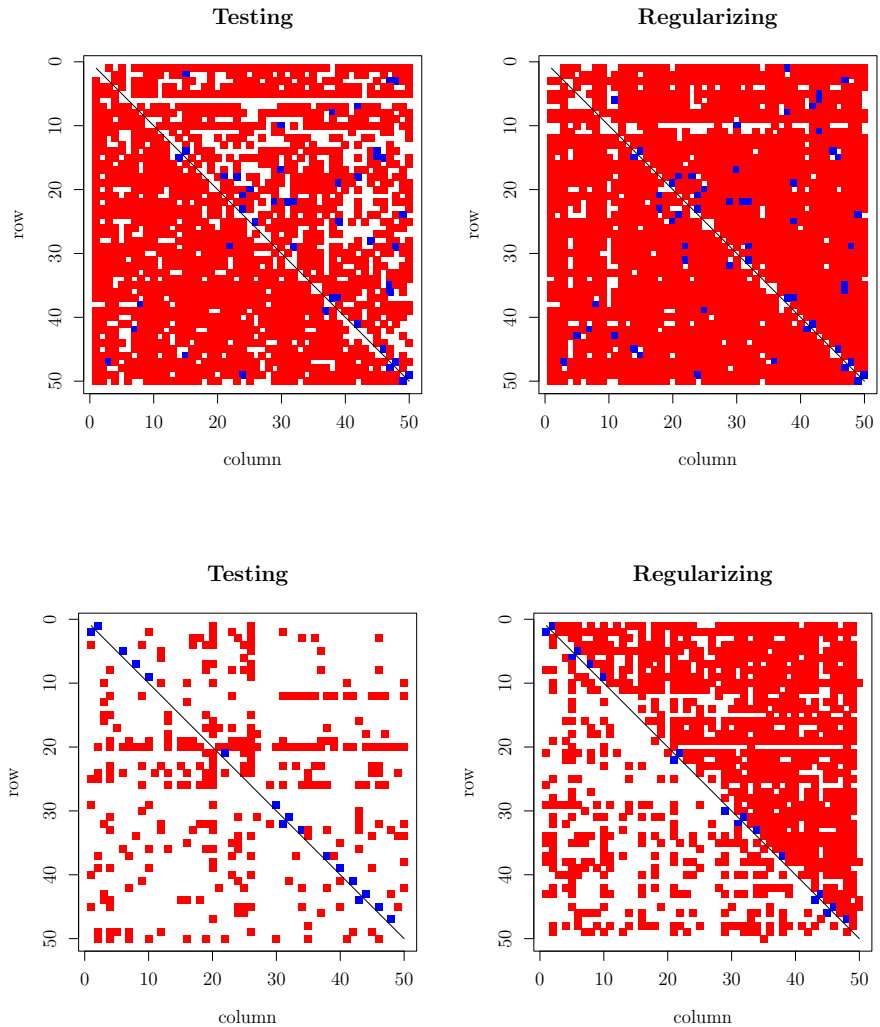

\begin{center}
    VARMA$(1,1)$ process\\
    \vspace{-0.5cm}
   \resizebox{0.75\textwidth}{!}{\input{n=512_4096_VARMA_1,1_p50_d=0.00}}\\
  VMA$(5)$ process\\
    \vspace{-0.5cm}
   \resizebox{0.75\textwidth}{!}{\input{n=512_4096_VARMA_0,5_p50_d=0.00}}
   \vspace{-0.75cm}
\end{center}    
    \caption{Detection of  non-zero partial coherences for the VARMA$(1,1)$ process with parameters as  in (a) and the VMA$(5)$ process with parameters as  in (d). The upper triangular part of every plot presents results for  $n=512$ and the lower triangular part for  $n=4096$. A bluish dot represents a not  discovery of a non-zero partial coherence, where the darker the blue color is the lower is the correct detection rate. A reddish dot represents a false discovery of a zero partial coherence, where  the darker the red color is the higher is the false detection rate. The left column of plots present  results for the testing procedure while  
     the right column of plots present results of the  regularizing procedure.}
    \label{fig.Example1}
\end{figure}

We visualize in Figure~\ref{fig.Example1} the results for the VARMA$(1,1)$ process with parameters as specified  in (a) and the VMA$(5)$ process with parameters as in (d)  and  for  $n=512$ and $n=4096$ observations. In this figure, a blank plot  describes the best possible situation while colored dots indicate some form of errors. Highly blue-saturated dots indicate that the underlying positive partial coherence is not correctly detected in most of the cases. Highly red-saturated dots indicate that a non existent partial coherence is identified  as positive.
Consider first the results for the VARMA$(1,1)$ process.
From Figure~\ref{fig.Example1}  we clearly   see the benefits of an increasing  sample size. The higher power of {\methoda} is recognized  by the far fewer blue dots appearing in the left part of Figure~\ref{fig.Example1} compared to those in the right part  of the same figure. Since red dots do not appear in the plots corresponding to {\methoda}, it seems that this method does not produce any  systematic error in identifying non existent partial coherences. To elaborate, the three highest false detection rates for an individual partial coherence in both sample sizes together are about $3.9\%$, $ 1.6\%$ and  $1.4\%$. In contrast to this, {\methodb}  seems to have a systematic error in identifying  zero partial coherences  as non  zero. For the sample size $n=512$, the three largest false detection rates are $99.9\%$, $98.7\%$, and $98.7\%$. Even for the sample size of $n=4096$ observations, some zero partial coherences are wrongly detected by this method in $100\%$ of the cases. For the VMA$(5)$ processes, both approaches give reasonable results and no approach seems to have  a systematic error. However, {\methoda} outperforms {\methodb} in terms of power as it can be seen in this figure by the overall smaller number of blue  dots of   {\methoda} compared to those of {\methodb}.

Table~\ref{table.2} reports size and power results 
obtained using the simulation set up described in Section~\ref{sec.simulations} for the case $\delta=0.2$. 
\begin{table}[t]
\spacingset{1}
\centering
\begin{tabular}{|cc|cc|cc|cc|}
\hline
  & &\multicolumn{4}{c|}{\methoda} & \multicolumn{2}{c|}{\methodb}\\ 
 & &\multicolumn{2}{c}{$\alpha=0.05$} & \multicolumn{2}{c|}{$\alpha=0.1$} & \multicolumn{2}{c|}{} \\
 $p$& $n$ &  FDR & Power & FDR & Power & FDR & Power  \\ 
 \hline
 \multicolumn{8}{|c|}{DGP: VARMA $(1,1)$}\\ \hline
   \multirow{4}{1cm}{50}&190  & 0.00(0.02) & 0.37(0.02) & 0.00(0.02) & 0.40(0.02) & 0.00(0.02) & 0.32(0.02) \\ 
  &512  & 0.00(0.02) & 0.57(0.03) & 0.00(0.02) & 0.59(0.03) & 0.05(0.03) & 0.55(0.03) \\ 
  &2048  & 0.00(0.00) & 0.81(0.02) & 0.00(0.00) & 0.81(0.02) & 0.19(0.03) & 0.72(0.02) \\ 
  &4096  & 0.00(0.00) & 0.87(0.02) & 0.00(0.00) & 0.87(0.02) & 0.16(0.03) & 0.80(0.02) \\ 
   \hline
\multirow{4}{1cm}{100}&190    & 0.00(0.00) & 0.11(0.02) & 0.00(0.00) & 0.11(0.02) & 0.00(0.01) & 0.06(0.01) \\ 
  &512    & 0.00(0.00) & 0.30(0.02) & 0.00(0.00) & 0.31(0.02) & 0.04(0.02) & 0.36(0.03) \\ 
  &2048   & 0.00(0.00) & 0.53(0.02) & 0.00(0.00) & 0.54(0.02) & 0.07(0.02) & 0.62(0.02) \\ 
  &4096   & 0.00(0.00) & 0.63(0.02) & 0.00(0.00) & 0.64(0.02) & 0.09(0.01) & 0.71(0.01) \\ 
   \hline
\multirow{4}{1cm}{200}&190  & 0.07(0.14) & 0.41(0.02) & 0.07(0.15) & 0.42(0.02) & 0.00(0.00) & 0.21(0.02) \\ 
  &512    & 0.00(0.03) & 0.62(0.01) & 0.00(0.03) & 0.63(0.01) & 0.06(0.01) & 0.63(0.01) \\ 
  &2048   & 0.00(0.00) & 0.80(0.01) & 0.00(0.00) & 0.81(0.01) & 0.08(0.01) & 0.70(0.01) \\ 
  &4096   & 0.00(0.00) & 0.84(0.01) & 0.00(0.00) & 0.84(0.01) & 0.11(0.01) & 0.76(0.01) \\ 
   \hline
\multicolumn{8}{|c|}{DGP: VMA $(5)$}\\ \hline
\multirow{4}{1cm}{50} &190   & 0.00(0.00) & 0.39(0.04) & 0.00(0.00) & 0.39(0.04) & 0.00(0.01) & 0.26(0.05) \\ 
   &512   & 0.00(0.00) & 0.49(0.03) & 0.00(0.00) & 0.49(0.04) & 0.00(0.00) & 0.49(0.04) \\ 
   &2048  & 0.00(0.00) & 0.72(0.03) & 0.00(0.00) & 0.72(0.03) & 0.00(0.00) & 0.70(0.05) \\ 
   &4096  & 0.00(0.00) & 0.82(0.03) & 0.00(0.00) & 0.83(0.03) & 0.01(0.02) & 0.85(0.04) \\ 
   \hline
\multirow{4}{1cm}{100} &190   & 0.01(0.05) & 0.11(0.01) & 0.01(0.06) & 0.11(0.01) & 0.00(0.00) & 0.03(0.01) \\ 
   &512   & 0.00(0.00) & 0.18(0.01) & 0.00(0.00) & 0.19(0.01) & 0.00(0.00) & 0.16(0.02) \\ 
   &2048  & 0.00(0.00) & 0.39(0.01) & 0.00(0.00) & 0.40(0.01) & 0.00(0.00) & 0.39(0.02) \\ 
   &4096  & 0.00(0.00) & 0.54(0.01) & 0.00(0.00) & 0.55(0.01) & 0.00(0.00) & 0.53(0.02) \\ 
   \hline
\multirow{4}{1cm}{200} &190    & 0.00(0.00) & 0.04(0.00) & 0.00(0.00) & 0.04(0.00) & 0.00(0.00) & 0.02(0.00) \\ 
   &512    & 0.00(0.00) & 0.08(0.00) & 0.00(0.00) & 0.09(0.01) & 0.00(0.00) & 0.09(0.01) \\ 
   &2048   & 0.00(0.00) & 0.18(0.01) & 0.00(0.00) & 0.19(0.01) & 0.00(0.00) & 0.20(0.01) \\ 
   &4096   & 0.00(0.00) & 0.28(0.01) & 0.00(0.00) & 0.28(0.01) & 0.00(0.00) & 0.31(0.01) \\ 
  \hline
\end{tabular}
\caption{Empirical false discovery rate and power for the case $\delta=0.2$ and for different models, different levels, and different sample sizes.} \label{table.2}
\end{table}
\spacingset{1.9}

As it is seen from the above table,  for the case $\delta=0.2$, {\methoda} has in all situations an empirical FDR of $0\%$. The  reason for this  lies in  the construction of the test statistic and the behavior of $ G_d(t)q$ used in the determination of the threshold value $\widehat{t}$. To elaborate, if   $\delta>0$ and   due to the indicator function appearing in the definition of  $ T^{(u,v)}_n$, the test statistic accumulates under the null hypothesis, 
more  point mass around  zero. In fact, the larger $\delta$ is the less  test statistics  $ T^{(u,v)}_n$ will exceed $\delta$, i.e., the more $T_n^{(u,v)}$ values will be equal to zero.  The asymptotic distribution used  to determine critical values of this test, however, only forms 
an upper bound; also  see the Proof of Theorem~\ref{thm.test-pc}.  Hence  for $\delta>0$, $G_d(t)q$ becomes  a crude estimator of the expected number of falsely rejected null hypotheses and as consequence,  the FDR is forced towards zero. Nevertheless, even under such a  conservative behavior of the empirical FDR,    the empirical power of  {\methoda} outperforms in all cases the corresponding  power of  {\methodb}. 
\FloatBarrier

\section{Multiple Bandwidths}\label{sec.multiple.bandwidth}

In this section, we summarize the main  steps involved in the practical implementation of our inference procedure if different bandwidth are used for each time series. 
A possible adaptive bandwidth rule is given in
\cite{politis2011higher}. 

\begin{enumerate}
    \item[Step 1:] \ Select a filter $\Phi(z)$ to prewhiten  the vector time series at hand.
    \item[Step 2:] \ Select a kernel $K$ and an individual bandwidth (truncation lag)  $M_{u,v}$ for each (cross)-spectral density $f_{u,v}, u,v=1,\dots,p$. 
    \item[Step 3:]  Set $N_{1,u,v}=\lfloor M_{u,v}/N_{u,v}\rfloor$ and use the individual grid of frequencies $\mathcal{L}_{u,v}=\{\omega_l^\prime=l \pi/N_{1,u,v}   \in \mathcal{W} :l \in {1,\dots,N_{1,u,v}-1} \}$ to cover the frequency band $ {\mathcal W}$ of interest, where  $N_{u,v}=\log^{2/r}(M_{u,v})$ and  $r$ is determined by the decay behavior of the Fourier coefficients of the kernel $K$ used; see Assumption 2 in Section 6. Set $\mathcal{L}=\bigcup \mathcal{L}_{u,v}$
    \item[Step 4:] \  Estimate the inverse spectral density matrix $f^{-1}$ at the  frequencies $\omega_l^\prime$ for every $ l \in \mathcal{L}$. 
    \item[Step 5] \  Compute for all $(u,v) \in \mathcal{Q}$ and for all $l \in \mathcal{L}_{u,v}$, the de-biased estimator 
    $\hat \rho_{u,v}^{(de)}(\omega_l^\prime)$ and  the test statistic $T_n^{(u,v)}$, where 
\begin{align*}
     T_n^{(u,v)}=&\ind\big(\max_{l \in \mathcal{L}_{u,v}}  |\hat \rho^{(de)}_{u,v}(\omega_l^\prime)|> \delta\big) \times  \label{eq.test.statistic}\\
     & \max_{l \in \mathcal{L}_{u,v}}\Big\{ \frac{n}{M_{u,v}} \begin{pmatrix}
    \Re(\hat \rho^{(de)}_{u,v}(\omega_l^\prime)-\delta \exp(i \tilde \omega_l)) \\
    \Im(\hat \rho^{(de)}_{u,v}(\omega_l^\prime)-\delta \exp(i \tilde \omega_l)) \\
    \end{pmatrix}
    \hat V^{-1}_{(u,v)}(\omega_l^\prime)
    \begin{pmatrix}
    \Re(\hat \rho^{(de)}_{u,v}(\omega_l^\prime)-\delta \exp(i \tilde \omega_l)) \\
    \Im(\hat \rho^{(de)}_{u,v}(\omega_l^\prime)-\delta \exp(i \tilde \omega_l)) \\
    \end{pmatrix}\Big\}, \nonumber
\end{align*}
$\tilde \omega_l=\arg(\hat \rho^{(de)}_{u,v}(\omega_l^\prime))$, 
and
\begin{align*}
    \hat V^{-1}_{(u,v)}(\omega)=\frac{2}{C_{K_2}(1-|\hat\rho_{u,v}(\omega)|^2)^2}\begin{pmatrix}
    1-\Im(\hat\rho_{u,v}(\omega))^2 & \Re(\hat\rho_{u,v}(\omega)\Im(\hat\rho_{u,v}(\omega) \\
    \Re(\hat \rho_{u,v}(\omega)\Im(\hat\rho_{u,v}(\omega) & 1-\Re(\hat\rho_{u,v}(\omega))^2 \\
    \end{pmatrix}.
\end{align*}
   \item[Step 6:] \  Set $d_{u,v}=|\mathcal{L}_{u,v}|$,  $G_{d_{u,v}}(t)=1-(1-\exp(-t/2))^{d_{u,v}}$ and calculate  the threshold   
    $$\widehat t = \operatorname{inf} \{ 0 \leq t \leq 2 \log(dq) : \frac{\sum_{(u,v) \in \mathcal{Q}} G_{d_{u,v}}(t)}{\max(1,\sum_{ (u,v) \in \mathcal{Q}} \ind(T_n^{(u,v)} \geq t))} \leq \alpha \}.$$
    \item[Step 7:] \  For each  $(u,v) \in \mathcal{Q}$ reject $H_0^{(u,v)}$ if $T_n^{(u,v)} \geq \widehat t$.
\end{enumerate}

\bibliographystyle{apalike}
\bibliography{bib}

\begin{thebibliography}{}

\bibitem[Bach and Jordan, 2004]{bach2004learning}
Bach, F.~R. and Jordan, M.~I. (2004).
\newblock Learning graphical models for stationary time series.
\newblock {\em IEEE Transactions on Signal Processing}, 52(8):2189--2199.

\bibitem[Barber and Cand{\`e}s, 2015]{barber2015controlling}
Barber, R.~F. and Cand{\`e}s, E.~J. (2015).
\newblock Controlling the false discovery rate via knockoffs.
\newblock {\em The Annals of Statistics}, 43(5):2055--2085.

\bibitem[Barigozzi and Farn{\`e}, 2022]{barigozzi2021algebraic}
Barigozzi, M. and Farn{\`e}, M. (2022).
\newblock An algebraic estimator for large spectral density matrices.
\newblock {\em Journal of the American Statistical Association}.

\bibitem[Benjamini and Hochberg, 1995]{benjamini1995controlling}
Benjamini, Y. and Hochberg, Y. (1995).
\newblock Controlling the false discovery rate: a practical and powerful
  approach to multiple testing.
\newblock {\em Journal of the Royal Statistical Society: Series B
  (Methodological)}, 57(1):289--300.

\bibitem[Benjamini and Yekutieli, 2001]{benjamini2001control}
Benjamini, Y. and Yekutieli, D. (2001).
\newblock The control of the false discovery rate in multiple testing under
  dependency.
\newblock {\em The Annals of Statistics}, pages 1165--1188.

\bibitem[Brillinger, 2001]{brillinger2001time}
Brillinger, D. (2001).
\newblock {\em Time Series: Data Analysis and Theory}.
\newblock Classics in Applied Mathematics. Society for Industrial and Applied
  Mathematics.

\bibitem[Cai et~al., 2011]{cai2011constrained}
Cai, T., Liu, W., and Luo, X. (2011).
\newblock A constrained l1 minimization approach to sparse precision matrix
  estimation.
\newblock {\em Journal of the American Statistical Association},
  106(494):594--607.

\bibitem[Cai and Liu, 2016]{cai2016large}
Cai, T.~T. and Liu, W. (2016).
\newblock Large-scale multiple testing of correlations.
\newblock {\em Journal of the American Statistical Association},
  111(513):229--240.

\bibitem[Cai et~al., 2016]{cai2016estimating_overview}
Cai, T.~T., Ren, Z., and Zhou, H.~H. (2016).
\newblock Estimating structured high-dimensional covariance and precision
  matrices: Optimal rates and adaptive estimation.
\newblock {\em Electronic Journal of Statistics}, 10(1):1--59.

\bibitem[Chatterjee, 2014]{chatterjee2014superconcentration}
Chatterjee, S. (2014).
\newblock {\em Superconcentration and related topics}, volume~15.
\newblock Springer.

\bibitem[Chen et~al., 2013]{chen2013covariance}
Chen, X., Xu, M., and Wu, W.~B. (2013).
\newblock Covariance and precision matrix estimation for high-dimensional time
  series.
\newblock {\em The Annals of Statistics}, 41(6):2994--3021.

\bibitem[Dahlhaus, 2000]{dahlhaus2000graphical}
Dahlhaus, R. (2000).
\newblock Graphical interaction models for multivariate time series 1.
\newblock {\em Metrika}, 51(2):157--172.

\bibitem[Dette and Paparoditis, 2009]{dette2009bootstrapping}
Dette, H. and Paparoditis, E. (2009).
\newblock Bootstrapping frequency domain tests in multivariate time series with
  an application to comparing spectral densities.
\newblock {\em Journal of the Royal Statistical Society: Series B (Statistical
  Methodology)}, 71(4):831--857.

\bibitem[Eichler, 2008]{eichler2008testing}
Eichler, M. (2008).
\newblock Testing nonparametric and semiparametric hypotheses in vector
  stationary processes.
\newblock {\em Journal of Multivariate Analysis}, 99(5):968--1009.

\bibitem[Eichler, 2012]{eichler2012graphical}
Eichler, M. (2012).
\newblock Graphical modelling of multivariate time series.
\newblock {\em Probability Theory and Related Fields}, 153(1):233--268.

\bibitem[Einmahl and Mason, 1997]{einmahl1997gaussian}
Einmahl, U. and Mason, D.~M. (1997).
\newblock Gaussian approximation of local empirical processes indexed by
  functions.
\newblock {\em Probability Theory and Related Fields}, 107(3):283--311.

\bibitem[Fiecas et~al., 2019]{fiecas2019spectral}
Fiecas, M., Leng, C., Liu, W., and Yu, Y. (2019).
\newblock Spectral analysis of high-dimensional time series.
\newblock {\em Electronic Journal of Statistics}, 13(2):4079--4101.

\bibitem[Fiecas and Ombao, 2016]{fiecas2016modeling}
Fiecas, M. and Ombao, H. (2016).
\newblock Modeling the evolution of dynamic brain processes during an
  associative learning experiment.
\newblock {\em Journal of the American Statistical Association},
  111(516):1440--1453.

\bibitem[Gray, 2014]{gray2014central}
Gray, D. (2014).
\newblock Central european foreign exchange markets: a cross-spectral analysis
  of the 2007 financial crisis.
\newblock {\em The European Journal of Finance}, 20(6):550--567.

\bibitem[Hannan, 1970]{hannan1970multiple}
Hannan, E.~J. (1970).
\newblock {\em Multiple time series}.
\newblock John Wiley \& Sons.

\bibitem[Javanmard and Montanari, 2014]{javanmard2014confidence}
Javanmard, A. and Montanari, A. (2014).
\newblock Confidence intervals and hypothesis testing for high-dimensional
  regression.
\newblock {\em The Journal of Machine Learning Research}, 15(1):2869--2909.

\bibitem[Jung, 2015]{jung2015graphical}
Jung, A. (2015).
\newblock Learning the conditional independence structure of stationary time
  series: A multitask learning approach.
\newblock {\em IEEE Transactions on Signal Processing}, 63(21):5677--5690.

\bibitem[Koopmans, 1995]{koopmans1995spectral}
Koopmans, L.~H. (1995).
\newblock {\em The spectral analysis of time series}.
\newblock Elsevier.

\bibitem[Krampe and Margaritella, 2021]{krampe2021dynamic}
Krampe, J. and Margaritella, L. (2021).
\newblock Dynamic factor models with sparse var idiosyncratic components.
\newblock {\em arXiv preprint arXiv:2112.07149}.

\bibitem[Krampe and Paparoditis, 2021]{krampe2020Est}
Krampe, J. and Paparoditis, E. (2021).
\newblock Sparsity concepts and estimation procedures for high-dimensional
  vector autoregressive models.
\newblock {\em Journal of Time Series Analysis}, 42(5-6):554--579.

\bibitem[Krampe and Subba~Rao, 2022]{krampe2022inverse}
Krampe, J. and Subba~Rao, S. (2022).
\newblock Inverse covariance operators of multivariate nonstationary time
  series.
\newblock {\em arXiv preprint arXiv:2202.00933}.

\bibitem[Li et~al., 2023]{li2023transfer}
Li, S., Cai, T.~T., and Li, H. (2023).
\newblock Transfer learning in large-scale gaussian graphical models with false
  discovery rate control.
\newblock {\em Journal of the American Statistical Association},
  118(543):2171--2183.

\bibitem[Liu, 2013]{liu2013gaussian}
Liu, W. (2013).
\newblock Gaussian graphical model estimation with false discovery rate
  control.
\newblock {\em The Annals of Statistics}, pages 2948--2978.

\bibitem[Liu and Wu, 2010]{liu2010asymptotics}
Liu, W. and Wu, W.~B. (2010).
\newblock Asymptotics of spectral density estimates.
\newblock {\em Econometric Theory}, pages 1218--1245.

\bibitem[Medkour et~al., 2009]{medkour2009graphical}
Medkour, T., Walden, A.~T., and Burgess, A. (2009).
\newblock Graphical modelling for brain connectivity via partial coherence.
\newblock {\em Journal of Neuroscience Methods}, 180(2):374--383.

\bibitem[Politis, 2003]{politis2003adaptive}
Politis, D.~N. (2003).
\newblock Adaptive bandwidth choice.
\newblock {\em Journal of Nonparametric Statistics}, 15(4-5):517--533.

\bibitem[Politis, 2011]{politis2011higher}
Politis, D.~N. (2011).
\newblock Higher-order accurate, positive semidefinite estimation of
  large-sample covariance and spectral density matrices.
\newblock {\em Econometric Theory}, 27(4):703--744.

\bibitem[Priestley, 1988]{priestley1988spectral}
Priestley, M. (1988).
\newblock The spectral analysis of time series.

\bibitem[{R Core Team}, 2021]{R}
{R Core Team} (2021).
\newblock {\em R: A Language and Environment for Statistical Computing}.
\newblock R Foundation for Statistical Computing, Vienna, Austria.

\bibitem[Rosenblatt, 1985]{rosenblatt2012stationary}
Rosenblatt, M. (1985).
\newblock {\em Stationary sequences and random fields}.
\newblock Springer Science \& Business Media.

\bibitem[Rosuel et~al., 2022]{rosuel2021asymptotic}
Rosuel, A., Loubaton, P., and Vallet, P. (2022).
\newblock On the asymptotic distribution of the maximum sample spectral
  coherence of gaussian time series in the high dimensional regime.
\newblock {\em Journal of Multivariate Analysis}, page 105124.

\bibitem[Schneider-Luftman, 2016]{schneider2016p}
Schneider-Luftman, D. (2016).
\newblock p-value combiners for graphical modelling of eeg data in the
  frequency domain.
\newblock {\em Journal of Neuroscience Methods}, 271:92--106.

\bibitem[Schneider-Luftman and Walden, 2016]{schneider2016partial}
Schneider-Luftman, D. and Walden, A.~T. (2016).
\newblock Partial coherence estimation via spectral matrix shrinkage under
  quadratic loss.
\newblock {\em IEEE Transactions on Signal Processing}, 64(22):5767--5777.

\bibitem[Shao and Wu, 2007]{shao2007local}
Shao, X. and Wu, W.~B. (2007).
\newblock Local whittle estimation of fractional integration for nonlinear
  processes.
\newblock {\em Econometric Theory}, 23(5):899--929.

\bibitem[Sun et~al., 2018]{sun2018large}
Sun, Y., Li, Y., Kuceyeski, A., and Basu, S. (2018).
\newblock Large spectral density matrix estimation by thresholding.
\newblock {\em arXiv preprint arXiv:1812.00532}.

\bibitem[Trujillo et~al., 2017]{trujillo2017effect}
Trujillo, L.~T., Stanfield, C.~T., and Vela, R.~D. (2017).
\newblock The effect of electroencephalogram (eeg) reference choice on
  information-theoretic measures of the complexity and integration of eeg
  signals.
\newblock {\em Frontiers in Neuroscience}, 11:425.

\bibitem[Tugnait, 2022]{tugnait2021sparse}
Tugnait, J.~K. (2022).
\newblock On sparse high-dimensional graphical model learning for dependent
  time series.
\newblock {\em Signal Processing}, 197:108539.

\bibitem[{v}an~de Geer et~al., 2014]{deGeer2014}
{v}an~de Geer, S., B{\"u}hlmann, P., Ritov, Y., and Dezeure, R. (2014).
\newblock On asymptotically optimal confidence regions and tests for
  high-dimensional models.
\newblock {\em The {A}nnals of {S}tatistics}, 42(3):1166--1202.

\bibitem[Wang et~al., 2009]{wang2009shrinkage}
Wang, H., Li, B., and Leng, C. (2009).
\newblock Shrinkage tuning parameter selection with a diverging number of
  parameters.
\newblock {\em Journal of the Royal Statistical Society: Series B (Statistical
  Methodology)}, 71(3):671--683.

\bibitem[Wu, 2005]{wu2005nonlinear}
Wu, W.~B. (2005).
\newblock Nonlinear system theory: Another look at dependence.
\newblock {\em Proceedings of the National Academy of Sciences},
  102(40):14150--14154.

\bibitem[Wu and Wu, 2016]{wu2016performance}
Wu, W.-B. and Wu, Y.~N. (2016).
\newblock Performance bounds for parameter estimates of high-dimensional linear
  models with correlated errors.
\newblock {\em Electronic Journal of Statistics}, 10(1):352--379.

\bibitem[Wu and Zaffaroni, 2018]{wu2018asymptotic}
Wu, W.~B. and Zaffaroni, P. (2018).
\newblock Asymptotic theory for spectral density estimates of general
  multivariate time series.
\newblock {\em Econometric Theory}, 34(1):1--22.

\bibitem[Xu et~al., 2022]{yi-22}
Xu, H., Wang, D., Zhao, Z., and Yu, Y. (2022).
\newblock Change point inference in high-dimensional regression models under
  temporal dependence.
\newblock {\em https://arxiv.org/pdf/2207.12453.pdf}.

\bibitem[Zhang and Zhang, 2014]{zhang2014}
Zhang, C.-H. and Zhang, S.~S. (2014).
\newblock Confidence intervals for low dimensional parameters in high
  dimensional linear models.
\newblock {\em Journal of the Royal Statistical Society: Series B (Statistical
  Methodology)}, 76(1):217--242.

\bibitem[Zhang and Wu, 2021]{zhang2020convergence}
Zhang, D. and Wu, W.~B. (2021).
\newblock Convergence of covariance and spectral density estimates for
  high-dimensional locally stationary processes.
\newblock {\em The Annals of Statistics}, 49(1):233--254.

\end{thebibliography}

\end{document}